\documentclass[twocolumn,showpacs,amsmath,amssymb,nofootinbib,floatfix]{revtex4}
\usepackage{graphicx}
\usepackage{dcolumn}%
\pdfoutput=1
\usepackage{color}
\usepackage{amsmath}
\usepackage{amssymb}
\usepackage{bm}
\usepackage{setspace}

\begin{document}
\title{Multiscale approach to the physics of radiation damage with ions}
\author{Eugene Surdutovich$^{1,2}$ and Andrey V. Solov'yov$^{2,3}$}
\affiliation{
$^1$Department
of Physics, Oakland University, Rochester, Michigan 48309, USA\\
$^2$Frankfurt Institute for Advanced Studies,
Ruth-Moufang-Str. 1, 60438 Frankfurt am Main, Germany\\
$^3$On leave from A.F. Ioffe Physical Technical Institute, 
St. Petersburg, Russian Federation
}
\date{\today}
\begin{abstract}
The multiscale approach to the assessment of biodamage resulting upon irradiation of biological media with ions is reviewed, explained and compared to other approaches. The processes of ion propagation in the medium concurrent with ionization and excitation of molecules, transport of secondary products, dynamics of the medium, and biological damage take place on a number of different temporal, spatial and energy scales.
The multiscale approach, a physical phenomenon-based analysis of the scenario that leads to radiation damage, has been designed to consider all relevant effects on a variety of scales and develop an approach to the quantitative assessment of biological damage as a result of irradiation with ions. This paper explains the scenario of radiation damage with ions, overviews its major parts, and applies the multiscale approach to different experimental conditions. On the basis of this experience, the recipe for application of the multiscale approach is formulated. The recipe leads to the calculation of relative biological effectiveness.
\end{abstract}


\maketitle

\section{Introduction}
\label{intro}

The physics and chemistry of radiation damage caused by irradiation
with protons and heavier ions has recently become a subject of intense interest
because of the use of ion beams in cancer therapy~\cite{MSAreview,Radam09editorial,Kraft07,SchardtRMP10,Durante10}.
Ion-beam cancer therapy (IBCT) was first realised in the 1950s as proton-beam therapy after being suggested by R.~Wilson in 1946 because of the favourable shape of the depth-dose distribution due to the fundamental difference in the energy deposition profile between massive projectiles and massless photons. This shape is characterised by the Bragg peak, which is a sharp maximum in the linear energy transfer (LET) of ions at the end of their trajectories. Due to this key feature, IBCT allows a delivery of high doses into tumours, maximising cancer cell destruction, and simultaneously minimising the radiation damage to surrounding healthy tissue. The effectiveness of radiation with ions depends on the choice of ions; it can be described by three factors: the peak value of LET, the proximal plateau value of LET, and the size of a tail distal to the peak. Since the LET is proportional to the square of charge of the projectile, ions heavier than protons are expected to be more effective; however, the increase of LET in the plateau region and the increasing size of the tail hinder the usage of heavier ions and, as a result, carbon ions, besides protons, are the most clinically used modality~\cite{SchardtRMP10,Kraft07}.  Because of its high costs, there are only 43 centres for proton beam therapy in 16 counties around the world\footnote{As of August 2013~\cite{wikiCentres}.}. More proton centres are under construction. Although heavy ion therapy was adopted in the 1990s, there are only four clinical centres (in Germany, Italy, and Japan) where carbon ions are used~\cite{wikiCarbon}.

The Bragg peak occurs because the inelastic cross sections of interactions of projectiles with the molecules of the medium increase up to the maximum values as the speed of the projectile decreases. As a result, the deposition of destructive energy to the tissue per unit length of the ion's path is maximised within 1~mm of the ion's trajectory. The location of the Bragg peak depends on the initial energy of the ions. Typical depths for carbon ions (in liquid water representing tissue)
range from about 2.5 to 28~cm as the initial energy ranges from 100 to 430~MeV/nucleon~\cite{SchardtRMP10,Schardt,Schardt4,Pshen,epjd,Scif}. Hence, a deeply-seated tumour can be scanned with a well-focused pencil beam of ions with minimal lateral scattering.

Over the past 20 years, technological and clinical advances of IBCT have developed more rapidly than the understanding of radiation damage with ions. Although an empirical approach has produced exciting results for thousands of patients thus far, many questions concerning the mechanisms involved in radiation damage with ions remain open and the fundamental quantitative scientific knowledge of the involved physical, chemical, and biological effects is, to a significant extent, missing. Indeed, the series of works that elucidated the importance of low-energy (below ionisation threshold) electrons appeared in ca. 2000, while the treatment of patients at GSI\footnote{Gesellschaft f{\"u}r Schwerionenforschung, Darmstadt, Germany} started in 1997. The dominant molecular mechanism of a double strand break (DSB), the most important DNA lesion~\cite{Chatterjee93,hyd2}, still remains unknown. Even the significance of the relation of DNA damage (including DSBs) compared to the damage of other cellular components to the cell death is not entirely clear. This list can be continued. Besides IBCT, the mechanisms of biodamage due to irradiation with heavy ions have attracted attention in regards to radioprotection from galactic cosmic rays, especially during potential long-term space missions~\cite{SchardtRMP10}.

Over many decades of using radiation with photons, vast data relating the radiation damage to deposited dose were accumulated. These data are currently used to describe the biological damage due to ions~\cite{SchardtRMP10}. Nonetheless, there are substantial qualitative and quantitative differences between the effects of ions and photons on tissue.
The first difference is in the localisation of the dose distribution for ions distinguished from the mostly uniform dose distribution for photons. This feature reveals itself longitudinally (along the ion's path)  as the Bragg peak.  Radially (with respect to the ion's path) it shows up as the sharply decreasing (within several tens of nm) radial dose distribution, while the average distance between adjacent ions in clinically used beams are several hundreds of nm.

The second difference is a consequence of the first. Secondary particles such as electrons, free radicals, etc., produced as a result of the interaction (ionisation and excitation) of ions with the medium, emerge at the location of the Bragg peak in much larger number densities than those produced by photons, and their distribution is also non-uniform. These secondary particles largely cause the biological damage, and in order to assess the damage, it is important to distinguish the biological effects of the locally deposited dose and the local number density of secondary particles. In other words, the (radial) dose is not the only characteristic that determines the biological damage. For instance, clustered damage, more lethal than isolated damage, can be caused by several low-energy electrons, which are not associated with a large dose deposition. This qualitatively and quantitatively changes the effect of the radiation~\cite{MSAreview,SchardtRMP10,precomplex}.

There are also differences in the chemical interactions related to a different balance between free electrons, free radicals, and other agents for ions versus photons. These differences, for example, affect the resistivity of cells to radiation and thus are quite important for the assessment of radiation damage. Finally, the Bragg peak leads to thermomechanical effects, which stem from the non-uniformity of the radial dose deposition.


One of the most important questions in the foundation of science devoted to radiation damage with ions is the question about molecular mechanisms leading to DNA damage, or more generally, biodamage. While ``whether the biodamage leads to cell death?'' is a biological question, the question about the mechanisms of biomolecular damage belongs to the realms of physics and chemistry. The role of low energy (sub-15~eV) electrons has been especially emphasised in Refs.~\cite{Sanche05,DNA2,DNA3,SancheCh9.2012}. A number of quantum effects, such as dissociative electron attachment (DEA), formation of electronic and phononic polarons, are discussed in the context of the interaction of these electrons with biomolecules. DEA is deemed to be the leading mechanism for DNA single strand breaks (SSBs) at low energies, while a number of ideas, including the action of Auger electrons, in relation to the mechanism of double strand breaks (DSBs) has been suggested~\cite{DNA3,prauger}. The Auger effect along with intermolecular Coulombic decay (ICD) are discussed not only in relation to the mechanism of DSBs, but also as important channels for production of secondary electrons, especially in the presence of nanoparticles as sensitizers~\cite{Prise11}. Still more understanding is needed for the interaction of electrons of higher energies.

This paper is devoted to the overview of the main ideas of the multiscale approach to the physics of radiation damage that has the goal of developing knowledge about biodamage at the nanoscale and molecular level and finding the relation between the characteristics of incident particles and the resultant biological damage~\cite{MSAreview,pre}. This approach is unique in distinguishing essential phenomena relevant to radiation damage at a given time, space, or energy scale and assessing the resultant damage based on these effects. The significance of understanding the fundamental mechanisms of radiation damage in order to exploit this knowledge for  practical applications has inspired the European COST Action~\cite{nanoIBCT}, which supports collaborations of physicists, chemists, and biologists, studying these phenomena both theoretically and experimentally.

\begin{figure*}
\begin{centering}
\resizebox{1.6\columnwidth}{!}
{\includegraphics{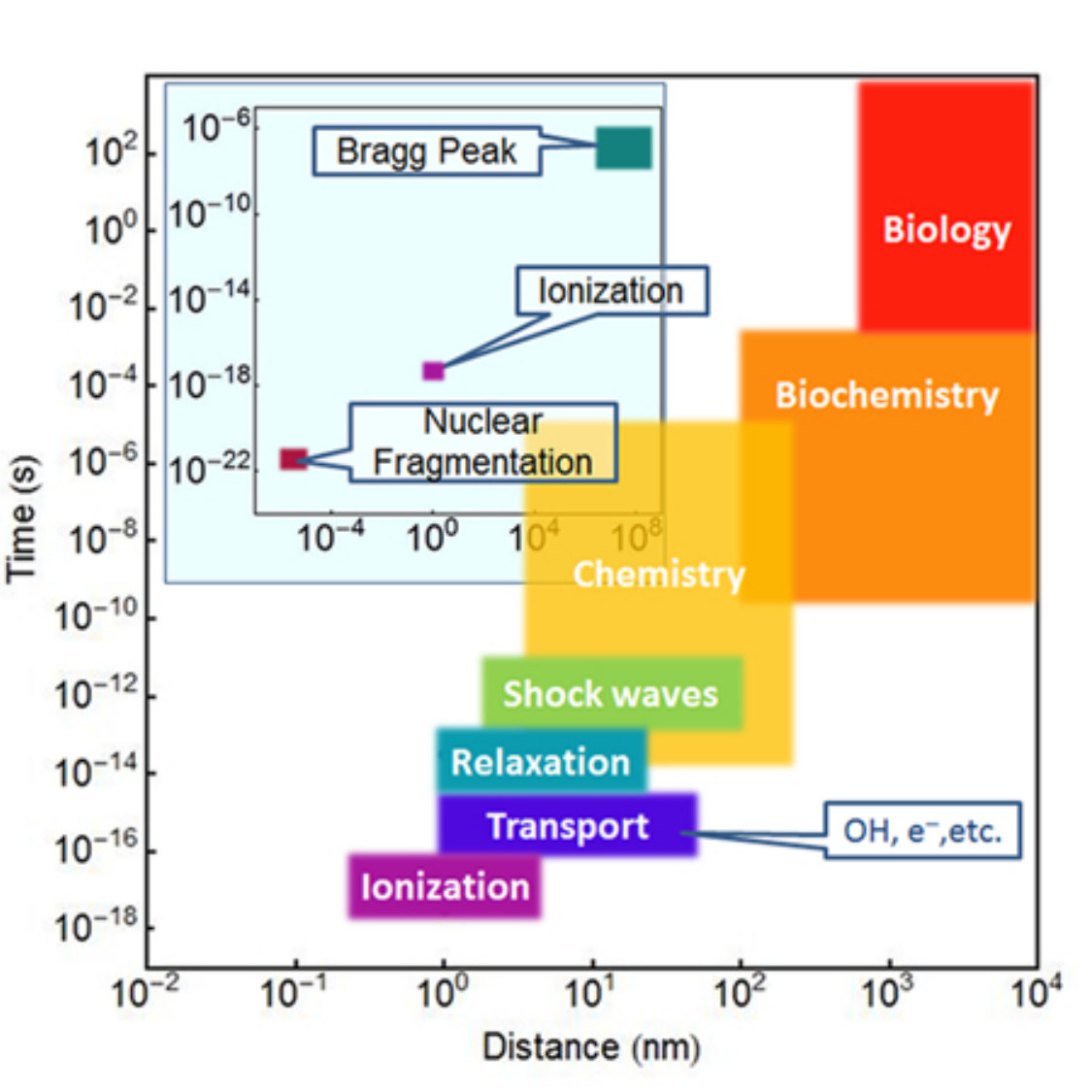}}
\caption{\label{fig.scales} (Colour online)
Features, processes, and disciplines, associated with radiation therapy shown in a space -- time diagram, which shows approximate temporal and spatial scales of the phenomena. The history from ionization/exciation to biological effects on the cellular level are shown in the main figure and features of ion propagation are shown in the inset.}
\end{centering}
\end{figure*}

The multiscale approach was formulated and then elaborated upon, as different aspects of the scenario were add-ed in a series of
works~\cite{epjd,Scif,precomplex,pre,epn,mutat,preheat,SYS,prehydro,natnuke,Pablo2012}. Its name emphasizes the fact that important interactions involved in
the scenario happen on a variety of temporal, spatial, and energy
scales. These scales are schematically shown in Fig.~\ref{fig.scales}.
From the very beginning, the approach was formulated as
phenomenon-based and was aimed at elucidating the physical,
chemical, or biological effects that are important or dominating on
each scale in time, space, and energy. {The practical goal of the multiscale approach is the calculation of relative biological effectiveness (RBE) \cite{SchardtRMP10,Kraft07,Alpen,Hall}, one of the key integral characteristics of the effect of ions compared to that of photons. The RBE is defined as a ratio of doses of photons and different projectiles leading to the same biological effect, such as killing a given percentage of cells in an irradiated region.
This is why the calculation of RBE is so important. Other characteristics, such as the oxygen enhancement ratio (OER), which compares the biological action of given projectiles to that at different aerobic or hypoxic conditions of irradiated targets.}

This paper is organised in the following way. In Sec.~\ref{msrd}, the scenario of radiation damage with ions is described. Section~\ref{ssec1} is devoted to the ion's transport in the medium. Section~\ref{ssec2} describes the applications of the random walk approach to the electron transport relevant for biodamage. Examples of DNA damage calculations are discussed there as well. These calculations are compared to experiments with plasmid DNA and foci studies. In Sec.~\ref{sec.thermo} thermomechanical effects are explored. Section~\ref{ssec5} is devoted to the evaluation of the probability for an irradiated cell to survive based on the calculation of clustered DNA damage. This is followed by the recipe for the assessment of radiation damage with ions using the multiscale approach starting from obtaining and analysing the LET dependence and ending by obtaining the RBE. The discussion is followed by conclusions.

\section{Multiscale scenario of radiation damage}
\label{msrd}

Radiation damage due to ionizing radiation is initiated by the ions incident on tissue. Initially, they have energy ranging from a few to hundreds of MeV. In the process of propagation through tissue they lose their energy in the processes of ionization, excitation, nuclear fragmentation, etc. Most of the energy loss of the ion is transferred to tissue\footnote{The only part that is not transferred is emitted as radiation. This part, in the case of ions interacting with tissue, is deemed to be insignificant.}. Naturally, radiation damage is associated with this transferred energy, and the dose (i.e., deposited energy density) is a common indicator for the assessment of the damage~\cite{MSAreview,SchardtRMP10,Alpen}. The profile of the dose deposition along the ion's path is characterised with a plateau followed by a sharp Bragg peak. The position of this peak depends on the initial energy of the ion and marks the location of the maximum radiation damage. In the process of radiation therapy, a tumour is being ``scanned'' with the Bragg peak\footnote{This scanning produces the so-called spread-out Bragg peak (SOBP).}  in order to deposit a large dose to the target and spare healthy tissues surrounding it.

However, the deposition of large doses in the vicinity of the Bragg peak does not explain how the radiation damage occurs, since projectiles themselves only interact with a few biomolecules along their trajectory and this direct damage is only a small fraction of the overall damage. It is commonly understood that the secondary electrons and free radicals produced in the processes of ionization and excitation of the medium with ions are largely responsible for the vast portion of the biodamage. 

Secondary electrons are produced during a rather short time of $10^{-18}-10^{-17}$~s following the ion's passage. The energy spectrum of these electrons has been extensively discussed in the literature~\cite{epjd,Scif,Pimblott2,Pimblott3,Pimblott} and the main result (relevant for this discussion) is that most secondary electrons have energy below 50~eV (more than 80\% for an ion energy\footnote{This value corresponds to the kinetic energy of ions near the Bragg peak.} of $0.3$~MeV/u) and only a few (less than 10\% for $0.3$~MeV/u-ions) 
have energy higher than 100~eV. Moreover, this is true for a very large range of ion energy. This has several important consequences. First, the ranges of propagation of these electrons in tissue are rather small, around 10~nm~\cite{Meesungnoen02}. Second, the angular distribution of their velocities as they are ejected from their original host, and as they scatter further, is largely uniform~\cite{Nikjoo06}; this allows one to consider their transport using a random walk approach~\cite{precomplex,prauger,pre,epjdisacc2011,epjdmarion}.

The next time scale $10^{-16}-10^{-15}$~s corresponds to the propagation of secondary electrons in tissue. These electrons (which start with about 45-50~eV energy) are called ballistic. In liquid water, the mean free paths of elastically scattered and ionizing 50-eV electrons are about 0.43 and 3.5~nm, respectively~\cite{Nikjoo06}.
This means that they ionize a molecule after about seven elastic collisions, while the probability of second ionization is small~\cite{epjd}. Thus, the secondary electrons are losing most of their energy within first 20 collisions and this happens within 1-1.5~nm of the ion's path~\cite{natnuke}. After that they continue propagating, elastically scattering with the molecules of the medium until they get bound or solvated electrons are formed. It is important to notice that these low energy electrons remain important agents for biodamage since they can attach to biomolecules like DNA causing dissociation~\cite{SancheCh9.2012,Sanche11}. The solvated electrons may play an important role in the damage scenario as well~\cite{hyd2,Kohanoff2012,SevReview}.

Additionally, the energy lost by electrons during the previous stage in the processes of ionization, excitation and electron-phonon interaction is transferred to the medi-um. As a result of this relaxation, the medium within about a $1-1.5$-nm cylinder (for ions not heavier than iron) around the ion's path becomes very hot~\cite{preheat,natnuke}. This cylinder is referred to as the hot cylinder. The pressure inside this cylinder increases by a factor of about $10^3$ compared to the pressure in the medium outside the cylinder. This pressure builds up by about $10^{-14}-10^{-13}$~s and it is a source of a cylindrical shock wave~\cite{prehydro}. This shock wave propagates through the medium for about $10^{-13}-10^{-11}$~s. Its relevance to the biodamage is as follows. If the shock wave is strong enough (the strength depends on the distance from the ion's path and the LET), it may inflict damage directly by breaking covalent bonds in a DNA molecule~\cite{natnuke}. Besides, the radial collective motion that takes place during this time is instrumental in propagating the highly reactive species such as hydroxyl radicals, just formed solvated electrons, etc. to a larger radial distance (up to tens of nm) thus increasing the area of an ion's impact.

 The assessment of the primary damage to DNA molecu-les and other parts of cells due to the above effects is done within the multiscale approach. This damage happens within $10^{-5}$~s from the ion's passage and consists of various lesions on DNA and other biomolecules. Some of these lesions may be repaired by the living system, but some may not and the latter may lead to cell death.
\begin{figure*}
\begin{centering}
\resizebox{1.6\columnwidth}{!}
{\includegraphics{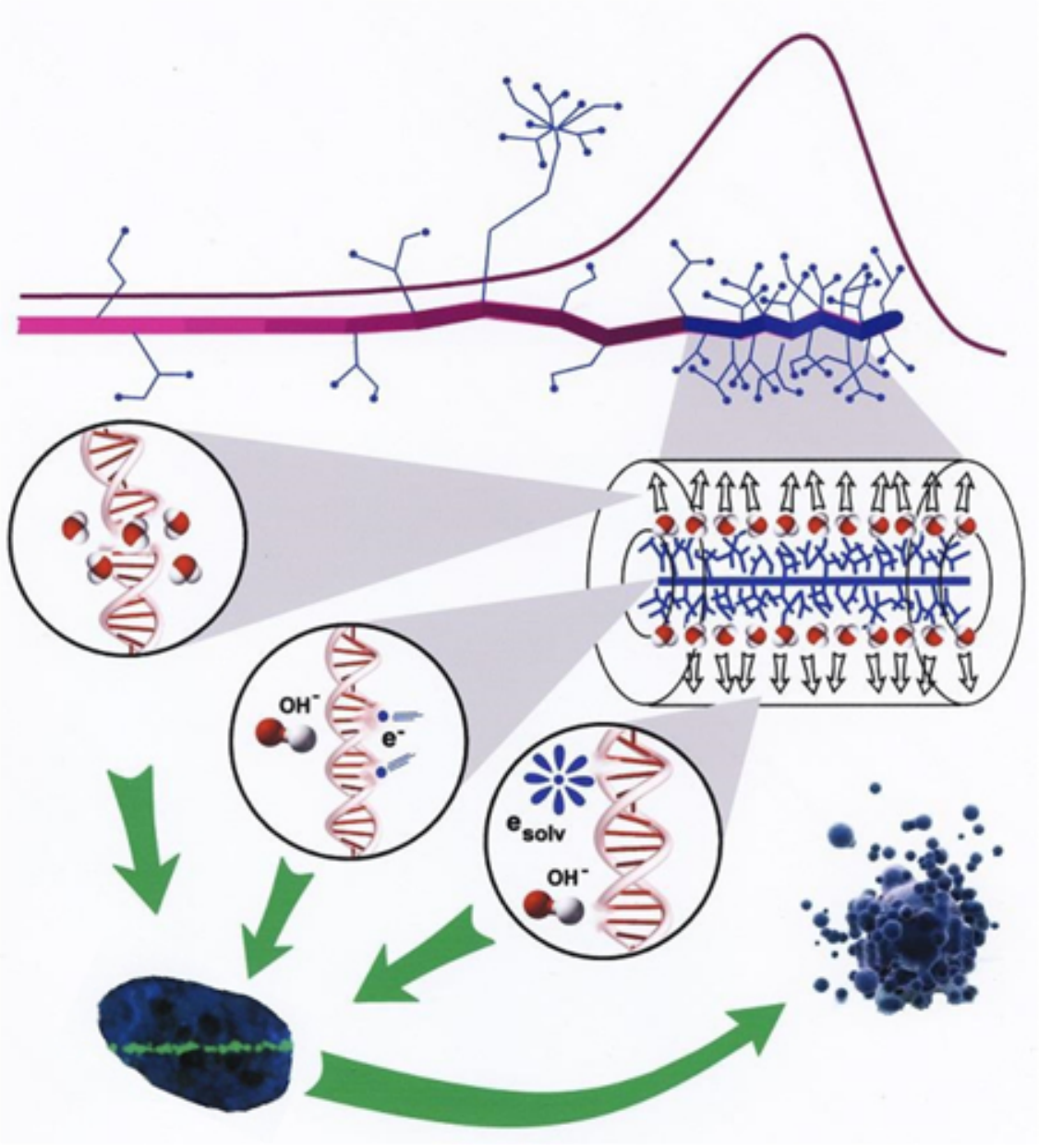}}
\caption{\label{fig.scen} (Colour online) The scenario of biological damage with ions. Ion propagation ends with a Bragg peak, shown in the top right corner. A segment of the track at the Bragg peak is shown in more detail. Secondary electrons and radicals propagate away from the ion's path damaging biomolecules (central circle). They transfer the energy to the medium within the hot cylinder. This results in the rapid temperature and pressure increase inside this cylinder. The shock wave (shown in the expanding cylinder) due to this pressure increase may damage biomolecules by stress (left circle), but it also effectively propagates reactive species, such as radicals and solvated electrons to larger distances (right circle). A living cell responds to all shown DNA damage by creating foci, in which enzymes attempt to repair the induced lesions. If these efforts are unsuccessful, the cell dies; an apoptotic cell is shown in the lower right corner. }
\end{centering}
\end{figure*}
The scenario described above is illustrated in Fig.~\ref{fig.scen}.

\section{Propagation of ions in tissue and primary ionization of the medium}
\label{ssec1}

\subsection{The main characteristics of ion's propagation in the medium}
\label{sec.ions1}

The scenario starts with the traverse of an ion through tissue. Ions enter
the medium with a sub-relativistic energy (for therapy, the carbon ion energy ranges through 100--420~MeV/nucleon and the proton energy can be up to 250~MeV, while the ions of galactic cosmic rays are much more energetic).
Then, the ions lose energy propagating in the tissue. This process is described by the stopping power, $S$, of the medium, equal to $-dE/dx$, where $E$ is the kinetic energy of the ion and $x$ is the longitudinal coordinate. For projectiles such as protons or heavier ions, there is not much difference between the location of the energy loss by projectiles and that absorbed by the medium {\em longitudinally}, i.e., along the ion's path\footnote{This is so because the energy is mostly transferred to electrons and other secondary particles, whose longitudinal ranges are many times smaller than the characteristic scale of $x$.}. Therefore, the linear energy transfer (LET), i.e., the energy absorbed by the medium per unit length of the projectiles's trajectory becomes similar to the stopping power. Hence, the terms ``LET'' and ``stopping power'' are used synonymously. The energy loss occurs due to ionization of the medium, nuclear fragmentation in collisions with nuclei, excitations of the medium, etc. The LET profile for ions is characterized by a plateau followed by the sharp Bragg peak, where the LET reaches its maximum. The tail is caused by the energy loss of the lighter products of nuclear fragmentation, such as protons, neutrons, $\alpha$-particles, etc.

{The behaviour of the LET is explained by features of inelastic cross sections of the projectile in the medium. The Bragg peak in the stopping power of massive charged particles is described by the Bethe - Bloch formula~\cite{Bethe,Bloch1,Bloch2}.
\begin{eqnarray}
-\frac{d E}{d x} = \frac{4\pi n_e z^2 e^4}{m V^2}\left[\ln{ \frac{2 m V^2}{\langle I \rangle (1-\beta^2)}} - \beta^2\right]~,
\label{bethebloch}
\end{eqnarray}
where $m$ and $e$ are the mass and charge of electron, $V$ is the velocity of the projectile, $\beta=V/c$ ($c$ is the speed of light in vacuum), $z |e|$ is the charge of projectile, $n_e$ is the number density of electrons in the target, and $\langle I \rangle$ is the mean excitation energy of its molecules.

This formula provides the dependence of the stopping power on the energy of the ion and practically depends on a single parameter, the mean excitation energy. This parameter for liquid water is chosen empirically somewhere between 70 and 80~eV~\cite{Pshen,Molina}. The use of such a non-physical parameter is sufficient for the calculations of the position of the Bragg peak and its shape, and Eq.~(\ref{bethebloch}) is used in many Monte Carlo (MC) simulations~\cite{Pshen} for that purpose. This parameter, however, hides all physical processes such as ionization and excitation of the medium, even though these same processes are important for the understanding of the scenario of radiation damage. Therefore, it is better to use a different approach, which uncovers the physics integrated in the empirical parameter. In Refs.~\cite{epjd,Scif,pre}, the singly-differentiated (with respect to the secondary electron energy) ionization cross sections of water molecules in the medium has been employed as a physical input. This allowed not only describing the features of the Bragg peak, but also obtaining the energy spectrum of secondary electrons, which are very much involved in subsequent radiation damage.}

\subsection{Singly-differentiated cross sections of ionization}
\label{sec.sdcs}

The total ionization cross section, $\sigma_t$, differentiated with respect to secondary electron kinetic energy, $W$, i.e., singly-differentiated cross section (SDCS)
is the main quantity in our analysis.  Besides the kinetic energy of secondary electrons and the properties of water molecules, the SDCS depends on the velocity $V$ of the projectile and its charge, $z|e|$.

\subsubsection{Calculation of the SDCS using a parametric semiempirical approach}

In Refs.~\cite{epjd,Scif,nimb}, the semi-empirical Rudd's expression for the calculation of SDCS has been used. This analytic expression, containing a number of parameters, is a combination of the experimental data and calculations within the plane-wave Born approximation and other theoretical models~\cite{Rudd92}. Since this model was developed for non-relativistic protons, it had to be modified  to include heavier ions at relativistic velocities. 
The original SDCS is given in the following form~\cite{Rudd92}:
\begin{eqnarray}
\label{sdcs}
\frac{d\sigma_t}{dW}= z^2 \sum\limits_i \frac{4\pi a_0 N_i}{I_i}
\left(\frac{I_0}{I_i}\right)^2 \times \\ \nonumber
\frac{F_1(v_i)+F_2(v_i)\omega_i}{\left(1+\omega_i\right)^3\left(1+\exp(\alpha
  (\omega_i-\omega^{\rm max}_i)/v_i)\right)}~,
\end{eqnarray}
\noindent where the sum is taken over the electron shells of the water
molecule, $a_0 = 0.0529$~nm is the Bohr radius, $I_0=13.6$~eV, $N_i$ is the shell occupancy, $I_i$ is the ionization
potential of the shell, $\omega_i=W/I_i$ is the dimensionless
normalised kinetic energy of the ejected electron, $v_i$ is the
dimensionless normalised projectile velocity given by
\begin{eqnarray}
v_i=\sqrt{\frac{m V^2}{2 I_i}}~.
\label{vi}
\end{eqnarray}
When $V \ll c$, $V=\sqrt{\frac{2E}{M}}$ (where $M$ is the mass of a projectile), and, hence $v_i=\sqrt{\frac{m}{M}\frac{E}{I_i}}$.
When $V$ approaches $c$, the definition of $v_i$, given by~(\ref{vi}),  holds, however, the projectile's velocity $V$ is given by $\beta c$, where
$\beta^2=1-1/\gamma^2=1-(Mc^2/(Mc^2+E))^2$, and $\gamma$ is the Lorentz factor of the projectile.

Functions $F_1$ and $F_2$ in (\ref{sdcs}) are given by
\begin{eqnarray}
F_1(v) = A_1 \frac{\ln(1+v^2)}{B_1/v^2+v^2} + \frac{C_1 v^{D_1}}{1+E_1
  v^{D_1+4}}~,
\label{F1}
\end{eqnarray}
and
\begin{eqnarray}
F_2(v) = C_2 v^{D_2} \frac{A_2 v^2+B_2}{C_2 v^{D_2+4}+A_2 v^2+B_2}~.
\label{F2}
\end{eqnarray}
The fitting parameters $A_1$ ... $E_1$, $A_2$ ... $D_2$, and $\alpha$ depend on the medium. In Ref.~\cite{Rudd92}, they are given for water vapour. The comparison of positions of Bragg peaks for different initial carbon ion energies with those measured in experiments provided sufficient material for refitting of these parameters for liquid water medium~\cite{Scif}.   These parameters are listed in Table~\ref{outer}~\cite{Scif}.
\begin{table*}
\caption{Fitting parameters and ionization energies for three outer and two inner shells of water molecules in a liquid water environment~\cite{Scif}.}  {\begin{tabular}{llcccccccccc}
    \hline\noalign{\smallskip}Shells &Ionization energies (eV)& $A_1$&$B_1$&$C_1$&$D_1$&$E_1$&
    $A_2$ & $B_2$ & $C_2$ & $D_2$ & $\alpha$
    \\ \noalign{\smallskip}\hline\noalign{\smallskip}Outer: $1b_1$, $3a_1$, $1b_2$ &10.79, 13.39, 16.05 & 1.02 & 82 & 0.5
    &$-0.78$&0.38&1.07&14.5&0.61&0.04&0.64\\ Inner: $2a_1$, $1a_1$ &
    32.3, 539.0 &1.25&0.5 & 1.0 & 1.0 & 3.0 & 1.1 & 1.3 & 1.0 & 0.0 &
            0.66\\
     \noalign{\smallskip}\hline
\end{tabular}}
\label{outer}
\end{table*}
\noindent The cut-off energy $\omega^{\rm max}$ is given by
\begin{equation}
\omega^{\rm max}_i = 4v_i^2-2v_i-\frac{I_0}{4I_i}~,
\end{equation}
\noindent where the first term on the right-hand side represents the free-electron limit, the second term represents a correction due to electron binding, and the third term gives the correct dependence of the SDCS for $v_i\ll 1$ \cite{Rudd92}. For $v_i\gg 1$, Eq.~(\ref{sdcs}) should asymptotically approach the relativistic Bethe - Bloch formula~(\ref{bethebloch}). This is accomplished when $F_1$, given by~(\ref{F1}), is replaced by the following expression,
\begin{equation}
F_1(v) = A_1 \frac{\ln(\frac{1+v^2}{1-\beta^2})-\beta^2}{B_1/v^2+v^2}
+ \frac{C_1 v^{D_1}}{1+E_1 v^{D_1+4}}~.
\label{F1rel}
\end{equation}
\noindent Indeed, the asymptotic behaviour of~(\ref{F1rel}) at $v \gg 1$ is given by
\begin{equation}
\frac{A_1}{v^2}\left[ \ln(\frac{v^2}{1-\beta^2})-\beta^2\right]~,\nonumber
\label{F1relasy}
\end{equation}
which, after being substituted to Eq.~(\ref{sdcs}) and the understanding that $\frac{dE}{dx}\sim \sum_i \int (W+I_i) \frac{d\sigma_t}{dW}dW$, leads to Eq.~~(\ref{bethebloch}).
The correction of Eq.~(\ref{F1rel}) reveals itself as an increase of the cross section at high energies.

\subsubsection{Calculations of SDCS based on the energy-loss function}

An alternative method has been used in Ref.~\cite{Pablo2012}, where the dielectric formalism based on the experimental measurements of the energy-loss function (ELF) of the target medium, ${\rm Im}\left(-1/\epsilon({\cal E},q)\right)$, where $\epsilon({\cal E},q)$ is the complex dielectric function, and $\hbar q$ and $\cal E$ are the momentum and energy transferred in the electronic excitation, respectively~\cite{landau8,Lindhard54}. This formalism allows obtaining the SDCS not only for liquid water but for a real biological medium containing sugars amino acids, etc. If the ELF is experimentally known, many-body interactions and target physical state effects are naturally
included in these calculations.

According to that formalism, the macroscopic (nonrelativistic) SDCS for ionization of the electronic shell $i$ is given by,
\begin{eqnarray}
 \frac{d \sigma_i(W,{\cal E})}{dW}= \frac{e^2}{n \pi \hbar^2}\frac{M z^2}{E} \int_{q_-}^{q^+}\frac{d q}{q} {\rm Im}\left[\frac{-1}{ \epsilon (q, I_i+W)} \right]_i~,
\label{eq1b}
\end{eqnarray}
where $q_{\pm}=\sqrt{2M}(\sqrt{E} \pm \sqrt{E-{\cal E}})$. Equation~(\ref{eq1b}) can be used for
different charged projectiles by properly taking into
account their charge state, or for electrons by introducing
an exchange term in the integrand and imposing the correct
integration limits.

Since Eq.~(\ref{eq1b}) requires the contribution of each
electronic shell of the target to its ELF, and the latter is usually
measured for all the excitations and ionizations of the
electronic system in the optical limit ($q=0$), the algorithm for obtaining the data at $q>0$ and splitting this ELF into different electronic shells is needed in addition to the experimentally measured ELF.

The optical ELF bioorganic condensed compounds and liquid water are rather similar and can be parameterized with a single-Drude
function~\cite{Tan04}
\begin{eqnarray}
{\rm Im}\left[\frac{-1}{ \epsilon (q=0,{\cal E})} \right]=\frac{a(Z){\cal E}}{\left({\cal E}^2-b(Z)^2\right)^2+c(Z)^2{\cal E}^2}~,
\label{eq2b}
\end{eqnarray}
where $a(Z)$, $b(Z)$, and $c(Z)$ are the functions of the mean atomic number of the target $Z$, corresponding to the height, position, and width~\cite{Tan04}.  While $b(Z)$ and $c(Z)$ are parametric functions, $a(Z)$ is obtained by imposing the f-sum rule~\cite{Altarelli74}, linked to the number of electrons in the target, $Z$, also accounting for the contribution from the inner shells, as explained in Ref.~\cite{Tan04}. Using this approach the ELF of an arbitrary bioorganic compound can be estimated, even in the case where no experimental data exist. A wide variety of extension algorithms for extrapolation of
optical-ELF to $q>0$ are available due to extensive research~\cite{RGM12}. In Ref.~\cite{Pablo2012} a simple quadratic dispersion relation introduced by Ritchie and Howie~\cite{Ritchie}, with its parameters for liquid water~\cite{RGM12}, has been used.


The issue of splitting of the ELF into contributions from different shells has been studied for liquid water~\cite{Dingfelder,Emfietzoglou03}, providing parameterizations of the ELF split in ionization and excitation arising from each different shell. In Ref.~\cite{Pablo2012} a specially designed approximation has been applied to split the ELF for biomolecules. In order to describe the outer-shell ionization of biomolecules, the mean value of their binding energies, ${\bar I}$, is calculated\footnote{The relevant data are available for some biological molecules, such as the DNA bases and the sugar-phosphate backbone~\cite{Paretzke03} and some amino acids~\cite{Terrisol06} and others~\cite{YKKim04}.}. It is then assumed that the outer-shell electrons will be ionized if the transferred energy satisfies ${\cal E}> {\bar I}$. Then, the ejected electron energy is $W={\cal E}-{\bar I}$. In Ref.~\cite{Pablo2012}, SDCSs are calculated and compared with other calculations and experiments for protons interacting with water, adenine, and benzene.

{ The total ionization cross sections (TICS) can also be estimated for different biomolecules relevant for IBCT~\cite{Pablo2012}.
For example, in Fig.~\ref{fig.pablo}a~\cite{Pablo2012} the macroscopic TICSs 
are calculated for proton impact in five representative biological materials relevant for cancer therapy: liquid water, dry DNA (C$_{20}$H$_{27}$N$_7$O$_{13}$P$_2$), protein, lipid, and the cell nucleus. Their atomic compositions and densities can be found in the ICRU Report~46~\cite{ICRU46} and other sources, and a reasonable value of their mean binding energies can be estimated from the values of their molecular components, such as the water molecule, DNA bases and backbone, and amino acids~\cite{Dingfelder,Paretzke03,Terrisol06}. The experimental data for water vapour~\cite{Wilson84,Rudd85,Rudd86} are also shown. They agree well with the calculations of Ref.~\cite{Pablo2012} above 100 keV, where the first Born approximation is applicable without further corrections. From these results, it is plausible that all the biological targets different from water have a larger ionization probability than water. One can also see that the TICS of a cell nucleus is only slightly larger than that of liquid water, and that protein has a slightly larger TICS than the rest of the biomaterials.
\begin{figure}
\resizebox{\columnwidth}{!} {\includegraphics{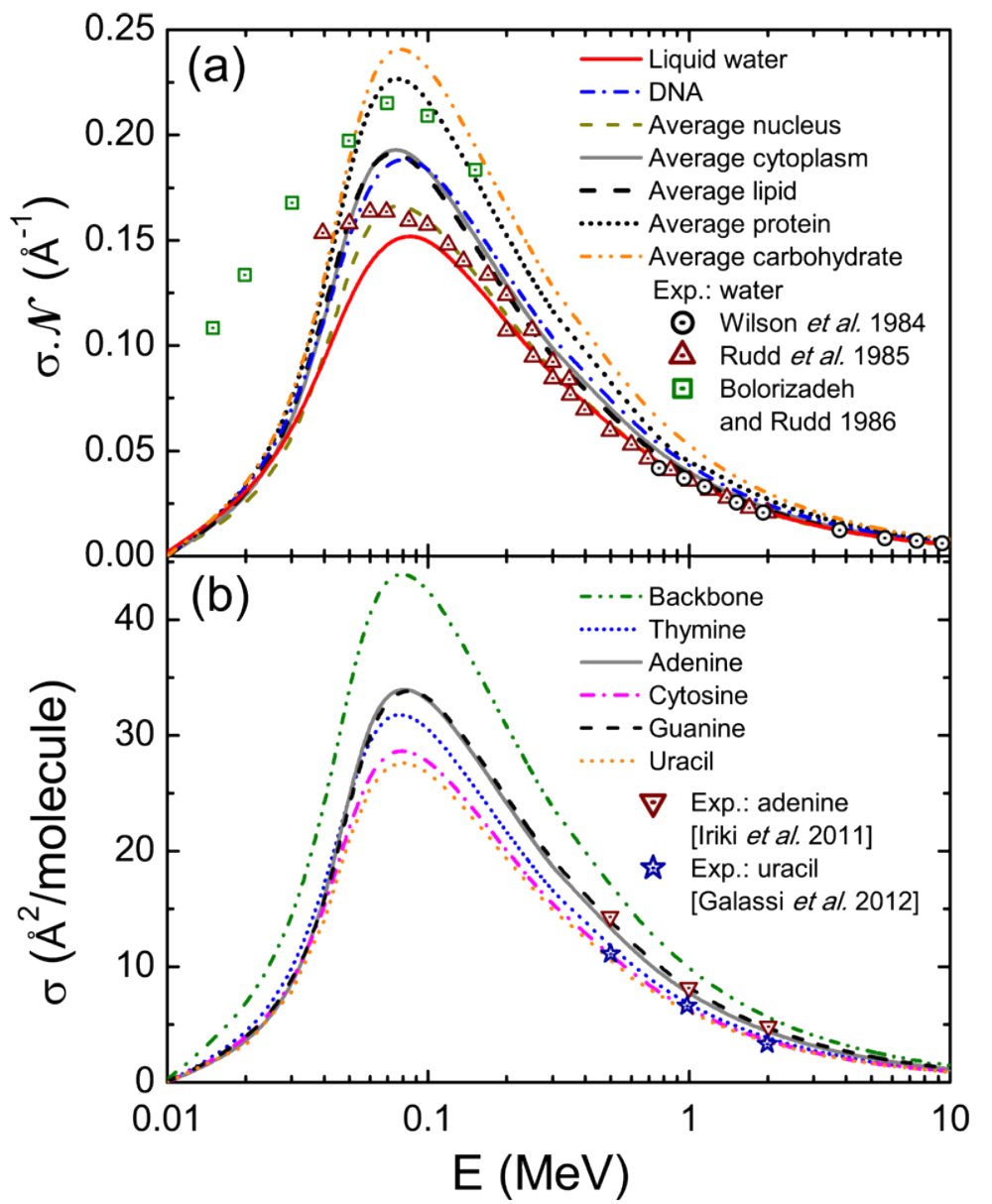}}
\caption{\label{fig.pablo} (Colour online) (a) Calculated macroscopic TICS for proton impact in liquid water, DNA, protein, lipid, and cell
nucleus. (b) Calculated microscopic TICS for proton impact
in the DNA components adenine, cytosine, guanine, thymine,
and sugar-phosphate backbone. Symbols represent experimental
data~\cite{Pablo2012}. ${\cal N}$ is the molecular
density of the target. }
\end{figure}

In Fig.~\ref{fig.pablo}b~\cite{Pablo2012} the microscopic TICS per molecule for proton impact in the DNA molecular components, such as adenine, cytosine, guanine, thymine, and sugar-phosphate backbone are shown. Their atomic composition can be easily found in the literature, and their mean binding energies were estimated from quantum chemistry calculations~\cite{Paretzke03,Terrisol06}. Also shown are experimental data at high energies for adenine~\cite{Itoh11}, which are in excellent agreement with the predictions of Ref.~\cite{Pablo2012}.
This method allows one to estimate the ionization probability of each constituent of the DNA molecule, which gives important information on the sensitivity of each one to radiation damage. According to these results, the DNA backbone is the most probable part of the DNA to be ionized by proton impact (a similar behaviour was previously observed for electron impact in Ref.~\cite{Paretzke03}; also, recent theoretical estimates~\cite{Simons07} point towards sugar-phosphate C-O bond cleavage due to interaction with low energy electrons) and, between bases, adenine and guanine are the most sensitive to proton impact ionization. This fact could have important implications in the DNA damage, since it seems that single or double strand breaks could be more probable than base damage, or that regions of the DNA with a bigger concentration of adenine or guanine would be more likely damaged by radiation than other parts of the genome, attending to direct ionization effects.
}


Much more information can be obtained with this me-thod, such as the number of emitted electrons, the average energy of electrons, SDCS and TICS for other biological targets and projectiles. This model, using little input information and physically motivated approximations, can provide useful information about the ion impact ionization of a huge number of relevant biological targets, for which data are lacking, both experimentally and theoretically. This model can be easily extended to ions heavier than protons, in different charge states, as well as to electron impact ionization, by introducing appropriate corrections, such as the description of the electronic structure of the ion, or exchange and relativistic  corrections for electrons.

\subsection{The position of the Bragg peak}

The stopping cross section, defined as
\begin{eqnarray}
\sigma_{\rm st}=\sum\limits_i \int^{\infty}_0
(W+I_i)\frac{d\sigma_{t,i}}{dW}dW~,
\label{eq6}
\end{eqnarray}
where the sum is taken over all electrons of the target, gives the average energy lost by a projectile in a single
collision, which can be further translated into energy loss within an
ion's trajectory segment, $d x$:
\begin{eqnarray}
\frac{dE}{dx}=-n\sigma_{st}(E)~.
\label{eq7}
\end{eqnarray}
This quantity is known as the stopping power~\cite{Alpen,Molina}. As was discussed above in Sec.~\ref{sec.ions1}, for ions this quantity is similar to the linear energy transfer (LET).

\begin{figure}
\resizebox{\columnwidth}{!} {\includegraphics{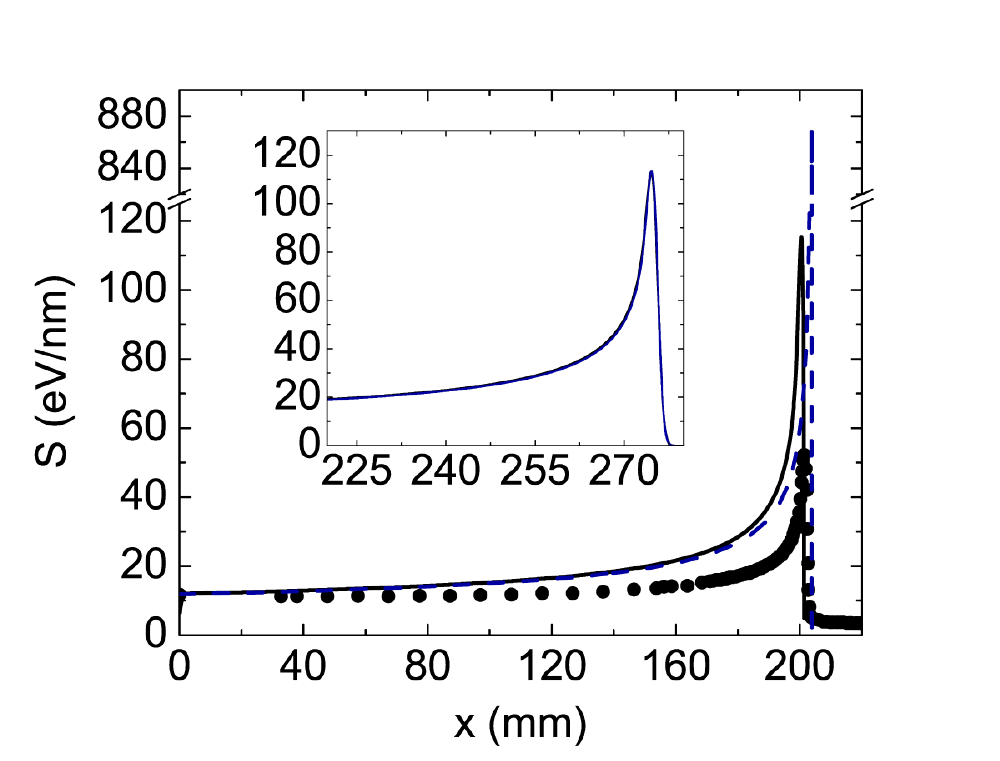}}
\caption{\label{fig4} (Colour online) The dependence of the LET on depth with the Bragg peak, plateau, and tail for carbon ions in liquid water. The calculations (solid line) are done for ions with the initial energy of 330~MeV/u and with use of Eqs.~(\ref{eq6}) and (\ref{eq7}). Experimental results~\cite{Schardt} for the same energy are shown with dots. The dashed line depicts the LET dependence without the effect of energy straggling. Two almost coinciding curves in the inset show the agreement between the analytical calculations and MC simulations~\cite{Pshen} for 420~MeV/u carbon ion projectiles with straggling being included. }
\end{figure}

The LET found from Eq.~(\ref{eq7}) is a function of the kinetic energy
of the ion rather than the ion's position along the path in the
medium.  The dependence of LET (and other quantities) on this
position, however, is more suitable for cancer therapy applications.
Integrating inverse LET, given by~(\ref{eq7}), yields
\begin{eqnarray}
x(E)=\int^{E_{0}}_E \frac{dE'}{|dE'/dx|}~,
\label{eq8}
\end{eqnarray}
\noindent where $E_0$ is the initial energy of the projectile. We
obtain the correspondence between the position of the ion along the
path and its energy.  
This allows one to obtain all quantities of interest in terms of $x$ rather than $E$. The depth dependence of the average
LET as a function of $x$ is shown in Fig.~\ref{fig4}.
The calculations of the LET include the effects that were discussed above, such as SDCS calculated using semi-empirical parameterization~(\ref{sdcs}), modified for relativistic energies~(\ref{F1rel}) with the use of the effective charge described below in Sec.~\ref{sec.chargetransfer}. The effect of energy straggling due to multiple ion scattering, described in the Sec.~\ref{sec.straggling} is also taken into account. This effect explains why the height of the Bragg peak decreases with the increasing initial energy of ions and thus increasing depths of the corresponding Bragg peaks. The contribution of non-ionization processes, such as excitation of neutral molecules, are also included in these calculations. In order to accomplish this, the excitation cross sections for proton projectiles~\cite{Dingfelder2000} were scaled using the ratio of the effective charges for carbon and proton at a given energy $E$.

In Fig.~\ref{fig4}, our calculated LET is compared with 
the experimental results~\cite{Schardt}. 
As can be seen from the figure, the experimental dots at the Bragg peak are systematically lower than the calculated curve, the difference being due to in the nuclear fragmentation component, which has not been included in the analytical calculations. It is feasible to include it, as has been done in Ref.~\cite{RafaIBCT2011} for protons, if the appropriate fragmentation cross sections are known. 

As confirmed by MCHIT MC simulations~\cite{Pshen}, nuclear fragmentation reactions become important for heavy-nuclei beams and deeply-located tumours. For example, both experimental data~\cite{Schardt} and MCHIT calculations~\cite{Pshen} indicate that more than 40\% of primary 200 MeV/u $^{12}$C$^{6+}$ nuclei undergo fragmentation before they reach the Bragg peak position, and this fraction exceeds 70\% for a 400 MeV/u $^{12}$C$^{6+}$ beam. As a result of nuclear reactions the beam is attenuated. New projectiles such as protons, neutrons, and $\alpha$-particles are formed. Since these particles are lighter than the incident ions, after fragmentation they carry a larger portion of the energy and their penetration depths are larger than that of the original ions~\cite{Schardt}. This results in a tail after the Bragg peak also seen in Fig.~\ref{fig4}.

\subsection{Charge transfer effect}
\label{sec.chargetransfer}
The incident ions are usually stripped of all electrons, but as they slow down they pick electrons off and their charge reduces. The dependence of the charge of ions on their velocity has been suggested by Barkas~\cite{Barkas63}, where the following empirical formula for the effective charge, $z_{eff}$, is introduced,
\begin{equation}
z_{eff}=z(1-\exp(-125\beta z^{-2/3}))~,
\label{zeff}
\end{equation}
where $z$ is the charge of the stripped ion. This formula is a result of studies of energy loss of ions in emulsions. More detailed descriptions of charge transfer effects have became available recently~\cite{ChargeFluct}. These studies allow one to not only estimate the effective charge of the ion, but also find its fluctuations. These fluctuations are important since LET increases proportionally as $z^2$ and if LET becomes large enough, qualitative differences related to thermomechanical effects may become substantial (see Sec.~\ref{sec.thermo} below).

Regardless of the method of the calculation of the effective charge,
in order to find the stopping power and estimate the secondary electron spectra (in the first approximation) $z$ in Eq.~(\ref{sdcs}) should be
replaced by an effective charge $z_{eff}$ which decreases with decreasing energy making the ionization cross section effectively smaller. In Refs.~\cite{epjd,Scif} the parameterization~(\ref{zeff}) was used.
The effective charge given by this expression slowly changes at high projectile velocity, but rapidly decreases in the vicinity of the Bragg peak.  As a result, charge transfer significantly affects the height of the Bragg peak, and only slightly shifts its position towards the projectile's entrance. { This happens because the stopping cross section as a function of velocity has a sharp peak as velocity decreases. At the same time $\sigma_{st}$ is proportional to $z_{eff}^2$. If the latter decreases with decreasing $V$, the Bragg peak shifts towards the direction of the beam's entrance to the tissue. For instance, with the account for charge transfer, for carbon ions the Bragg peak occurs at $E=0.3$~MeV/u rather than at $E=0.1$~MeV/u.}

\subsection{The effect of ion scattering}
\label{sec.straggling}

It will become clear below, in Sec.~\ref{sec.plasmid} and Sec.~\ref{sec.surv}, that tracks of ions emerging from clinically used accelerators do not interfere, i.e., the effects of a single ion do not spread far enough to reach the area affected by adjacent ions. Therefore, it is usually sufficient to study a single ion interacting with tissue and then combine these effects relating the action of the beam with the dose.
Even though the Bragg peak is a feature of every ion's LET, each peak cannot be observed separately. Since each of the projectiles in the beam experiences its own multiple scattering sequence, peaks for different ions occur at a slightly different spatial location and only the Bragg peak, averaged over the whole beam, is observed experimentally. Therefore, in order to compare the shape of the Bragg peak with experiments, the whole ion beam should be considered.

In Ref.~\cite{epjd}, the Bragg peak for an ion beam was obtained via introduction of the energy-loss straggling due to ion scattering. The energy straggling, described by a semi-analytical model~\cite{Kundrat07}, is given by
\begin{eqnarray}
\lefteqn{\left\langle \frac{dE}{dx}(x)\right\rangle =}\\ \nonumber
&&\frac{1}{\lambda_{str}\sqrt{2\pi}}\int_{0}^{x_0}
\frac{dE}{dx}(x')\exp \left[-\frac{(x'-x)^{2}}{2\lambda_{str}^{2}}\right]dx'~,
\label{eqstrg}
\end{eqnarray}
where $x_0$ is a maximum penetration depth of the projectile and $\lambda_{str}=0.8$~mm is the longitudinal-straggling standard deviation computed by Hollmark {\em et al.}~\cite{Hollmark04} for a carbon ion of that range of energy.  The Bragg peak shown in Fig.~\ref{fig4} was calculated using Eq.~(\ref{eqstrg}).

\subsection{Energy spectra of secondary electrons}
\label{sec.enspec}

The most important effect that takes place during the propagation of the ion in tissue is the ionization of the medium. This is how, when, and where the secondary electrons, the key player in the scenario of radiation damage, are produced. The information, required for the understanding of phenomena related to secondary electrons, is the number of electrons produced per unit length of the ion's trajectory and their energy distribution.
This section is devoted to the analysis of the electron energy distributions obtained from ionization cross sections discussed above.

The emission of electrons in collisions of protons with atoms and molecules has been under theoretical and experimental investigation for { decades}~\cite{Rudd92,Inokuti,Pimblott,Doerner2004}. The quantity of interest is the probability to produce $N_e$ secondary electrons with kinetic energy $W$, in the interval $d W$, emitted from a segment $\Delta x$ of the trajectory of a single ion at the depth $x$ corresponding to the kinetic energy of the ion, $E$. This quantity is proportional to the singly-differentiated cross ionization section (SDCS)\footnote{The SDCS are integrated over full solid angle of electron emission}, discussed in Sec.~\ref{sec.sdcs}.
\begin{eqnarray}
 \frac{d N_e(W,E)}{dW}= n\Delta x \frac{d\sigma_t}{dW}~.
\label{eq1a}
\end{eqnarray}
where $n$ is the number density of molecules of the medium (for water at standard conditions $n \approx 3.3\times 10^{22} {\rm cm}^{-3}$).
Equation~(\ref{eq1a}) relates the energy spectrum of secondary electrons to the SDCS regardless of the method, by which the latter are obtained.

\begin{figure}
\resizebox{1.05\columnwidth}{!}{ \includegraphics{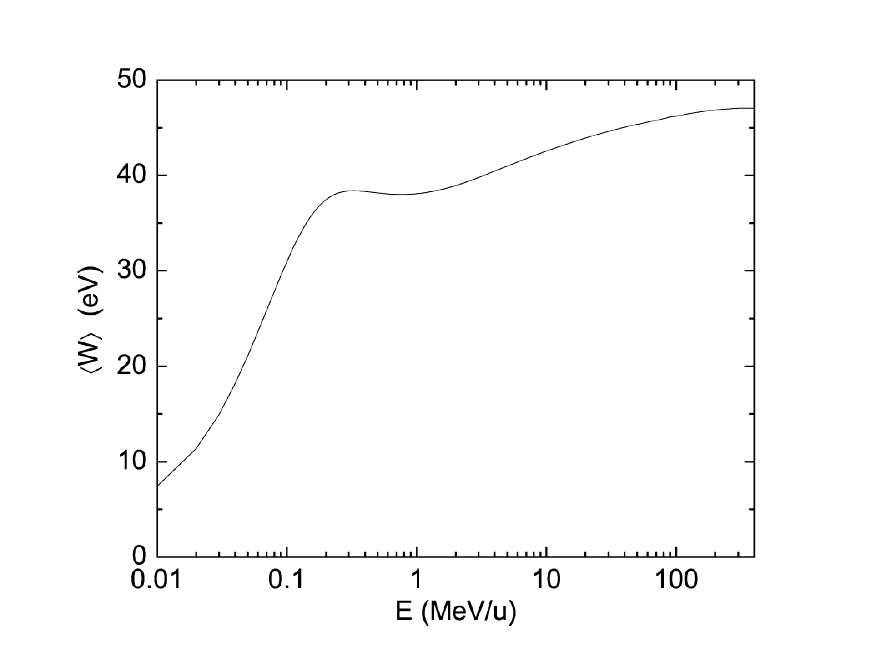} }
\caption{\label{WaveC} Average energy of secondary electrons produced
  as the result of impact ionization as a function of kinetic energy (per nucleon)
  of $^{12}$C$^{6+}$ ions.}
\end{figure}

 One important characteristics that can be obtained from the SDCS is the average energy of the secondary electrons, $\langle W \rangle$, which is given by
\begin{eqnarray}
{\langle W \rangle}(E)=\frac{1}{\sigma_t}\int^{\infty}_0 W\frac{d\sigma_t}{dW}dW~.
\label{eq5}
\end{eqnarray}
\noindent The dependence of the average energy of electrons on the energy of the projectile, given by the result of integration (\ref{eq5}) for liquid water medium is shown in Fig.~\ref{WaveC}. Notice, that this figure is different from similar figures of Refs.~\cite{epjd,pre}, where the calculations were done with parameters for water vapour.
This dependence indicates that the energy of secondary electrons is somewhere below 50~eV for the whole range of the ion's energy and it levels out as the the energy of projectiles increases. There are several consequences from this. First, since the dependence of $\langle W \rangle$ on the ion's energy $E$ on a relevant range of projectile energies (0.3 -- 400~MeV/u) is weak for the large range of the ion's energy, the number of produced secondary electrons is largely proportional to the value of LET, more precisely to the electronic component of the LET, $S_e$, that excludes nuclear stopping. Indeed, if the ion is destroyed in a nuclear collision, ionization due to its debris should be discussed instead; if it survives then its ionizing capabilities do not change too much, unless it slows down considerably; then, its stopping power may change correspondingly. Second, the expression for $\langle W \rangle$ is independent of the charge of the projectile, e.g., the difference between, say, protons and iron ions is in their values of $S_e$, i.e., in the number of secondary electrons, but not in their relative energy spectra. Therefore, the difference between the effects of these different ions will be in the number of secondary electrons produced by these ions per unit length of path. Third, most of the secondary electrons are capable of ionizing just one or two water molecules; thus, there is no significant avalanche ionization effect~\cite{epjd}.  This can be explained by a simple estimate. Since the average energy of secondary electrons in the vicinity of the Bragg peak is about 40~eV (somewhat below this value), the maximum average energy that can be transferred to the next generation secondary electron is just
$(40-I_i)/2$, which is about 15~eV for the outermost electrons, an
energy barely enough to cause further ionization. Finally, what is of crucial importance for the consideration of the next scale of electron propagation is that at sub-50~eV energies, the electrons' cross sections are nearly isotropic~\cite{Nikjoo06,Tung} and it is possible to use the random walk approximation in order to describe their transport~\cite{precomplex,pre,epjdisacc2011,epjdmarion}. This transport is described in the next section.

\section{Random walk approximation for the description of the secondary electron transport}
\label{ssec2}

{ The next stage of the scenario is related to secondary electrons ejected from the molecules of the medium as a result of ionization. As has been discussed above, most of these electrons have energies below 50~eV. They are called ballistic electrons until their energy becomes sufficiently small and coupling with phonons, recombination, and other quantum processes start dominating their transport. While the electrons are ballistic, their interactions with molecules can be described as a sequence of elastic and inelastic collisions. Many works, by and large using MC simulations, describe the transport of ballistic electrons. They are known as track structure codes~\cite{Nikjoo06}. Some of them describe chemical reactions in the medium including production of radicals and their propagation. However, regardless of how sophisticated these codes are, they do not contain the whole physical picture as will be shown below. In this section, a rather simple analytical approach is applied to the description of the propagation of ballistic electrons and its results are compared to MC simulations. It is  also demonstrated how to make sense of radiation damage based on these calculations.

The main mechanism of radiation damage by ballistic electrons is inelastic collisions with targets. A target in this discussion is a biomolecule, such as DNA. Therefore, the probability of biodamage is a combination of the number of electrons (or other secondary particles) colliding with a given segment of a biomolecule and the probability of a certain inelastic process on impact. The first part is described by the fluence of electrons or other particles on the target. Fluence is the integral of the flux of particles (the number of particles hitting a part of the target's surface per unit time) over the entire time after the ion's passage and over the surface of the target. In general, the fluence depends on the distance of the target from the ion's path and its geometrical orientation. It will be shown that this part can be calculated analytically with accuracy, sufficient for understanding the scenario of radiation damage. The second part, i.e., the probability of a certain inelastic process on impact, is more difficult to assess mainly because of the diverse variety of possible processes. However, there are plenty of data that allows one to make reasonable quantitative estimates for this probability.

Let us start with the calculation of fluence for a number of relevant configurations. It will be shown that important characteristics of the track structure such  as radial dose can also be calculated via fluence. The random walk approach~\cite{chandra} used for these problems allows one to make simple analytical calculations of fluence. The main requirement for the use of this approach is that the elastic and inelastic scattering of secondary electrons is isotropic. The anisotropy in the angular dependence of the cross sections for sub-50-eV electrons appears to be insignificant~\cite{Nikjoo06}. As was noted above, more than 80\% of secondary electrons satisfy this condition and only for less than 10\% of $\delta$-electrons with energies higher than 100~eV is this condition violated significantly. The effects of $\delta$-electrons will be considered in Sec.~\ref{sec.delta}. In Sec.~\ref{sec.radicals}, the transport and effects of radicals, whose role in radiation damage is quite substantial, will be discussed.

Sections~\ref{sec.fluence}--\ref{sec.fluence.twist} are devoted to the transport of sub-50-eV electrons. Moreover, unless specifically stated to the contrary, these secondary electrons are produced by carbon ions in the vicinity of the Bragg peak in liquid water. At this part of the ion's trajectory, while a 0.3-MeV/u carbon ion passes 1~$\mu$m along the path, a typical radius within which the secondary electrons propagate is about
1~nm~\cite{prehydro,natnuke}. This allows one to assume that the
electron diffusion is cylindrically symmetric with respect to the
ion's path. The electronic component of the LET, $S_e$, remains nearly constant along this 1~$\mu$m of ion's path described by the coordinate $\zeta$. Therefore, the number of ejected secondary electrons per unit length $\frac{dN_e}{d \zeta}$ is independent of ${\zeta}$.
A typical elastic mean free path of sub-50-eV electrons $l$ ranges between 0.1 and
0.45~nm~\cite{Nikjoo06,Tung}. 
Since the scale along the Bragg peak is measured in tens of $\mu$m, while the radial scale is only tens of nm, therefore one can assume $\zeta$ to be ranging from $-\infty$ to $+ \infty$.
}

\subsection{Calculation of the fluence of secondary electrons}
\label{sec.fluence}

\begin{figure}
\resizebox{1.0\columnwidth}{!}{ \includegraphics{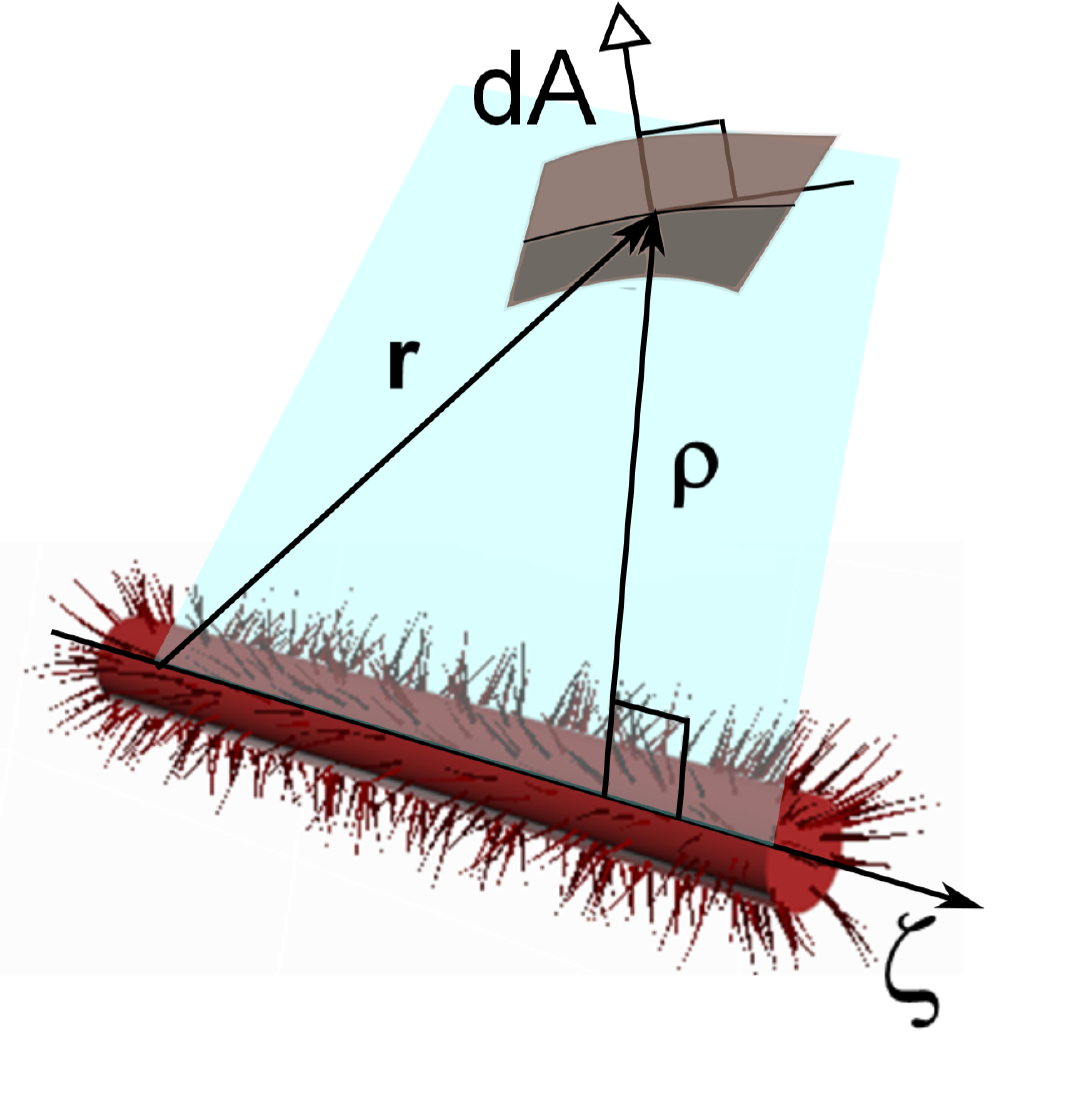} }
\caption{\label{fig.fluencegeo} (Colour online) Geometry for the general calculation of fluence through a segment of surface $d {\bf A}$. The ion path is along the axis. }
\end{figure}

In the three-dimensional axially symmetric propagation of ballistic
electrons from the axis, the key differential quantity is the flux of secondary electrons originating from a segment $d \zeta$ of the ion's path through an area $d{\bf A}$ located at a distance $\rho$ from the ion's path, as is shown in Fig.~\ref{fig.fluencegeo}. Vector {\bf r} connects the element $d \zeta$ with $d{\bf A}$.
This flux is given~\cite{epjdisacc2011,epjdmarion,chandra} by the following expression:
\begin{eqnarray}
\frac{dN_d({\bf r}, t)}{dt} = d{\bf A}\cdot D \nabla P(t, r)\frac{d
N_e}{d \zeta} d \zeta \nonumber\\= d{\bf A}\cdot D { \bf n_r}\frac{\partial
P(t, r)}{\partial
  r}\frac{d
N_e}{d \zeta}d \zeta~, \label{mult1t}
\end{eqnarray}
where, $D={\bar v} l/6$ is the diffusion coefficient, ${\bar v}$ is
the average speed of electrons, ${\bf n_r}$ is a unit vector in the
radial direction (from the segment to the center of the area $d{\bf
A}$), and
\begin{eqnarray}
P(t, r)=\left(\frac{3}{2\pi {\bar v}t
l}\right)^{3/2}\exp\left(-\frac{3 r^2}{2{\bar v} t l}\right)
\label{rwalk2t}
\end{eqnarray}
is the probability density to observe a randomly walking electron at
a time $t$ and a distance $r$ from the electron's origin.

The next step in the calculation of fluence is the integration of Eq.~(\ref{mult1t}) over time. In order to do this, we change variables from $t$ to the number of steps by secondary electrons $k$ using ${\bar v}t=kl$. We rewrite Eq.~(\ref{mult1t}), substituting (\ref{rwalk2t}), and switching from variable $t$ to $k$ as
\begin{eqnarray}
dN_A({\bf r})= \int \frac{dN_d(\vec r, t)}{dt} dt=d{\bf A}\cdot{\bf
n_r}\frac{d N_e}{d \zeta}d \zeta \int_{r/l}^\infty d k \nonumber\\ \times
\frac{r}{2 k}
     \left(\frac{3}{2 \pi k l^2}\right)^{3/2}
\exp\left(-\frac{3 r^2}{2 k l^2}-\gamma k\right)~.\label{mult4}
\end{eqnarray}
An attenuation exponential factor $e^{-\gamma k}$ is introduced in
order to take into account electrons falling out from the random
walk. The coefficient $\gamma$ is equal to the ratio of the cross section
of processes in which electrons stop being ballistic to the total cross section. An example of such a process is an inelastic collision of an electron with a water molecule after which the energy of the electron drops below a certain excitation or ionization threshold related to the molecules of the medium. This does not completely inactivate it as an agent of radiation damage since it may attach itself to a molecule and bring about its dissociation, but such electrons vanish from the picture of radial dose delivery.

{
The integration over $k$ in Eq.~(\ref{mult4}) is carried out from the minimal number of steps necessary to reach a distance $r$ to infinity. After that, the fluence, ${\cal F}(\rho)$, can be calculated as the integral over the surface of the target:
\begin{eqnarray}
{\cal F}(\rho)=\int_A dN_A({\bf r})= \int_A d{\bf A}\cdot{\bf
n_r}\frac{d N_e}{d \zeta}d \zeta \int_{r/l}^\infty d k \nonumber\\ \times
\frac{r}{2 k}
     \left(\frac{3}{2 \pi k l^2}\right)^{3/2}
\exp\left(-\frac{3 r^2}{2 k l^2}-\gamma k\right)~.\label{fluence.for}
\end{eqnarray}
Strictly speaking, the fluence given by Eq.~(\ref{fluence.for}) depends on more variables than just the distance between the target and the ion's path. These variables include the elastic mean free path of secondary electrons and more geometrical parameters. The mean free path corresponds to some energy between zero and 50~eV and thus energy averaging is achieved. After this averaging, the energy of electrons is assumed to be constant. In different works~\cite{precomplex,prauger,pre,epjdisacc2011,epjdmarion} this averaging was done according to the particular physical problem. However, it is the dependence on the distance $\rho$, kept in Eq.~(\ref{fluence.for}), that remains important for calculations of radiation damage. The application of this method to specific geometries that were considered in some of these works are demonstrated below.}

\subsection{Calculation of the radial dose}
\label{sec.radial}

The radial dose is an important quantity in the physics of IBCT since the dose distribution around the ion path is highly non-uniform. Starting from the works of Katz et al.~\cite{Katz67,Katz71,KatzRBE} the radial dose has been used for the assessment of radiation damage with ions. Since then the radial dose has been calculated since then using MC
\begin{figure}
\resizebox{1.0\columnwidth}{!}{ \includegraphics{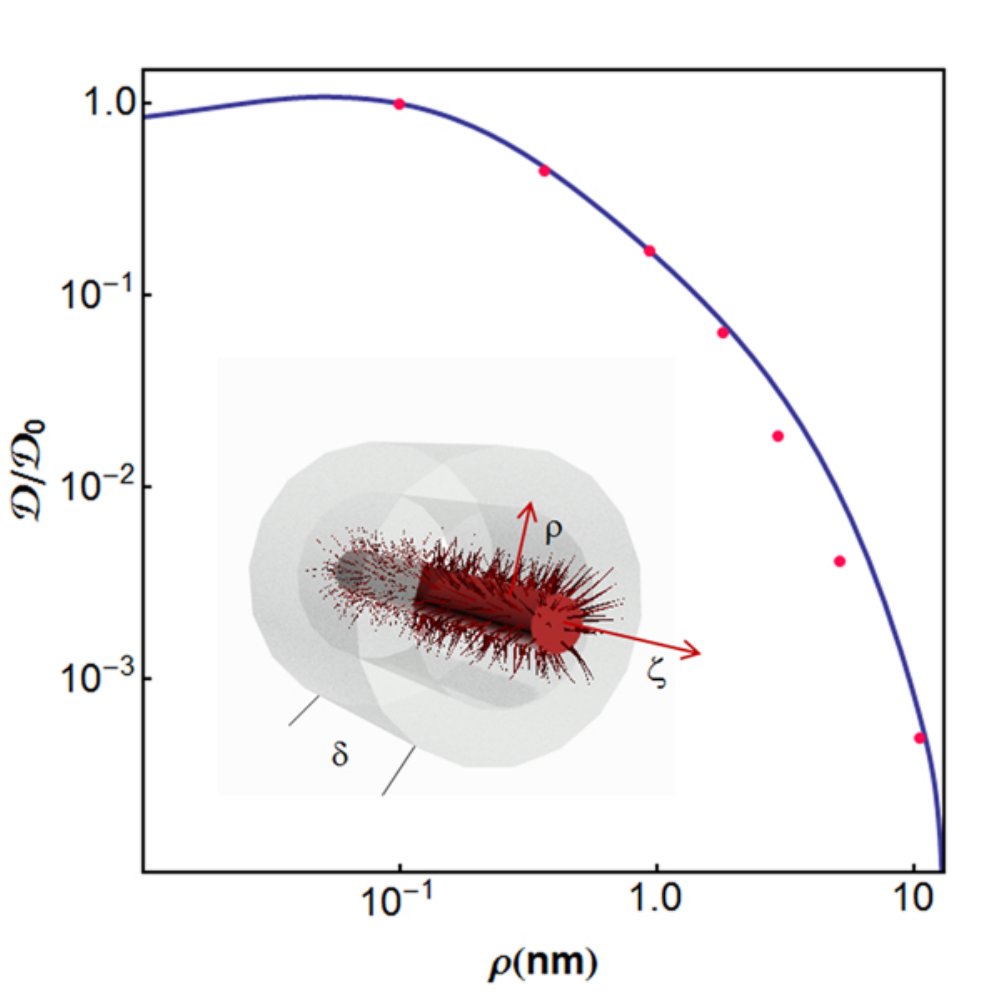} }
\caption{\label{fig.raddosegeo} (Colour online) The normalised radial dose determined using the random walk approximation (solid line) compared
with the results of Ref.~\cite{31} for 1-MeV protons (dots).
These dots are digitized from the solid line and thus only represent
a fragment of the data. These calculations were done
with l = 0.15~nm and $\gamma = 0.0006$~\cite{epjdisacc2011}. In the inset, the geometry for the calculation of radial dose. The ion path is along the axis. Secondary electrons propagate radially and the energy is deposited in the coaxial cylindrical shell of length $\delta$, inner radius $\rho$, and outer radius $\rho+d\rho$. }
\end{figure}
simulations~\cite{31,Cucinotta98,Plante}. In Ref.~\cite{epjdisacc2011}, it was shown that the radial dose, i.e., the locally absorbed energy density as a function of the distance from the ion's path, $\rho$, can be calculated  analytically using the random walk approach. 

This calculation is based on the application of Eq.~(\ref{fluence.for}) to the simplest geometry, where the target is a cylinder of radius $\rho$ and length $\delta$, coaxial with the ion's path, shown in the inset of Fig.~\ref{fig.raddosegeo}.
In Sec.~\ref{sec.app.raddose} it is shown how to calculate the number of secondary electrons, ${\cal F}_\delta (\rho)$, incident on such a surface.
Then, the number of ionization events in a shell between $\rho$ and $\rho+d\rho$ is proportional to the number of secondary electrons (of a given energy) incident on the inner cylindrical surface multiplied by the
probability of ionization per electron (effective area over the total
area). This is equal to the number of water molecules inside the
volume (number density times volume $n 2\pi \rho \delta d\rho$)
multiplied by the ionization cross section $\sigma$ and divided by
the total area of the cylindrical shell ($2 \pi \rho \delta$)
\begin{eqnarray}
d{\cal N}={\cal F}_\delta(\rho)\frac{n \sigma 2\pi \rho \delta d\rho}{2 \pi
\rho \delta} = {\cal F}_\delta(\rho){n \sigma  d\rho}~.\label{mult9do}
\end{eqnarray}
The energy deposited in this shell is equal to the product of this number of events and the average energy per event ${\bar W}$,
\begin{eqnarray}
d{\cal E}={\bar W} {\cal F}_\delta(\rho){n \sigma  d\rho}~.\label{mult10do}
\end{eqnarray}
Finally, the radial dose is the volume density of the deposited
energy, i.e., Eq.~(\ref{mult10do}) divided by the volume of the
shell $2\pi \rho \delta d\rho$:
\begin{eqnarray}
{\cal D}(\rho)={\bar W}\frac{{\cal F}_\delta(\rho){n \sigma  d\rho}}{2\pi
\rho \delta d\rho}
={\bar W}n\sigma\frac{dN_e}{d\zeta}{\cal Q}(\rho/l,\gamma)~,
\label{mult11do}
\end{eqnarray}
where the function ${\cal Q}(\rho/l,\gamma)$ is defined by Eq.~(\ref{mult5y}) in the Appendix.

Thus, the radial dose due to ions propagating in a medium in the
vicinity of the Bragg peak, obtained using a random walk
approximation, is given by Eq.~(\ref{mult11do}).
This dependence is studied analytically in some special cases~\cite{epjdisacc2011}. It compares reasonably well with the MC simulations of Ref.~\cite{31} at small and moderate distances from the ion's path, as shown in Fig.~\ref{fig.raddosegeo}. The radial dose calculated using the random walk as well as that obtained using MC simulations corresponds to the radial dose in a static medium where the effects of relaxation of the deposited energy are not included. This corresponds to the dose distribution up to $10^{-14}$~s when this relaxation takes place and leads to collective flow effects (discussed in Sec.~\ref{sec.flow}), which increase the volume around the ion's path where the energy is absorbed and thus decrease the radial dose.

\subsection{Targeting a twist of DNA with secondary electrons}
\label{sec.fluence.twist}

The first analytical calculation of biodamage using a random walk approach
was done in Ref.~\cite{pre}, where the dependence of the fluence through a twist of DNA, which was represented as a cylinder of size corresponding to one twist of a DNA molecule (radius of 1.15~nm and length of 3.4~nm), was calculated. A choice of a twist of a DNA molecule as a target is related to the types of DNA damage, such as single and double strand breaks (SSB and DSB) which are widely discussed in the literature~\cite{Chatterjee93,hyd2,Ward2}. The DSB is a severe lesion, which can still be repaired, but its contribution to the probability of cell death is significant. The DSB is defined as two SSBs of the opposite strands within 10 base pairs of each other, i.e., within a single twist of a DNA molecule.

\begin{figure}
\resizebox{\columnwidth}{!}{\includegraphics{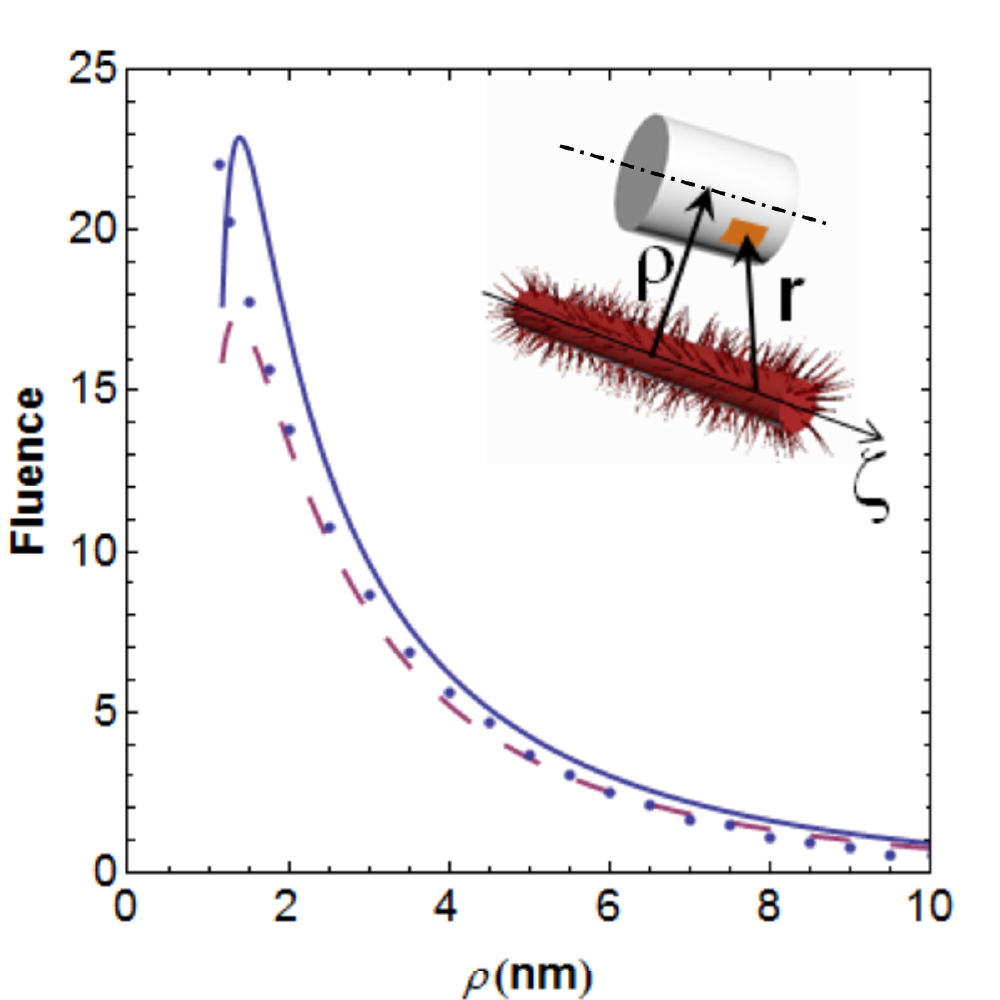} }
\caption{\label{fig.fluence} (Colour online) Fluences of secondary electrons produced by a single $^{12}$C$^{6+}$ ion in the vicinity of a Bragg peak through a cylinder enwraping a DNA twist are shown with respect to the distance of the cylinder from the ion's path. Two different orientations (parallel and perpendicular) are shown as well as MC simulations for the perpendicular case. In the inset, the geometry for the calculation of fluence through a cylinder enwraping a DNA twist is shown. In this figure, the cylinder is perpendicular to the ion's trajectory and symmetric with respect to the plane of incidence.}
\end{figure}

The probability of an SSB or a DSB in a given twist is related to the fluence of secondary electrons produced by the passing ion. Therefore, the first problem is to calculate the fluence of these electrons through a cylinder enwraping the twist. This cylinder may be arbitrarily oriented with respect to the ion's path. A perpendicularly (and symmetrically) oriented cylinder is shown in the inset of Fig.~\ref{fig.fluence}.

In Appendix A, it is shown how to apply Eq.~(\ref{fluence.for}) to different orientations of a cylindrical target. With expressions for $r^2$ and ${\bf n}_r\cdot d{\bf A}$, the integrations over $\zeta$ and the area of cylinder~(\ref{fluence.for}) give the fluence through the cylinder in two limiting cases of different orientation. The results are shown in Fig.~\ref{fig.fluence},
where, the fluence through a perpendicular cylinder is compared with MC simulations~\cite{epjdmarion}. The fluence through a parallel cylinder is larger than that of a perpendicular one by about 20\%. This allows one to average the fluence through the cylinder enwraping the DNA twist over its orientation with respect to the ion's path. This fluence can be extrapolated to the region $0<\rho<a$ , where $a=1.15$~nm is the radius of the cylinder enwraping the twist in order to estimate the probabilities of DNA damage in the ``whole'' space without limitations.

\subsubsection{Calculation of the number of SSBs per single ion}
\label{sec.ssb}

An estimate of the number of SSBs per unit length of the ion's trajectory can be obtained assuming that this number is proportional to the number of secondary electrons incident on a given twist of a DNA molecule. For example, for a straight segment of length $d \zeta$ of the ion's path, the number of SSBs is given by the integral,
\begin{eqnarray}
\frac{d N_{SSB}}{d \zeta}=\Gamma_{SSB}\int_0^\infty {\cal F}(\rho) n_t 2\pi \rho d \rho~,
\label{n.ssb}
\end{eqnarray}
where $\Gamma_{SSB}$ is the probability that an electron incident on a DNA twist induces a SSB and $n_t$ is the number density of DNA twists (i.e., cylinders). Since the spatial dependence of $n_t$ is unknown, it is reasonable (in the first approximation) to assume that it is constant. The fluence ${\cal F}(\rho)$ for carbon ions at the Bragg peak, obtained in Sec.~\ref{sec.fluence.twist}, can be substituted in the integral (\ref{n.ssb}).  This gives us an estimate of
\begin{eqnarray}
\frac{d N_{SSB}}{d \zeta}= \Gamma_{SSB} n_t \phi~,
\label{nssb.est}
\end{eqnarray}
where $\phi=\int_0^\infty {\cal F}(\rho) 2\pi \rho d \rho=1.1\times 10^3$~nm$^2$. The value of $\phi$ is obtained using a simple diffusion model that contains two parameters, the mean free path $l$ (assumed to be the same for all electrons) and the ratio of elastic and inelastic cross sections $\gamma$. The third input in this value is $\frac{dN_e}{d\zeta}$. This number can be calculated using ionization cross sections discussed in Sec.~\ref{sec.sdcs}. However, the number calculated from Eq.~(\ref{nssb.est}) only includes the electrons ejected in the primary ionization with projectiles and does not include ionizations due to secondary electrons. Since the low-energy electrons produced in the latter are important for biodamage, the number $\frac{dN_e}{d\zeta}$ and, therefore, both the fluence and $\phi$ are underestimated by a factor of about two~\cite{epjd}.

{
The probability of the production of a SSB by an electron incident on a twist of a DNA molecule, $\Gamma_{SSB}$, appears in Eq.~(\ref{n.ssb}) as well as in (\ref{nssb.est}). This probability can be estimated as the cross section for breaking an important covalent bond that leads to a SSB multiplied by the number of such bonds in a single DNA twist and divided by the lateral area of this DNA segment, represented above by a cylinder. However, the cross section for breaking a covalent bond is energy-dependent and the energy of secondary electrons varies from zero to about 50~eV. At low energies (below the ionization threshold) the cross section is deemed to be that of DEA, i.e., resonant attachment of the secondary electron to the molecule (formation of temporary negative ion) followed by dissociation (SSB). At higher energies of impact electrons, the cross sections contributing to $\Gamma_{SSB}$ are defined by the ionization cross sections provided that the formation of a cation leads to a strand break. These processes are being studied theoretically and experimentally~\cite{SancheCh9.2012,SevReview,Simons07,Fabrikant12,SevDQD,Gianturco1}.
Their typical cross sections vary, but the cross section for a SSB as a consequence of DEA for about 1-eV electrons can be up to 10~nm$^2$~\cite{SancheCH1,Sanche2006} per plasmid DNA, which can be converted to $3\times10^{-2}$~nm$^2$ per single twist and therefore $\Gamma_{SSB}=10^{-3}$. For higher energy electrons, ionization of a DNA molecule does not necessarily lead to a SSB and many pathways are being discussed. Nevertheless, reported SSB yields at higher electron energies are of the same order (if not higher) as those for low-energy electrons~\cite{SancheCh9.2012}, which once again gives $\Gamma_{SSB}\approx 10^{-3}$.


\subsubsection{Calculation of the number of DSBs}
\label{sec.dsb}

The estimate of the number of DSBs is more ambiguous than that of SSBs. This is mainly due to the lack of understanding of the mechanism of producing this lesion. Many works~\cite{Sanche05,DNA3} suggest that a DSB is the result of the action of a single electron that dissociatively attaches to a DNA molecule. The dissociative attachment is considered to be an important pathway of SSBs at very low energies and in about one out of five such incidences, a DSB takes place due to the interactions with the debris of a SSB~\cite{DNA3}. Alternatively, DSBs can be due to two separate SSBs on opposite strands. This may be possible if the number density of secondary electrons is high enough. 
It is also possible that double ionization events play a significant role~\cite{prauger}. Such events create a high local number density of low energy electrons at a considerable distance from the ion's path and if this occurs in the vicinity of a DNA twist, at least two of the three electrons involved in a double ionization event may be incident on the same twist. This depends on the values of the cross sections for double ionization. The probability of ICD-effects on DNA molecule and water molecules adjacent to it may also be an important factor~\cite{prauger,Becker}.

Regardless of the pathway for DSBs, for a given ion in a given medium, the ratio of yields of DSBs and SSBs (per unit length of ion trajectory) is fixed and {\em dose independent} unless tracks of different ions interact. Indeed, each ion's track is determined by the type of ion and an increased dose just means an increase in the density of ion tracks. Only after some critical value of dose is reached, do the tracks start overlapping. Only then can the dependencies of yields of SSBs and DSBs on dose become not proportional to each other. These conditions are not being observed in the analysed experiments or in therapy\footnote{In this section only effects of secondary electrons are discussed. The situation may be different when radicals are included, see Sec.~\ref{sec.radial}.}, however, if laser-driven ion beams are used~\cite{Bulanov}, track interaction effects may become important.

Therefore, the DSB yield can be calculated as a sum of two terms, the first of which represents the events where SSBs are converted to DSBs and the second accounts for DSBs due to separate electrons. In order to calculate the second term, the average number of SSBs per twist, ${\cal N}$, can be introduced as\footnote{This number is a part of the integrand of Eq.~(\ref{n.ssb}).}
\begin{eqnarray}
{\cal N}=\Gamma_{SSB}{\cal F}(\rho)~.
\label{caln1.ssb}
\end{eqnarray}
Then, the probability of a DSB due to two separate electrons in this twist is given by $\frac{1}{2}{\cal N}^2\exp{\left[-{\cal N}\right]}$. The second term in the DSB yield is given by the integration over the volume similar to Eq.~(\ref{n.ssb}). Thus, the estimate for DSBs is given by
\begin{eqnarray}
\frac{d N_{DSB}}{d \zeta}=\lambda \Gamma_{SSB} n_t \int_0^\infty {\cal F}(\rho) 2\pi \rho d \rho ~~~~~~~~~~~\nonumber\\+\frac{n_t}{2}\int_0^\infty {\cal N}^2\exp{\left[-{\cal N}\right]} 2\pi \rho d \rho,
\label{n.dsb}
\end{eqnarray}
where ${\cal N}$ is given by Eq.~(\ref{caln1.ssb}) and $\lambda$ is a fraction of SSBs converted to DSBs, i.e., the number of DSBs due to the action of a single electron.

At this point the phenomenon-based approach can be related to experiments. If real tissue is irradiated, one can only find the percentage of cells surviving. If this value is measured as a function of dose, the survival curve is obtained as a result. Many interactions on sub-cellular, cellular, or even at the organismic level may affect the survival curve. In {\em in vitro} experiments on cell cultures, elimination of some of these interactions allows, e.g., synchronizing cell cycles, control over the environment, etc. Still, there are no direct ways of relating cell death to, e.g., DSBs produced by secondary electrons. Therefore, the comparison with experiments on DNA molecules irradiated with ions is the most appropriate.

\subsubsection{Comparison with experiments on plasmid DNA}
\label{sec.plasmid}

Of all the experiments investigating DNA molecules irradiated with ions, the study of plasmid DNA is the most valuable, since there are reasonably reliable ways to distinguish the intact molecules from those with a SSB and from those with a DSB. Another important feature is that the effects of DNA damage observed in these experiments are not affected by the biological effects of repair that take place in living cells. This allows for a more pure comparison.

An undisturbed plasmid is a closed loop of a supercoiled DNA molecule~\cite{plasmid}. This loop contains a given number of base pairs, e.g., in experiments described in Ref.~\cite{Thomas1} plasmid DNA pBR322 irradiated with carbon ions contains 4361~bp. The characteristic size of this molecule is about 100~nm. If such a molecule experiences a SSB, it becomes ``circular'' or just a loop without the supercoil structure. A DSB makes the plasmid ``linear'' since both of its strand are broken. These structural conformations can be distinguished using electrophoresis or high-performance liquid chromatography~\cite{SancheCh9.2012,Thomas1}. This allows the measuring of SSB and DSB yields experimentally. In one of the experiments described in Ref.~\cite{Thomas1}, plasmid DNA was dissolved in a 600~mmol/l solution of mannitol in water. Mannitol serves as a radical scavenger so their contribution to DNA damage may be neglected. This is adequate for the theoretical treatment (Sections~\ref{sec.ssb} and \ref{sec.dsb}), which only includes secondary electrons.

The results of experiments of Ref.~\cite{Thomas1} are shown in Fig.~\ref{fig.cf48.cor} with dots. They represent the probabilities for two outcomes after an irradiation with carbon ions at the spread-out Bragg peak. The first outcome (open squares) is for the plasmid to become open circular (not supercoiled), associated with a SSB. There is a reported problem with the quality of the data resulting in the probability corresponding to SSBs not starting from zero at a zero dose~\cite{Thomas1}. This means that some plasmids are either not supercoiled to begin with or appear as such in the electrophoresis. This probability remains elevated by about the same value throughout the dose range. In order to compare these data with our calculations, the zero-level probability of the SSB yield was subtracted in order to ``clean'' the data. These data points are shown with filled squares. The second outcome is for the plasmid to become linear, associated with a DSB and is shown with filled circles. These probabilities (filled squares and circles) are monotonically increase with dose with the SSB dependance being slightly non-linear. In order to explain these data using the multiscale approach, let us start with the dose dependance.
\begin{figure}
\resizebox{1.0\columnwidth}{!}{ \includegraphics{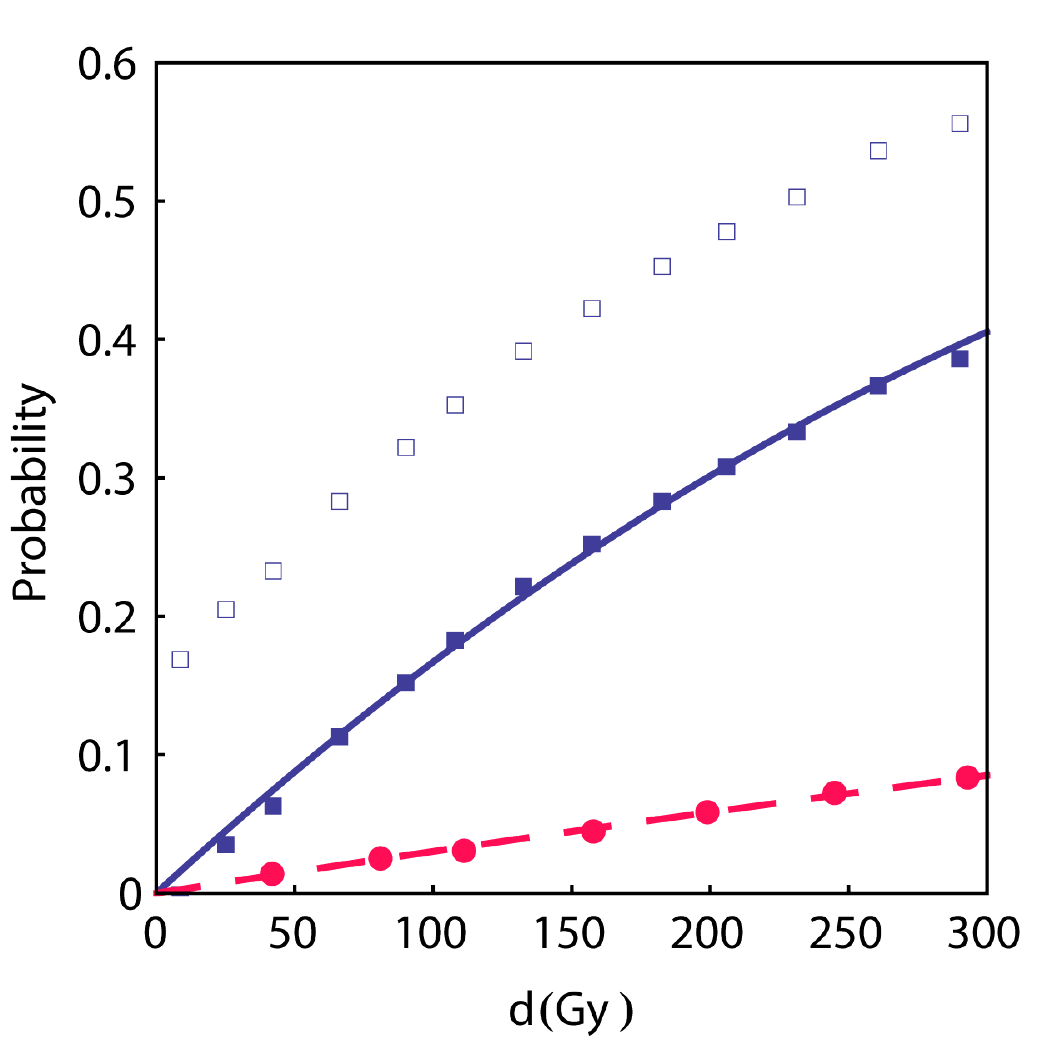} }
\caption{\label{fig.cf48.cor} (Colour online) Probabilities for SSBs and DSBs induced in plasmid DNA by secondary electrons as a function of dose. Dots correspond to experiments~\cite{Thomas1}: open squares to the original SSBs, filled squares to the ``cleaned'' SSBs, and filled circles to DSBs. Calculated probabilities are shown with lines. Solid line corresponds to the probability of SSBs calculated using Eq.~(\ref{ssb.plasmid}). The dashed line depicts the probability for DSBs calculated using Eq.~(\ref{dsb.plasmid}).}
\end{figure}

When a beam of carbon ions is incident on the plasmid solution, there is a dose-dependent probability that $\nu$ ions will traverse through a plasmid. This probability is given by the Poisson distribution:
\begin{eqnarray}
P_{\nu}=\frac{N_{ion}^\nu}{\nu !}\exp{\left[-N_{ion}\right]}~,
\label{one.ion.pois}
\end{eqnarray}
where $N_{ion}$ is the average number of ions passing through the cross sectional area of a plasmid, $A_p\approx 7.8\times 10^3$~nm$^2$ \cite{plasmid}. The average number of ions passing through this area is equal to the ratio of this area to the average area per ion. The average area per ion, ${\cal A}$, can be calculated if a uniform distribution of ions in the beam is assumed. Then the dose is equal to the LET (which is associated with the average for the Bragg peak stopping power due to ionization processes, ${\bar S}_e$) divided by the average area per ion, i.e., $d=\frac{{\bar S}_e}{\cal A}$. Then, $N_{ion}$ is given by:
\begin{eqnarray}
N_{ion}=\frac{A_p}{\cal A}=\frac{A_p}{{\bar S}_e}d~.
\label{nions}
\end{eqnarray}
In Ref.~\cite{Thomas1}, the average LET over the spread-out Bragg peak, ${\bar S}_e$ is $189\pm15$~eV/nm. This includes energy straggling effect along the ion's trajectory.
Then, the number of SSBs that are likely to be induced in a plasmid, i.e., SSB yield per plasmid is given by the sum,
\begin{eqnarray}
Y_{SSB,e}=\frac{d N_{SSB}}{d \zeta} {\bar x}_p \sum_{\nu=1}^\infty \nu P_{\nu}~,
\label{ssb.plasmid}
\end{eqnarray}
where ${\bar x}_p$ is the average length of an ion's path through the plasmid. The subscript ``e'' indicates that this yield is only due to secondary electrons. Each term of this sum is a product of the number of SSBs per unit length of trajectory of a single ion, the length of this trajectory through the plasmid and the number of ions traversing the plasmid. The length of a trajectory, ${\bar x}_p$, is equal to the average chord length of a sphere, representing a plasmid, which is about 0.78 of its diameter. If exactly $\nu$ ions pass through the plasmid, this length is multiplied by $\nu$. This is the first term in the sum of Eq.~(\ref{ssb.plasmid}). Then the factor $P_{\nu}$ gives the probability that $\nu$ ions are passing through the target. Hence, the whole sum multiplied by ${\bar x}_p$ determines the average length of tracks through the plasmid.

The sum in Eq.~(\ref{ssb.plasmid}) does not include interactions of different ions that could occur if trajectories of two or more ions are so close that the same twist of a DNA molecule could be hit with electrons originating from the different ions. The probability of such an interference can be estimated. Since the range of 50-eV electrons in liquid water is about 10~nm, the two ion's trajectories must be within 20~nm, for the interference to occur. Then the estimate is obtained from Eqs.~(\ref{one.ion.pois}) and~(\ref{nions}) with $\nu=2$ and $A_p=\pi\times 10^2$~nm$^2$. For the maximal dose of 300~Gy used in Ref.~\cite{Thomas1} the resulting probability is $5\times 10^{-6}$. This number is very small compared to the probability that one ion will pass through the plasmid at this dose (equal to 0.3) or even that two ions will pass through it (equal to 0.02). Therefore, the interference term in Eq.~(\ref{ssb.plasmid}) can be neglected.

The only term of  Eq.~(\ref{ssb.plasmid}) that depends on dose is $P_{\nu}$, therefore the dose dependence of the yield is contained in the sum $\sum_{\nu=1}^\infty \nu P_{\nu}$. This dependence is not unique for the yield of SSBs. The same sum appears in all calculations, provided that the damage due to each ion is localised in its track and the tracks do not interfere. The dependence of this sum on dose is asymptotically exponential at large values of $N_{ion}$. This means that on a semi-logarithmic plot the dose dependence will be asymptotically a straight line. This will be seen below in the analysis of survival curves in Sec.~\ref{sec.surv}.

The numbers relevant to the experiments of Ref.~\cite{Thomas1}, such as $A_p=7.8\times 10^3$~nm$^2$ and ${\bar S}_e=189$~eV/nm, substituted to Eq.~(\ref{nions}) give $N_{ion}=2.6\times 10^{-4} d$ with the dose in Gy. This means that even at the highest dose of 300~Gy used in Ref.~\cite{Thomas1} $N_{ion}\ll 1$. However, Ref.~\cite{Thomas1} gives the dose dependence of the {\em probability} of a SSB per plasmid rather than yield. This probability is given by Poisson statistics,
\begin{eqnarray}
P_{SSB,e}=Y_{SSB,e}\exp\left[-Y_{SSB,e}\right]~~~~~~~~~\nonumber\\
+\frac{1}{2}Y_{SSB,e}^2\exp\left[-Y_{SSB,e}\right]~,
\label{prob.SSB.plas}
\end{eqnarray}
where the first term corresponds to a single SSB in the plasmid DNA and the second term corresponds to two SSBs on the same strand.
The fit of Eq.~(\ref{prob.SSB.plas}) to the probability of the SSB dependence on dose, shown in Fig.~\ref{fig.cf48.cor}, gives $\frac{d N_{SSB}}{d\zeta} {\bar x}_p=0.12$. If ${\bar x}_p\approx 75$~nm, $\frac{d N_{SSB}}{d\zeta} \approx 1.6~\mu$m$^{-1}$, then comparing this with Eq.~(\ref{nssb.est}) and taking $n_t=5.6\times 10^{-2}$~nm$^{-3}$, we obtain an estimate for $\Gamma_{SSB}=1.9\times 10^{-3}$, which is larger than the the value estimated in Sec.~\ref{sec.ssb} on the basis of experimental results by the factor of 1.9.

Now the comparison for DSBs can be made. Similar to Eq.~(\ref{ssb.plasmid}), the number of DSBs induced in a plasmid (a DSB yield per plasmid) is given by the sum,
\begin{eqnarray}
Y_{DSB,e}=\frac{d N_{DSB}}{d \zeta} {\bar x}_p \sum_{\nu=1}^\infty \nu P_{\nu}~,
\label{dsb.plasmid}
\end{eqnarray}
and the probability of a DSB per plasmid is given by,
\begin{eqnarray}
P_{DSB}=Y_{DSB,e}\exp\left[-Y_{DSB,e}\right]~.
\label{prob.DSB.plas}
\end{eqnarray}
A fit of Eq.~(\ref{prob.DSB.plas}) to the probability of the DSB dependence on dose (for Ref.~\cite{Thomas1}) gives $\frac{d N_{DSB}}{d\zeta} {\bar x}_p=0.015$. The substitution of ${\bar x}_p\approx 75$~nm gives $\frac{d N_{DSB}}{d\zeta} \approx 0.2~\mu$m$^{-1}$. Then comparing this with Eq.~(\ref{n.dsb}) and taking $n_t=5.6\times 10^{-2}$~nm$^{-3}$, we obtain an estimate for $\lambda=0.15$, which is in reasonable agreement with the values between 0.1 and 0.2 for different electron energies~\cite{Sanche05,DNA3,SancheCh9.2012}. Thus, the comparison of our model for the effect of secondary electrons with the results of Ref.~\cite{Thomas1} for a plasmid DNA solution in the presence of radical scavengers is reasonable.

Some comments regarding these calculations should be made. First, as has been noted in Sec.~\ref{sec.ssb}, the number of secondary electrons is underestimated. This happens because in our calculations only the electrons ejected by ions were included, missing those ejected in the process of secondary ionization by electrons. The correction for this number will increase fluence, but will not affect the dose dependence. Since the actual fluence will then be larger (by the factor of about two~\cite{epjd}), $\Gamma_{SSB}$ will be smaller (by the same factor, i.e., closer to $10^{-3}$ (see Sec.~\ref{sec.ssb}).
The second issue is that the treatment of a supercoiled plasmid as an object with uniformly distributed chromatin may be a little far-fetched.  Also, if a plasmid suffers a single strand break, its size increases by a factor larger than two and then it may be a target for another ion. Nevertheless, the comparison that was just made is quite reasonable and encouraging for further steps in the assessment of radiation damage.

\subsection{Damage of plasmid DNA in the presence of free radicals}
\label{sec.radicals}

Free radicals play a very important role in DNA damage~\cite{Chatterjee93,hyd2}. Their role has been especially emphasized in the case of irradiation with photons, where they are the main instrument of the so-called indirect damage of DNA~\cite{sev97}. However, even in the context of IBCT, where direct mechanism involving secondary electrons are so important,
multiple experiments~\cite{Thomas1} indicate that the damage due to radicals exceeds that due to direct electrons.

The damage done by radicals can be calculated in the same fashion as the damage due to electrons if their fluence, ${\cal F}_r$, and the probability of producing a strand break on impact with a radical, $\Gamma_r$, are known. However, the analysis of damage in a plasmid DNA solution in pure water, where the action of radicals is not abated (studied in Ref.~\cite{Thomas1}) shows that the picture of the dose dependence is quite different from the one for secondary electrons that was discussed above. The main difference of this picture is a strong dependence on dose for the same conditions as in the experiment with mannitol, compare Figs.~\ref{fig.cf48.cor} and \ref{fig.cf45.lin}. In Eq.~(\ref{ssb.plasmid}) the probability of a SSB per segment of an ion's trajectory, $\frac{d N_{SSB}}{d \zeta}$, is independent of dose. The dependence on dose comes from the probability $P_{\nu}$ that a certain number of ions traverse through the plasmid. This probability does not change in the case of radicals while the dose dependence does.

In Fig.~\ref{fig.cf45.lin}, measured SSB and DSB probabilities depending on dose are shown with dots. It is obvious that in this case the curves are not proportional to each other. This means that the interference term, absent in Eq.~(\ref{ssb.plasmid}), plays an important role in the case of radicals. One way to explain this phenomenon is to infer that the radicals are distributed much more broadly than secondary electrons. Since a uniform distribution of ions in the beam is assumed, it is reasonable to assume (in the first approximation) that radicals are also distributed uniformly, such that their number density is proportional to the dose. Then the probability that they inflict a SSB on a plasmid is
\begin{eqnarray}
P_{SSB,r}=
{\cal N}_{r}\exp{\left[-{\cal N}_{r}\right]}
+\frac{1}{2}{\cal N}_{r}^2\exp{\left[-{\cal N}_{r}\right]}~,
\label{n.ssbrad}
\end{eqnarray}
where ${\cal N}_{r}=\Gamma_r {\cal F}_r$ is the average number of SSBs due to a given number density of radicals per plasmid. The second term includes the events when two SSBs take place on the same strand or are too far from each other to cause a DSB. This, however, is not sufficient since secondary electrons are still present in the experimental results shown in Fig.~\ref{fig.cf45.lin}. Then, the probability of a SSB is given by
\begin{eqnarray}
P_{SSB}=P_{SSB,r}(1-P_{SSB,e})~~~~~~~~~~~~~~~~~~~\nonumber\\
+(1-P_{SSB,r})P_{SSB,e}+\frac{1}{2}P_{SSB,r}P_{SSB,e}~.
\label{n.ssbradel}
\end{eqnarray}
This expression can be compared with the experiment and a fit gives the value of ${\cal N}_{r}=0.012 d$, where the dose is in Gy. The results of this comparison are shown with a solid line in Fig.~\ref{fig.cf45.lin}. They reasonably agree with the experiment at least for doses less than 250~Gy. 

\begin{figure}
\resizebox{1.0\columnwidth}{!}{ \includegraphics{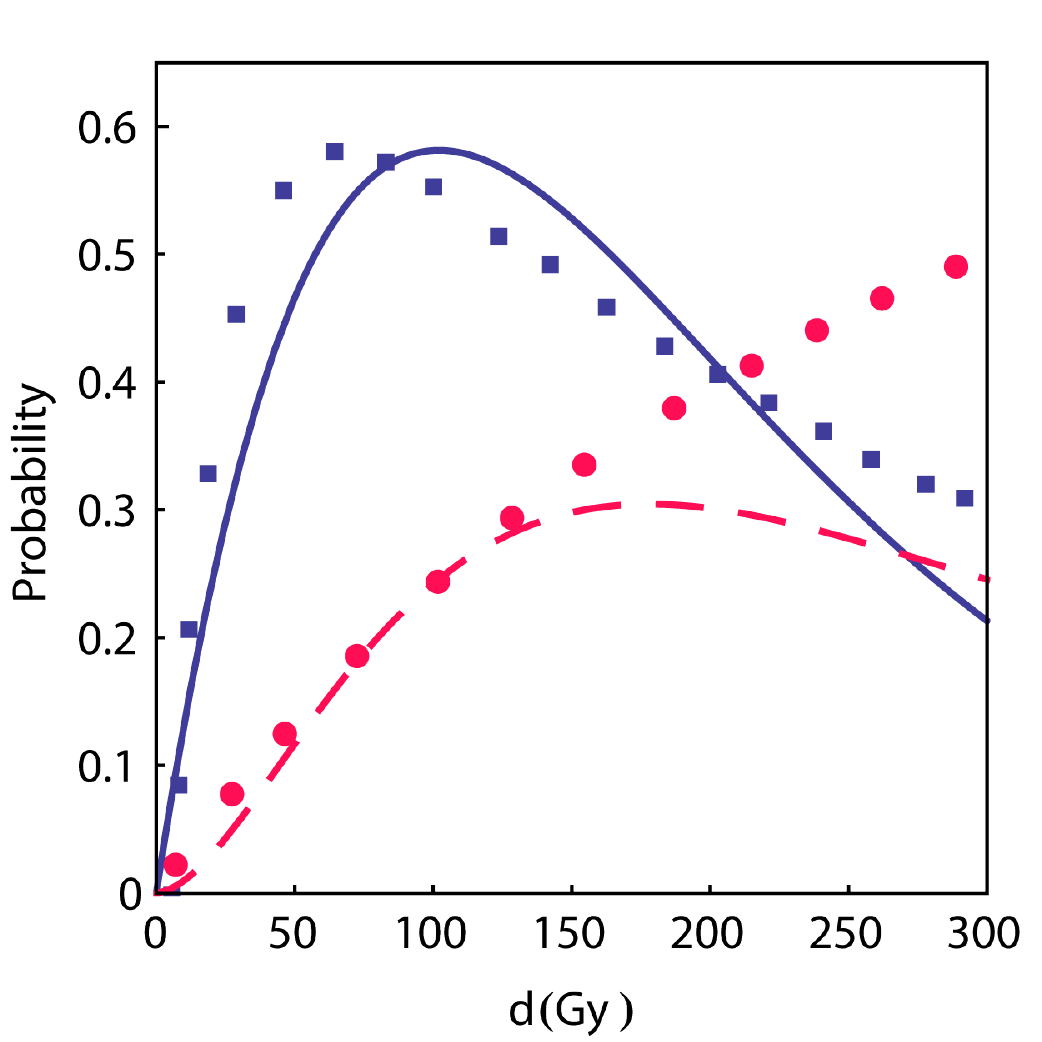} }
\caption{\label{fig.cf45.lin} (Colour online) Probabilities for SSBs and DSBs induced in plasmid DNA with secondary electrons and radicals as a function of dose. Dots correspond to experiments~\cite{Thomas1}: squares to the ``cleaned'' SSBs, and circles to DSBs. Calculated probabilities are shown with lines. Solid line corresponds to the probability of SSBs calculated using Eq.~(\ref{n.ssbradel}). The dashed line depicts the probability for DSBs calculated using Eq.~(\ref{n.dsbradel}).}
\end{figure}
The expression for DSBs is obtained similarly:
\begin{eqnarray}
P_{DSB,r}=\frac{1}{2}{\cal N}_{r}^2\exp{\left[-{\cal N}_{r}\right]}
\label{n.dsbrad}
\end{eqnarray}
and
\begin{eqnarray}
P_{DSB}=P_{DSB,e}(1-P_{DSB,r}-\frac{1}{2}P_{SSB,r}P_{SSB,e})~~~~\nonumber \\ +P_{DSB,r}(1-P_{DSB,e}-\frac{1}{2}P_{SSB,r}P_{SSB,e})~~~~\nonumber \\ +\frac{1}{2}P_{SSB,r}P_{SSB,e}(1-P_{DSB,e}-P_{DSB,r})~.~~
\label{n.dsbradel}
\end{eqnarray}
The results for DSBs are shown in Fig.~\ref{fig.cf45.lin} with a dashed line. Once again, a reasonable agreement for doses less than about 150~Gy can be observed. As the dose increases further, the higher order effects that are not included in Eqs.~(\ref{n.ssbrad}-\ref{n.dsbradel}) contribute to the number of observed DSBs and this number is thus underestimated.

The comments after previous section are still relevant, but it is also important to add to them the discussion about the spatial distribution of the fluence of radicals. It has been assumed to be uniform, but it was not discussed why it could be such. A possible mechanism that can be much more effective than diffusion is the collective transport due to shock waves to be discussed in Sec.~\ref{sec.shock}. One inference from the comparison of the results for secondary electrons and radicals, i.e., the comparison of Figs.~\ref{fig.cf48.cor} and \ref{fig.cf45.lin} is that the effect of radicals on DSBs is quite substantial. 
}
{

\subsubsection{Comparison with repair foci observations}
\label{sec.foci}

It is also possible to apply Eqs.~(\ref{n.dsb}) and (\ref{n.dsbradel}) to the observed phosphorylated histone variants H2AX, which accumulate near DNA DSBs in cell nuclei. These accumulations are called foci and their distribution along the ion's track can be translated to the number of DSBs per unit length of trajectory.

Reference~\cite{Niklas} reports the foci distributions in the nuclei of human lung adenocarcinoma epithelial cells (A549) after they were irradiated with carbon ions. The 
cross sectional area of an  A549 nucleus is 83~$\mu$m$^2$~\cite{A549}. Given the approximate number of base pairs in human DNA  ($3.2\times 10^9$) and assuming an uniform distribution of chromatin, the DNA twist density is $n_t=6.5\times 10^{-4}$~nm$^{-3}$. The average distance between the projections of foci on the ion's trajectory observed in Ref.~\cite{Niklas} is 2.2~$\mu$m,
 i.e., $dN_{DSB}/d\zeta\approx 0.5$~$\mu$m$^{-1}$. The integral fluence, 
\begin{eqnarray}
\int_0^\infty {\cal F}(\rho) 2\pi \rho d \rho~, \nonumber
\end{eqnarray}
is in this case smaller, since the carbon ions interacting with cells were of energy 52~MeV/u, which is far from the Bragg peak. The value of LET (not measured in Ref.~\cite{Niklas}) can be calculated using the methods developed in Sec.~\ref{ssec1}. It is about 50~eV/nm, so the integral fluence (linear with respect to LET) can be estimated to be about 25\% of the value used in Sec.\ref{sec.ssb}, i.e., about 290. Then assuming only the action of secondary electrons $\Gamma_{SSB}$ can be estimated from Eq.~(\ref{dsb.plasmid}). It turns out to be $6.5\times10^{-3}$, which is about 3.4 times higher than our estimate from experiments with plasmid DNA. The discrepancy can be attributed to the unaccounted action of radicals. The radicals are included in the analysis of cell survival in Sec.~\ref{sec.surv}. Here, it is pertinently to give an estimate for the number of produced radicals to be (far from the Bragg peak) about two times larger than the number of secondary electrons ejected by ions (if at least one radical is produced by a secondary electron). Therefore, it is plausible that the production of DSBs by electrons and radicals combined can explain the number $dN_{DSB}/d\zeta\approx 0.5$~$\mu$m$^{-1}$ observed experimentally.
}

\subsection{Accounting for $\delta$-electrons}
\label{sec.delta}

The effects due to secondary electrons with energies of 100~eV and above or the so-called $\delta$-electrons should be discussed separately. These particles cannot be included in the diffusion model because their cross sections are strongly peaked in the forward direction, their mean free paths exceed 1~nm and they lose their energy ionizing the medium and are capable of producing a number of extra electrons and creating a cluster-damage site. In order to estimate corrections due to $\delta$-electrons, several quantities pertinent to these particles need to be analysed.

The first is the mean free path. According to Ref.~\cite{Nikjoo06}, both the elastic and inelastic mean free paths of 100-eV electrons are about 1~nm. If such an electron is ejected in the most likely direction according to the binary interaction model~\cite{Rudd92}, about $70^\circ$, this electron will start losing energy within 1~nm of the ion's trajectory. Even if it produces more electrons than a sub-50-eV electron, they will not spread much further than them. 
Because of the kinematic limit, for an ion in the Bragg peak region, energies of ejected secondary electrons are below 0.7~keV. These electrons with elastic mean free path of about 4.5~nm are emitted in the forward direction, and it can be shown that the maximal distance between the first collision and the ion's path is 1.6~nm and it is reached by the electrons of energies $400-500$~eV. There is no way that further transport can carry further generations of electrons far beyond the 10~nm distance off the ion's path. In addition, the probability of producing a 400-eV secondary electron is only about 0.02 of that producing a 50-eV electron. Therefore, even though $\delta$-electrons are not included in the random walk approach, the location of their effect is by and large overlapped with that of sub-50~eV secondary electrons. The number of electrons ejected as the consequence of ionization by $\delta$-electrons can be estimated from energy conservation and these electrons have already been effectively included in the random walk, since $dN_e/d\zeta$ was obtained from the value of the stopping power, $S_e$.

Still another possibility exists for $\delta$-electrons to affect the discussed scenario. If a much more energetic electron, i.e., with energy larger than 20~keV, then with the mean free path of the order of 100~nm, it can cause damage elsewhere. Moreover, these electrons cannot be ejected in the Bragg peak region, since the required ion energy must be over 9~MeV. The probability of such events is very small; it is less than that of emitting a 50-eV electron by a factor over $10^6$. Therefore, this possibility is realised so rarely that it can be neglected.

\section{Thermomechanical effects}
\label{sec.thermo}

{Thus far, the energy loss by incident ions, the transport of produced secondary particles, and the radiation damage induced by these particles have been discussed. The transport, described by diffusion or MC simulations, is that of the ballistic electrons, radicals, etc. in a static medium. This transport does not include the whole physical picture because propagating secondary particles transfer the energy further, making the medium hot and dynamic.

Energy relaxation in the medium has been studied in Ref.~\cite{preheat}, where the inelastic thermal spike model was applied to liquid water irradiated with carbon ions. This model has been developed to explain track formation in solids irradiated with heavy ions and it studies the energy deposition to the medium by swift heavy ions through secondary electrons~\cite{Toul01,Toul02,Toul03,Toul04,Toul05,20,22,24,26,27,28}. In this model, the electron-phonon coupling (strength of the energy transfer from electrons to lattice atoms) is an intrinsic property of the irradiated material. }

The application of the inelastic thermal spike model to liquid water predicted that the temperature increases by 700-1200~K inside the hot cylinder by $10^{-13}$~s after the ion's traverse~\cite{preheat}. However, within this model, only coupled (between electrons and atoms of the medium) thermal conductivity equations are solved, while the further dynamics of the medium is missing. This dynamics is the consequence of a rapid pressure increase inside the hot cylinder around the ion's path up to $10^4$~atm, while the pressure outside of it is about atmospheric. Since the medium is liquid, this pressure difference prompts rapid expansion, resulting in a shock wave, which has been analysed in Refs.~\cite{prehydro,Vilnius,nimbnuke,natnuke}.

\subsection{Hydrodynamic expansion on the nanometre scale}

The problem of the expansion of the medium driven by the high pressure inside the hot cylinder is in the realm of hydrodynamics and it has been thoroughly analyzed in Ref.~\cite{prehydro}. It has been shown that the expansion is cylindrically symmetric. If the ratio of pressures inside and outside of the hot cylinder is high enough, as happens for large values of LET, the cylindrical expansion of the medium is described as a cylindrical shock wave, driven by a ``strong explosion''~\cite{LL6}. For an ideal gas, this condition holds until about $t=1$~ns, but in liquid water the shock wave relaxes much sooner. In Ref.~\cite{Vilnius} the molecular dynamics simulations of liquid water expansion showed that the shock wave weakens by about 0.5~ps after the ion's passage.


The hydrodynamic problem describing the strong explosion regime of the shock wave is self similar.
Its solution, as well as its mechanical features and limitations, is very well described in Refs.~\cite{LL6,Zeldovich,Chernyj}. In Ref.~\cite{prehydro},  the solution for the cylindrical case has been reproduced and analyzed in order to apply it for the nanometre-scale dynamics of the DNA surroundings. In this section, only the results pertinent to the further discussion of biodamage are presented.

The self similar flow of water and heat transfer depend on a single
variable, $\xi$. This variable is a dimensionless combination of the radial distance, $\rho$, from the axis, i.e., the ion's path, the time $t$ after the ion's passage, the energy dissipated per
unit length along the axis, which is equal to the LET per ion, $S_e$, and the density of undisturbed water, $\varrho=1$~g/cm$^3$. This combination is given by
\begin{equation}
\xi=\frac{\rho}{\beta\sqrt{t}}\left[\frac{\varrho}{S_e}\right]^{1/4}~,
\label{zeta}
\end{equation}
where $\beta$ is a dimensionless parameter equal to 0.86 for $\gamma=C_P/C_V=1.222$~\cite{prehydro}.
The radius and the speed of the wave front are given by
\begin{equation}
R=\rho/\xi=\beta \sqrt{t}\left[\frac{S_e}{\varrho}\right]^{1/4}
\label{radius}
\end{equation}
and
\begin{equation}
u=\frac{dR}{dt}=\frac{R}{2t}=\frac{\beta}
{2\sqrt{t}}\left[\frac{S_e}{\varrho}\right]^{1/4}~,
\label{speed}
\end{equation}
respectively. It is also worthwhile to combine Eqs.~(\ref{speed}) and (\ref{radius}) and obtain the expression of the speed of the front in terms of its radius $R$,
\begin{equation}
u=\frac{\beta^2}
{2 R}\left[\frac{S_e}{\varrho}\right]^{1/2}~.
\label{speedrho}
\end{equation}
Using Eq.~(\ref{speedrho}), pressure $P$ at the wave front can be obtained as
\begin{eqnarray}
P=\frac{2}{\gamma+1}\varrho u^2=\frac{1}{\gamma+1} \frac{\beta^4}
{2}\frac{S_e}{R^2}~.
\label{hydeqsBC}
\end{eqnarray}
Then, one can solve the hydrodynamic equations in order to obtain the expressions for speed, pressure, and density in the wake of the shock wave, i.e., behind the wave front~\cite{prehydro}.


The following intriguing questions have been raised in Refs.~\cite{preheat,prehydro}. What can such a shock wave do to biomolecu-les such as DNA located in the region of its propagation through the medium; can it cause biodamage by mechanical force? The forces acting on DNA segments were predicted to be as large as 2~nN, which is more than enough to break a covalent bond, causing a strand break; however, these forces are only acting for a short time and it remained unclear whether this is sufficient to cause severe damage to DNA molecules. The other question is: how significant can the transport due to the collective flow of this expansion be compared to the diffusion of secondary particles?

\subsection{Investigations of the effects of shock waves using simulations}
\label{sec.shock}

There are several effects of shock waves directly or indirectly related to biodamage or cell death. The first effect is the direct thermomechanical damage of a DNA molecule as a result of interaction with the shock wave, and it can be explored using molecular dynamics (MD) simulations~\cite{natnuke}. The second effect is a similar rupture of covalent bonds in water molecules leading to the extra production of radicals. This effect is still being studied. The third effect is the propagation of reacting species as a result of collective motion initiated by the shock wave. The fourth effect, which is also under investigation, is related to the damage of a cell membrane due to action of the shock wave~\cite{membraneilia}. Irreversible damage of the cell membrane may be lethal for the cell. In Ref.~\cite{natnuke}, the effects of the interaction of the shock wave formed in a liquid water medium following the traverse of an ion through this medium at different values of LET have been studied. The values of the LET were used as parameters describing the energy propagated by the shock wave, and the physical conditions in which the action of the shock wave is significant or even dominant for radiation damage assessment were analysed.

\begin{figure}
\resizebox{1.0\columnwidth}{!}{ \includegraphics{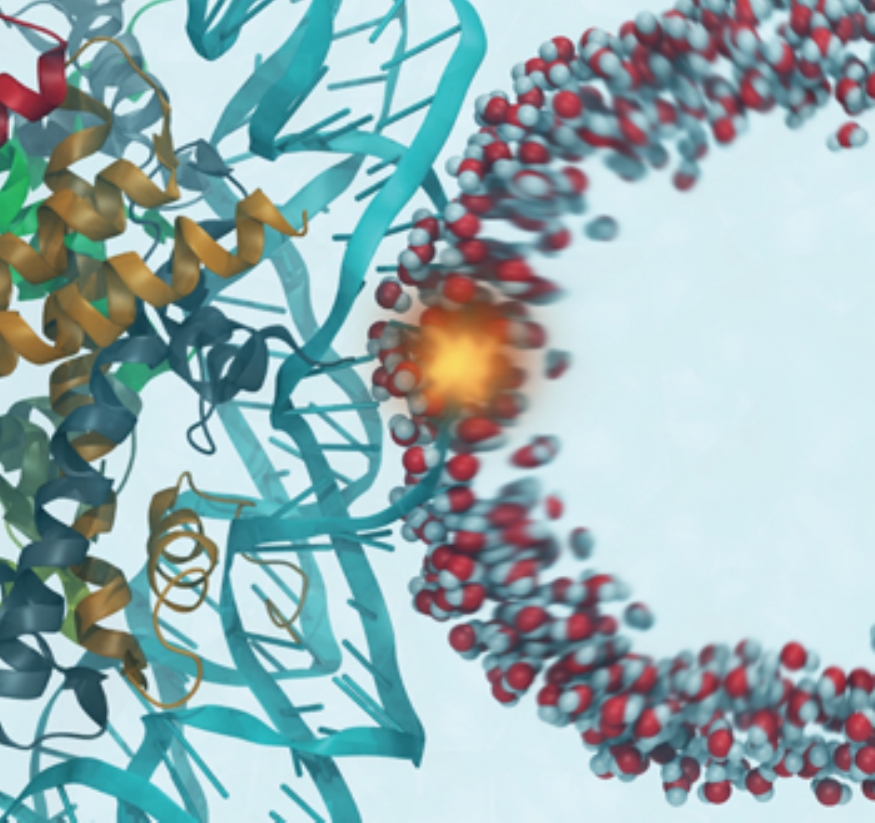} }
\caption{\label{fig.nucl} (Artistic view, colour online) The cylindrical shock wave front in water (on the right; the ion's path is the axis of this cylinder, perpendicular to the figure plane) interacts with a DNA segment on the surface of a nucleosome (on the left). The bright dot indicates the place where interactions occur. The medium is very dense following the wave front and is rarefied in the wake.}
\end{figure}

In eukaryotic cells, DNA molecules are packed into chromatin fibers. A nucleosome, a histone-protein octamer wrapped about with a DNA double helix, is the primary structural unit of chromatin. Therefore, the MD simulations were focused on the interaction of the cylindrical shock wave originating from ion's path with a fragment of a DNA molecule situated on the surface of a nucleosome. The artistic picture of this interaction is shown in Fig.~\ref{fig.nucl}~\cite{natnuke}.
The ion's path is perpendicular to the paper plane as well as the axis of the nucleosome disk. The simulations were done for four values of LET, 0.9, 1.7, 4.7, and 7.2~keV/nm, corresponding to the predicted values of LET at the Bragg peak for carbon, neon, argon, and iron ions, respectively~\cite{ChargeFluct}. Among these, carbon ions are currently the most used ions for heavy-ion therapy. Iron ions are important for  space-mission safety assessments. Between these are neon and argon ions, which are being considered for medical applications; they are used in a number of experimental studies~\cite{SchardtRMP10}.

\subsection{Simulations of the direct shock wave effect}
\label{sec.mdsim}

The estimate for the radius of the hot cylinder can be obtained from the analysis of the diffusion of secondary electrons from the ion's path.
This radius is associated with the average radius within which secondary electrons lose most of their energy~\cite{prauger}. At a time $t$, a secondary electron originating from the ion's path is most likely to be situated at a distance from the path, equal to $
{\bar \rho}=\int \rho P(t, r)d^3r$, 
where $P(t, r)$ is given by Eq.~(\ref{rwalk2t}). This integral is equal to $l \sqrt{\pi k/6}$, where $l$ is the average elastic mean free path of electrons ejected at the ion's path and $k$ is an average number of elastic collisions they undergo before they lose energy in inelastic collisions. For $S_e=0.9$~keV/nm, the estimate for ${\bar \rho}$ is 1~nm and this was taken to be the radius of the hot cylinder.

The simulations show a noticeable distortion
due to the shock wave at 10~ps after the expansion
starts. 
This distortion comprises the rupture of the secondary structure of the most exposed parts of the DNA molecule, manifested by the nonnative orientation of DNA nucleotides. Many hydrogen bonds are broken, and the bases are located outside the DNA double helix. However, these distortions are reversible, while our main interest is in the investigation of more permanent covalent bond breaking events.


In order to study whether the covalent bonds in the DNA backbone can be broken during the shock wave action, the energy temporarily deposited to these bonds was calculated. If this energy exceeded the binding energy of a given bond, it was assumed that thermomechanical stresses in the DNA fragment were sufficiently high to break the bond. The corresponding binding energies are referred to as thresholds for breaking the DNA backbone covalent bonds; they are between 3 and 6~eV~\cite{Range}. Even though the thresholds may be lower (even as low as 0.3~eV) in the environment as a consequence of the ion's passage~\cite{Kohanoff2012}, high thresholds were kept in order to obtain conservative estimates for the direct action of the shock wave.

\begin{figure}
\resizebox{1.15\columnwidth}{!}{ \includegraphics{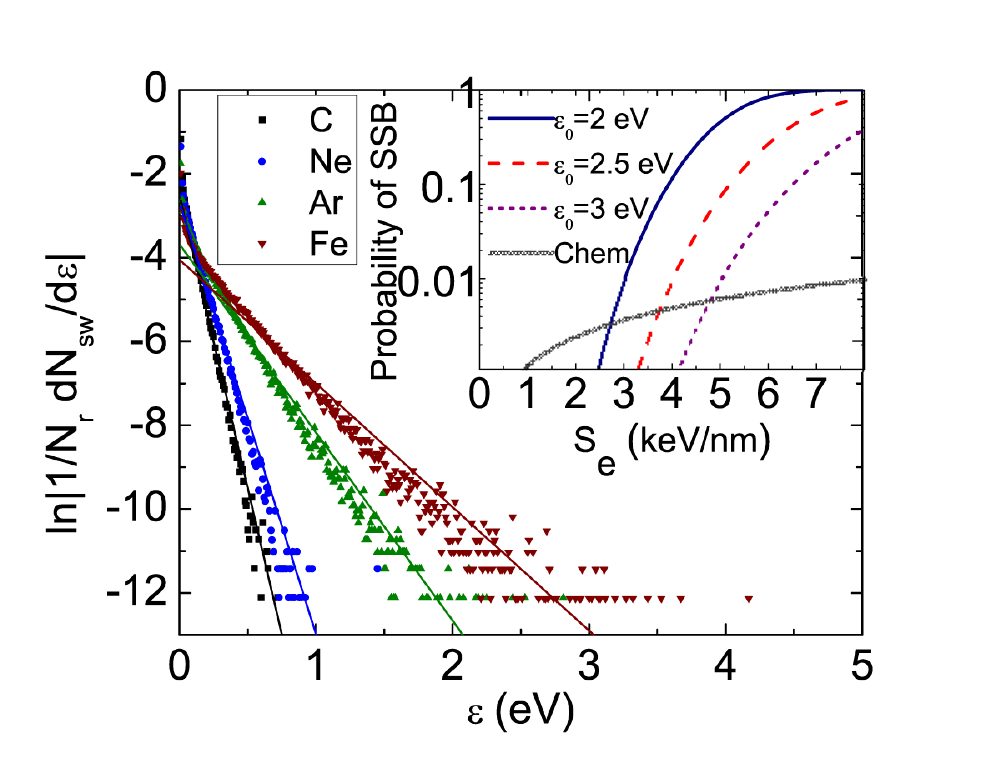} }
\caption{\label{fig.spec} (Colour online) The dependence of the logarithm of the normalised number of the covalent bond energy records for the selected DNA backbone region per 0.01~eV energy interval on the bond energy for four values of LET: 0.9, 1.7, 4.7, and 7.2~keV/nm, corresponding to the Bragg peak values for ions of carbon, neon, argon, and iron, respectively. Straight lines correspond to the fits of these distributions. In the inset, the dependence of the probability of producing at least one SSB in a 3-base-pair segment of a DNA molecule located between 1.5 and 2.2~nm from the ion's path, on LET.
The shock wave probability lines correspond to the estimates done using Eqs.~(\ref{trian2}-\ref{spc3}).}
\end{figure}

The analysis of MD simulations performed for four values of LET (0.9, 1.7, 4.7, and 7.2~keV/nm) gives the distributions of the bond energy records. These records can be represented by a histogram that assigns to every interval of energy $(\varepsilon, \varepsilon+\delta \varepsilon)$, the number of records corresponding to the bond energies from this interval. For each value of LET, the bond energy distribution was constructed. These distributions (normalised to the total number of records $N_{r}$ for each value of LET) are shown in Fig.~\ref{fig.spec}, where $\ln(1/N_{r}dN/dE)$ is plotted vs. the corresponding energy interval.

For the most part, these distributions correspond to Boltzmann distributions with different temperatures and they can be fitted as
\begin{eqnarray}
\frac{1}{N_0}\frac{d N_{sw}}{d \varepsilon}=\frac{1}{N_0}\frac{\delta N_{sw}}{\delta \varepsilon}
 =\frac{1}{k_B T}\exp\left[{-\frac{\varepsilon}{k_B T}}\right]~,
\label{spc1}
\end{eqnarray}
where $\delta \varepsilon=0.01$~eV is the width of the energy bin, $\delta N$ is the number of records with energy between $\varepsilon$ and $\varepsilon+\delta \varepsilon$, deposited in selected covalent bonds, the normalisation constant $N_0=2.17\times 10^4$ and the temperature $T$ are parameters; $k_B$ is the Boltzmann constant. Both parameters, $N_0$ and $T$, are determined from the fitting of the distributions obtained from the MD simulations.

$T$ is the temperature corresponding to the thermal parts of the distributions (for the four values of LET) shown as linear fits in Fig.~\ref{fig.spec}. The values of $T$ are 870, 1130, 2580, and 3970~K, indicating that the temperature increase above the temperature of the medium before the interaction with the shock wave, $T_0=$310~K (corresponding to 
a biological system), is directly proportional to $S_e$,
\begin{eqnarray}
T-T_0=\alpha S_e~,
\label{spc2}
\end{eqnarray}
where $\alpha=494$~K$\cdot$nm$\cdot$keV$^{-1}$.

The average number of breaks can be estimated by integrating Eq.~(\ref{spc1}) over energies $\varepsilon$, exceeding a chosen threshold $\varepsilon_0$:
\begin{eqnarray}
{N_{sw}}=\int_{\varepsilon_0}^\infty \frac{d N_{sw}}{d \varepsilon} d \varepsilon=N_0 \exp\left[-\frac{\varepsilon_0}{k_B T}\right]~.
\label{trian2}
\end{eqnarray}
Since the parameters $N_0$ and $T$ are fitted, this procedure allows us to predict the number of these over-threshold bond energy records for any value of LET. These numbers correspond to the number of bond breaks caused by the ion's passage.

In order to compare the number of strand breaks due to the shock wave action, to chemical effects, the probability of a strand break based on bond-breaking events has to be calculated. From the predicted number of bond breaks $N_{sw}$, Poisson statistics yields the probabilities for the exact number, $\nu$, of strand breaks to occur, $
P(\nu)=\exp\left(-{N_{sw}}\right)N_{sw}^\nu/\nu !$.
Then, the probability $P_{sw}$ of at least one strand break in a given segment of a DNA molecule is equal to $1-P(0)$, i.e.,
\begin{eqnarray}
P_{sw}=1-\exp\left(-{N_{sw}}\right)~.
\label{spc3}
\end{eqnarray}
The dependence of this probability on LET for different thresholds is shown in the inset of Fig.~\ref{fig.spec}, where this probability is compared to the probability of producing a single strand break owing to chemical effects in a similar DNA segment located at the same distance from the ion's path. The probability of SSBs due to chemical effects, $P_{ch}$, is estimated using the argument of Section~\ref{sec.ssb} and is given by,  
\begin{eqnarray}
P_{ch}=\kappa \frac{S_e}{\rm S_{e,0}}~,
\label{chem}
\end{eqnarray}
where $\kappa=(1.1\pm0.5)\times10^{-3}$ and $S_{e,0}=0.9$~keV/nm.

{}The inset of Fig.~\ref{fig.spec} predicts that for a given threshold, the shock wave breaking effect starts at a certain critical value of LET. After that, the probability of direct breaking increases with increasing LET steeply, readily overcoming chemical effects that include interactions of DNA molecules with free radicals, secondary electrons, solvated electrons, etc. The inset of Fig.~\ref{fig.spec} indicates that  bond breaking due to the shock wave mechanism starts (for the 3-eV threshold) at $S_e\approx 4$~keV/nm and by $5$~keV/nm it becomes the dominant effect in radiation damage. Two smaller thresholds of 2 and 2.5~eV are shown for comparison. This means that for heavier than Ar ions propagating in tissue, the bond breaking in DNA molecules located within about 2~nm of the ion's path will primarily be due to the direct effect of shock waves. The radius of dominance of this effect increases with further increasing LET~\cite{natnuke}.


\subsection{Transport of reactive species by the radial collective flow}
\label{sec.flow}

The study done in Ref.~\cite{prehydro} suggests that a considerable collective radial flow emerges from the hot cylinder region of medium.
The maximal mass flux density carried by the cylindrical shock wave is given by $\varrho_f u$, where $\varrho_f= \frac{\gamma+1
}{\gamma-1 }\varrho$ is the matter density on the wave front. This expression is proportional to $u$ and its substitution from Eq.~(\ref{speedrho}) yields that the mass flux is proportional to the $\sqrt{S_e}$. This flux density is inversely proportional to radius $\rho$ and is linear with respect to the $\sqrt{S_e}$. It sharply drops to zero in the wake of the wave along with the density. A sharp rarefaction of the volume in the wake of the wave follows from the results of Ref.~\cite{prehydro}. This is the effect of cavitation on a nanometer scale and due to this effect the water molecules of the hot cylinder along with all reactive species formed in this cylinder are pushed out by the radial flow. Such a mechanism of propagation of reactive species, formed within the hot cylinder, is competitive with the diffusion mechanism, studied in MC simulations done using track structure codes~\cite{Nikjoo06}.

Intriguingly, the cylindrical shock wave accomplishes the transfer of reactive species such as hydroxyl and solvated electrons, which play important roles in chemical DNA damage~\cite{hyd2,SancheCh9.2012,Kohanoff2012} much more effectively than the diffusion mechanism.
Indeed, the time at which the wave front reaches a radius $\rho$ can be derived from Eq.~(\ref{radius}) as it is equal to $(\rho^2/\beta^2) \sqrt{\varrho/S_e}$. This time has to be compared to diffusion times, which can be estimated for different reactive species as $\rho^2/D$, where $\rho$ is the distance from the ion's path and $D$ is the corresponding diffusion coefficient. The ratio of these times is equal to $(\sqrt{\varrho/S_e})D/\beta^2$. For all relevant species, the diffusion coefficient is less than $10^{-4}$~cm$^2$/s~\cite{laverne89}. Therefore, the above ratio is less than $10^{-3}/\sqrt{S_e {\rm(keV/nm)}}$, which is much less than unity even for protons.
For instance, for carbon ion projectiles, the wave front reaches 5~nm from the path in 2.8~ps after the ion's traverse, while hydroxyl radicals reach the same distance via the diffusion mechanism 
in about 9~ns, a more than 3000 times longer time. In fact, the lifetime of hydroxyl free radicals is shorter than 5~ns~\cite{laverne89,hyd2,Alpen}, therefore the shock wave transport may be the only means to deliver hydroxyl radicals to distances farther than 3.5~nm of the ion's path.

The collective flow is expected to play a significant role in the transport of reactive species at values of LET that are large enough to produce a shock wave, even if this wave is not sufficiently strong to cause covalent bond ruptures. The analysis shows that even at small values of LET, typical for the plateau region in the LET dependence on depth (well before the Bragg peak), a shock wave is formed; however it damps and becomes acoustic at radii under 10~nm. At $S_e=0.9$~keV/nm shock waves propagate further than 10~nm.

Thus, the effects following the local heating of the medium in the vicinity of an ion's path are quite striking. The MD simulations of a shock wave on a nanometer scale, initiated by an ion propagating in tissue-like medium, demonstrate that such a wave generates stresses, capable of breaking covalent bonds in a backbone of a DNA molecule located within 1.5~nm from the ion's path when the LET exceeds 4~keV/nm and this becomes the dominating effect of strand breaking at\footnote{These values correspond to conservative estimates ($\varepsilon_0=3$~eV)~\cite{natnuke}. They may be much lower if the actual thresholds appear to be smaller~\cite{Kohanoff2012}.} $S_e\gtrsim 5$~keV/nm.  The LET of $\sim4-5$~keV/nm corresponds to the Bragg peak values for ions close to Ar and heavier in liquid water. Besides the dramatic effects at such high values of LET, it was found that weaker shock waves produced by carbon ions or even protons transport the highly reactive species, hostile to DNA molecules, much more effectively than diffusion.

The notion of thermomechanical effects represents a paradigm shift in the understanding of radiation damage due to ions and requires re-evaluation of the relative biological effectiveness. This is due to the collective transport effects for all ions and direct covalent bond breaking by shock waves for ions heavier than argon. These effects will also have to be considered for high-density ion beams, irradiation with intensive laser fields, and other conditions prone to causing high gradients of temperature and pressure on a nanometer scale.

\section{Estimation of radio-biological effects}
\label{ssec5}

The essence of results obtained in sections~\ref{sec.ssb} and~\ref{sec.dsb} is that for a given ion, the numbers of SSBs and DSBs per unit length of the ion's path can be calculated. Or, alternatively, for a given DNA twist, the probability of the above lesions can be calculated. However, those calculations are still far from predicting whether the cell containing a given segment of DNA molecule will die or survive. {This question is largely in the realm of biology, because of a variety of biological mechanisms, which are activated following the creation of a lesion. Nearby proteins are engaged in DNA repair and may or may not be successful. Such an activity is marked by the appearance of the so called foci that can be observed experimentally~\cite{Jakob,Jakob1,Niklas}. These protein foci remain visible until the repair is finished. If a lesion cannot be repaired the cell containing this DNA molecule is likely to die. There is a plethora of biological studies directed at determining the probabilities of a successful DNA repair depending on the extent of the damage.}

It is established that a simple SSB is most likely to be fixed within minutes after this lesion is produced. DSBs can also be fixed, however, with smaller probability and there is also a chance that its repair (e.g., the non-homolo-gous end joining (NHEJ) type of DSB repair~\cite{apoptosis06}) may not be successful. The probability of repair is even smaller for multiply-damaged sites also known as clustered DNA lesion or complex DNA damage. A clustered DNA lesion is defined as the number of DNA lesions, such as DSBs, SSBs, abasic sites, damaged bases, etc., that occur within about two helical turns of a DNA molecule so that, when repair mechanisms are engaged, they treat a cluster of several of these lesions as a single damage site~\cite{Ward1,Ward2,Goodhead94,Lynn1,Lynn2,Lynn11}.
Let us start our discussion with the analysis of this type of damage in order to arrive at a prediction of cell death/survival caused by biodamage of a certain complexity that can be quantified by the formalism described above. 

\subsection{Assessment of the complex DNA damage}
\label{sec.complex}

When a DSB is induced due to secondary electrons, as discussed in Sections~\ref{sec.dsb} and~\ref{sec.plasmid}, there is a substantial probability (between 0.1 and 0.2 for plasmid DNA) for a DSB to occur as a result of the interaction of the molecule with a single electron. However, it is difficult to expect a clustered DNA damage site to be caused by a single electron or another secondary particle, since the distance between lesions in such a site can be too large (more than 5~nm). Therefore, in Refs.~\cite{SYS,precomplex}, the complexity of DNA damage has been quantified by defining a cluster of damage as a damaged portion of a DNA molecule by several independent agents, such as secondary electrons or radicals. Then, it is reasonable to expect that the probability that the electrons or radicals induce clustered damage is related to the fluence of these agents on a given DNA segment in the same sense as the probabilities of other types of lesions, such as SSBs or DSBs, as discussed in Sections~\ref{sec.dsb} and~\ref{sec.plasmid}.

Therefore, it is natural to start with the calculation of the number of clustered damage sites, produced by an ion, per unit length of its trajectory, $\frac{d N_C}{d \zeta}$, similar to what was done for SSBs and DSBs in  Eqs.~(\ref{n.ssb}) and (\ref{n.dsb}). DNA molecules are on the surface of nucleosomes and the latter is modelled as a cylinder of radius 5.75~nm. Then, an element of its lateral surface that enwraps two twists of a DNA molecule serves as a target for secondary electrons and radicals. This segment of the surface is 2.3~nm wide (along the axis of the cylinder) and 6.8~nm long (along the cylinder's circumference). For definiteness, this cylinder is taken to be perpendicular to the ion's path, its axis to be at a distance $\rho$ from the path, and the cylinder to be symmetric with respect to the plane containing the path and vector ${\bf \rho}$.

\begin{figure}
\resizebox{1.0\columnwidth}{!}{ \includegraphics{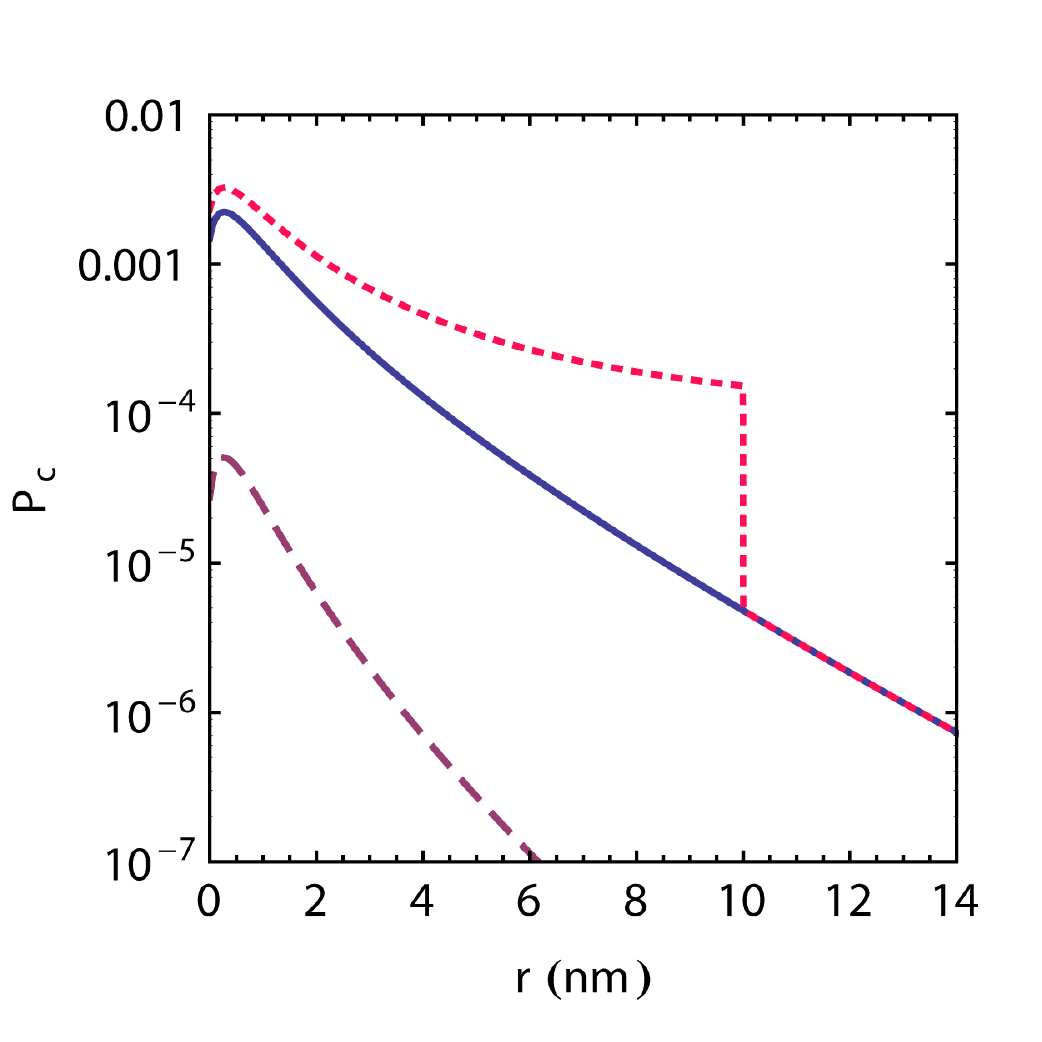} }
\caption{\label{fig.complex} (Colour online) The dependence of probabilities for complex damage to be induced by secondary electrons on the distance from the nucleosome to the ion's path. The solid line represents the degrees of complexity larger than one; the dashed line represents those larger than two. The dotted line shows the inclusion of the effect of radicals uniformly distributed inside a 10-nm cylinder. This curve is plotted with parameters discussed in Sec.~\ref{sec.surv}. }
\end{figure}

First, the fluence of secondary electrons on such a target, ${\cal F}_c$, using Eq.~(\ref{fluence.for}) has to be calculated. The geometry for this problem is similar to that for the calculation of the fluence on a cylinder perpendicular to the ion's path, considered in Sec.~\ref{sec.fluence.twist}.
The details for arranging the integration are given in the Appendix. The integration gives ${\cal F}_c(\rho)$, which when multiplied by $\Gamma_{SSB}$ gives the average number of SSBs for a given DNA segment, ${\cal N}_c=\Gamma_{SSB}{\cal F}_c(\rho)$. Then it is assumed, as was suggested in Refs.~\cite{SYS,precomplex,epjdmarion}, that the degree of complexity of damage is given by the number of agents that cause it\footnote{In the first approximation, it is assumed that all lesions comprising a clustered damage site occur with the same probability as a SSB.}. The sum of probabilities,
\begin{eqnarray}
P_c(\rho)=\sum_{\nu=2}^\infty\frac{{\cal N}_c^\nu}{\nu !}\exp{\left[-{\cal N}_c\right]}~,
\label{nrob.comp}
\end{eqnarray}
gives the probability, $P_c(\rho)$, that the damage complexity at a given site is larger than or equal to two. This probability is shown in { Fig.~\ref{fig.complex}}.

The next step is the integration of this probability over the volume of the cell nucleus with the number density of such sites. More precisely, the integration over $\rho$ is done from zero to infinity, since ${\cal F}_c(\rho)$ rapidly decreases with $\rho$ and one does not have to worry about reaching the limits of the cell nucleus. This integral,
\begin{eqnarray}
\frac{d N_c}{d \zeta}= n_s \int_0^\infty P_c(\rho) 2\pi \rho d \rho=n_s \psi~,
\label{n.comp}
\end{eqnarray}
where $n_s$ is the number density of sites, gives the number of clustered damage sites per unit length of the ion's trajectory.
Equation~(\ref{n.comp}) can be numerically integrated similar to Eqs.~(\ref{n.ssb}) and~(\ref{n.dsb}). For $\Gamma_{SSB}=2\times 10^{-3}$ (approximately corresponding to that found in Sec.~\ref{sec.plasmid},
$\psi=\int_0^\infty P_c(\rho) 2\pi \rho d \rho=3.2\times 10^{-2}$~nm$^2$ (these calculations are done for carbon ions near the Bragg peak).
However, it is very inconvenient to to deal with the introduced number density $n_s$. If one notices that each nucleosome geometrically contains about 5.3 targets for clustered damage sites, $n_s$, can be exchanged for the number density of nucleosomes $n_n$.
With this consideration, Eq.~(\ref{n.comp}) becomes,
\begin{eqnarray}
\frac{d N_c}{d \zeta}=(5.3 n_n) \int_0^\infty P_c(\rho) 2\pi \rho d \rho =n_n \Psi,
\label{n.comp.est}
\end{eqnarray}
where $\Psi=0.17$~nm$^2$ and $n_n$ is the number density of nucleosomes in cell nucleus.  This number density depends on the type of cells and, in principle, on the cell cycle. In our estimates it is assumed that $n_n$ is uniform throughout the cell nucleus.
In Sec.~\ref{sec.surv}, Eq.~(\ref{n.comp.est}) is related to the dose and thus, the survival curves leading to the calculation of the RBE are obtained.

However, the estimate given by Eq.~(\ref{n.comp.est}) does not include the effect of radicals, which may be significant for the overall assessment of radiation damage. In order to include it, the $\rho$-dependent distribution of the probability of inducing a SSB by radicals, i.e., ${\cal N}_r$, introduced in Sec.~\ref{sec.radicals}, is needed. Since this distribution is not known, one can start with a uniform distribution of radicals as was done for the {\em in vitro} experiment in Sec.~\ref{sec.radicals}, but only within a certain radius from the ion's path. This implies that the reactive species, formed in the nearest proximity to the path, are transported by the shock wave and their number density is nearly uniform inside the cylinder that enwraps the decayed shock wave. The estimate of this radius for carbon ions at the Bragg peak is 10~nm (see Sec.~\ref{sec.flow}). Then ${\cal N}_r(\rho)$ can be added to ${\cal N}_c(\rho)$ and the sum can be substituted into Eq.~(\ref{n.comp}) for $\rho < 10$~nm. Of course, in reality the boundary of radical propagation beyond the estimated radius will not be sharp (as shown if Fig.~\ref{fig.complex}) because of diffusion, but the estimate of the softness of the boundary is not essential.
The value of ${\cal N}_r(\rho)$ may be affected by environmental conditions in the tissue. As is known, the fixation of damage due to radicals depends on the presence of oxygen at the damage site~\cite{hyd2,Alpen}. This means that even before the enzymatic repair mechanisms are engaged, the radical-induced damage may be fixed if oxygen is not present, i.e., in hypoxic conditions. Then, the value of ${\cal N}_r(\rho)$ is effectively reduced. The study of such a reduction for different concentrations of oxygen leads to the calculation of the oxygen enhancement ratio (OER), another important parameter for optimization of IBCT.

\subsection{Obtaining the survival curves}
\label{sec.surv}

A survival curve is the dependence of the probability of cell survival on the absorbed dose of radiation. On one side, it relates the goal with the means, i.e., it predicts the dose that is necessary in order to achieve cell deactivation with a desired probability. On the other side, it allows comparing different modalities (photons, protons, heavier ions, etc.) and thus allows one to optimize the choice of therapy. This comparison is achieved via the calculation of the ratio of doses of different projectiles necessary to achieve the same probability of cell survival. The ratio of the dose due to photons to that for other projectiles is called the relative biological effectiveness (RBE).

The assessment of RBE for ions, from the point of view of the multiscale approach, starts from the calculation of survival curves for a given type of cell irradiated with a given type of ion. This means that for a given type of cell and a given dose the probability of cell survival (or death) has to be calculated. In the previous section, the probability of cell death was related to the probability of inducing a DNA lesion of a given complexity, so that it is unlikely to be repaired with proteins. In this section, the accomplishments of previous section are applied to the calculation of survival curves.

The conditions of irradiation are the same as in Sec.~\ref{sec.complex}, i.e., the doses are small and the ion tracks are not going to interfere. Then, for a given type of cell and a given dose, the number of ions that traverse a cell nucleus can be calculated. This, similar to our experience with plasmid DNA in Sec.~\ref{sec.plasmid}, gives us the dose dependence. The average number of complex DNA lesions in the cell nucleus is given by the following expression, similar to Eqs.~(\ref{ssb.plasmid}) and (\ref{dsb.plasmid}),
\begin{eqnarray}
Y_c=\frac{d N_c}{d \zeta} {\bar x}_{nc} \sum_{\nu=1}^\infty \nu P_{\nu}(d)~,
\label{death.cell}
\end{eqnarray}
where $P_{\nu}(d)$, given by Eq.~(\ref{one.ion.pois}) is the probability that $\nu$ ions traverse the cell nucleus and ${\bar x}_{nc}$ is the average distance of the ion's traverse through the cell nucleus. The average number of traversing ions, $N_{ion}$, is given by Eq.~(\ref{nions}) with $A_p$ replaced with the cross section of the cell nucleus. The value of $\frac{d N_c}{d \zeta} $ is taken from Sec.~\ref{sec.complex}.

It is worthwhile to apply this method to the calculation of the survival curve for A549 cells irradiated with $\alpha$-particles at ${\bar S}_e=115$~eV/nm, studied in Ref.~\cite{Heuskin}. The nucleosome number density is estimated to be $2.2\times 10^{-4}$~nm$^{-3}$. The average number of ions traversing such a nucleus gives $N_{ion}\approx 4 d$, where the dose is in Gy. This is a much larger number than that in the case of plasmids even for doses not exceeding 2~Gy. This means that a goodly number of terms in Eq.~(\ref{death.cell}) has to be retained.

\begin{figure}
\resizebox{1.0\columnwidth}{!}{ \includegraphics{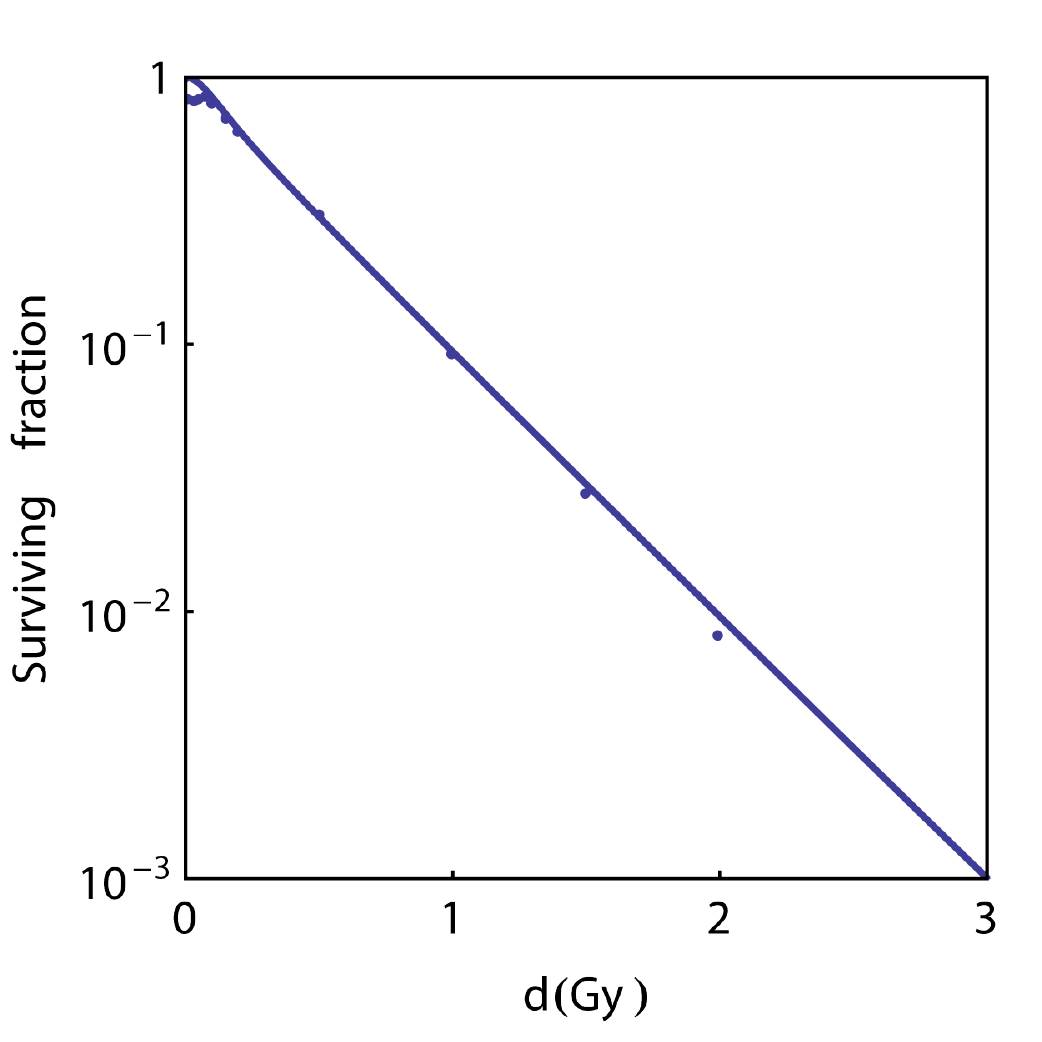} }
\caption{\label{fig.surv} (Colour online) Survival curves for A549 cells irradiated with $\alpha$ particles. Solid line is calculated using Eq.~(\ref{prob.surv}). The dots represent the experimental data~\cite{Heuskin}.}
\end{figure}

Equation (\ref{death.cell}) gives the number of clustered damage sites per cell nucleus. Since each site of this kind is assumed to be lethal for the cell, the probability of cell death, $\Pi_{d}$ is given by
\begin{eqnarray}
\Pi_{d}=1-\exp\left[-Y_{c}\right]~.
\label{prob.death}
\end{eqnarray}
This means that the probability is unity less the probability of zero clustered damage sites occurring in the cell nucleus, which is given by the second term in Eq.~(\ref{prob.death}). Finally, the probability of cell survival is given by unity less the probability of cell death, i.e., by that second term of Eq.~(\ref{prob.death}):
\begin{eqnarray}
\Pi_{surv}=1-\Pi_{d}=\exp\left[-Y_{c}\right]~.
\label{prob.surv}
\end{eqnarray}
This probability depends on dose and this dependence, shown in  Fig.~\ref{fig.surv}, is the survival curve for A459 cells irradiated with $\alpha$-particles with ${\bar S}_e=115$~eV/nm. This curve is compared to the survival curves for the same cells in the same conditions reported in Ref.~\cite{Heuskin}.
The dose dependence is very close to exponential (nearly a straight line in a semi-logarithmic plot); it is determined by Eq.~(\ref{death.cell}) and is universal for the radiation conditions considered in this work, i.e., where the interaction between ion tracks is absent. This corresponds to a large variety of observed survival curves for different cells and projectiles~\cite{Heuskin,Elsaesser,Weyrather99}. From the model point of view this corresponds to the so-called single-hit model described in Ref.~\cite{Alpen}. Enzymatic repair may affect the behaviour of this dependence only at very low doses, which may explain a slight discrepancy seen in Fig.~\ref{fig.surv}.

The calculated survival curve depends on the following numbers, which were either used as parameters or were determined from the comparison with experiments: the number of secondary electrons produced by the ion per unit length,  $dN/d\zeta$, calculated in Sec.~\ref{sec.enspec} and dependent on LET, the probability for an electron incident on DNA to induce a SSB per one DNA twist, $\Gamma_{SSB}$, the fluence of radicals on a DNA twist, ${\cal F}_r$, or the number of produced radicals and the average radius of their propagation from the ion's trajectory that allow one to estimate it, the probability for radicals to induce SSBs, $\Gamma_r$, the size of the nucleosome and their number density, $n_n$ (assumed to be uniform) in a given cell, and the lethal degree of the lesion complexity. Some of the parameters in this list can be found from the literature, estimated or calculated (rather accurately) for given media, cells, and projectiles. Some of them remain unknown for now, but further research may clarify their values. It is noteworthy that a complicated problem of the calculation of the RBE can be reduced to the search of several microscopic parameters that can be determined theoretically or experimentally.

\subsection{The recipe for obtaining the RBE}
\label{sec.recipe}

In this section, the multiscale approach is summarised in a recipe for a phenomenon-based assessment of radiation damage that results in the calculation of survival curves and RBE. Let us imagine that a certain type of cell is irradiated with certain ions. Here are the steps necessary for finding the location of damage and the RBE in the irradiated region.

First, it is desirable to know the composition of the medium. Cross sections of ionization, excitation, and nuclear fragmentation will affect the shape of the LET curve. Section~\ref{sec.sdcs} gives recipes for determining some of these cross sections for water and more complex media. Second, using Eqs.~(\ref{eq6}--\ref{eqstrg}), the LET dependence on energy and longitudinal coordinate can be obtained. This gives the location of the Bragg peak, its height, and other features of the LET curve. The conditions related to secondary particles should be assessed: the energy spectrum of secondary electrons, their average energy, and other features provide the grounds for inference on what methods can be used for the calculation of their transport and energy transfer.
If the value of the LET (at least at some section of the ion's propagation) is higher than 4~keV/nm, shock wave effects may dominate the scenario of biodamage.

Third, the cells should be thoroughly investigated. In this work, the cell nucleus was by and large discussed as a target, however there could be conditions in which other parts of a cell, such as the cell membrane, cytoplasm, mitochondria, and other organella are targets, whose damage may be lethal to the cell. If the cell nucleus' DNA is the target, it is important to know how it is distributed. Any information on the structure of chromatin, size of nucleosomes, their number density, etc., is important for the description of the target.

Fourth, as soon as the target is described it is important to calculate the fluence of secondary particles, such as secondary electrons and radicals, on this target. The random walk has been used in this and other works related to the multiscale approach to describe transport of electrons, but it can be calculated using more sophisticated methods. As for radicals, they are carried by the collective flow of the shock wave and diffuse through the medium. More research is needed to describe their transport. The damage probability due to radicals depends on their production and transport, both of which depend on the LET. High temperatures inside the hot cylinder and consequent shock waves contribute to these processes. The ultimate effect of radicals depends on the hypoxic/aerobic conditions in the medium. If the effectiveness is known, the OER can be determined for the given conditions.

Fifth, the average number of DNA lesions of interest (that could be lethal) per unit length of the ion's trajectory should be calculated. This implies the integration and averaging of damaging effects in the radial direction with respect to the ion's trajectory. Then, knowing the size of the cell nucleus (or other target), the dose dependence using Eqs.~(\ref{nions}) and (\ref{death.cell}) can be determined. The survival curve can then be calculated using Eq.~(\ref{prob.surv}). The comparison of the survival curve with that for x-rays gives the RBE for a given location. This location is described with the value of the LET and the depth coordinate that corresponds to it. This means that these calculations predict the RBE (and OER) at the Bragg peak, plateau, and the tail of the LET-depth dependence. Then, if the tissue can be scanned to produce a spread-out Bragg peak, the calculations can be superimposed.

Enzymatic repair mechanisms play an important role for the overall damage assessment. They may be included in several places. For instance, their effectiveness determines the definition of ``lethal'' damage, e.g., a DSB or degree of complexity (size of a clustered damage site). Probabilities such as $\Gamma_{SSB}$ and $\Gamma_r$ can also be adjusted or calculated using quantum mechanical methods.

\subsection{Multiscale approach vs. other models for the assessment of radiation damage with ions}
\label{disc}

The expertise for the assessment of radiation damage historically comes from the times when x-rays (photons) were used as the only projectiles. For x-rays, the dose distribution is practically uniform.  Therefore, it is not accidental that the dose has been chosen as the main parameter for prediction of radiation damage. Treatment plans had to deliver certain doses to certain locations in order to achieve the desired results.\footnote{The optimisation related to reducing dose deposition in healthy regions
and treatment partitioning is left aside.} A vast majority, if not all, of other existing models that ``calculate'' the survival curves are based on an empiric formula,
\begin{eqnarray}
-\ln{\Pi_s}=\alpha d + \beta d^2~, \label{rev1}
\end{eqnarray}
where $\alpha$ and $\beta$ are coefficients. Several features of these
curves have been discussed, one of which is the ratio of
$\alpha/\beta$. If this ratio is large, the survival curve is
``steep'' and more like a straight line\footnote{For example, for the survival curve, shown in Fig.~\ref{fig.surv}, $\alpha=1$ and $\beta=0$, which is typical for cells irradiated with ions.}; if it is small then it is a ``shouldered'' curve. A series of models suggested since
1955~\cite{Alpen,MKModel,MKModel09,RobertMech,Goodhead93,DNA5,Goodhead06} provided a phenomenological explanation to this dependence and developed approaches to the calculation of RBE. The coefficients $\alpha$ and $\beta$ depend on the kind of cells, on the cell cycle, on the access of oxygen to the irradiated cells and other factors.

For many practical purposes, an experimentally
obtained curve, given by Eq.~(\ref{rev1}) is {\em sufficient}
information for the evaluation of radiation damage, and it has been
used for many years for treatment planning and optimization.
Atomic or molecular interactions are not mentioned in those models; these and more information are hidden in the purely empirical coefficients $\alpha$ and $\beta$. Since the dose distribution is uniform, it is possible to solve all practical problems without atomic/molecular physics, since it brings up too many difficult questions involving interactions with biomolecules that seem irrelevant as compared to the biological unknowns related to repair mechanisms.

Particle projectiles change this picture. As was shown above, the dose distribution around each particle's path is highly nonuniform. The track structure and the consequent damage are much more complicated.
A solution to this problem was suggested by the Katz approach in
which the radial dose distribution is calculated  and related to the
inactivation of sub-cell-nucleus
targets~\cite{Katz67,Katz71,Cucinotta99,KatzRBE}. The quality factor
of radiation was introduced in order to relate the survival curve
parameters to a given type of radiation, differentiating between track
types, inactivation modes, the structural complexity of targets,
etc. The eventual goal of the Katz model was to calculate the RBE.
Nevertheless, the biological relation of the radial dose distribution with the cell survival probability was done based on the survival curves for x-rays, without analyzing particular physical processes, i.e., the empiric coefficients $\alpha$ and $\beta$ remain central to this approach.

The Local Effect Model (LEM), developed at GSI, calculates
the RBE assuming that the biological effect of radiation is entirely
determined by the spatial distribution of the radial dose inside the
cell nucleus. It relates the response of biological systems,
following ion irradiation, to the corresponding response after x-ray
irradiation~\cite{LEM96,SchardtRMP10}.
Corrections for the quality
of damage was included in a later version of the
LEM~\cite{SchardtRMP10}. This model operates on the schematic level
using Eq.~(\ref{rev1}) with empirical coefficients $\alpha$ and
$\beta$. The LEM solves technical problems related to the
optimization of treatments, leaving no place for {\em ab initio}
approaches and physical, chemical, or biological effects in general;
even a consideration of DNA lesions such as DSBs is beyond the scope
of the LEM~\cite{SchardtRMP10,Radam09editorial}.

The calculation of survival curves shown in Sec.~\ref{sec.surv} demonstrates a new way of relating the physical parameters with biological outcomes thus fulfilling the goal of the multiscale approach and predicting the RBE.
The multiscale approach is unique in relating the biological consequences of radiation to the actual physical, chemical, and biological effects.

\section{Conclusions and outlook}
\label{concl}
The multiscale approach to the assessment of radiation damage with ions has been reviewed. It was demonstrated that the main difference from other approaches is the in-depth focusing on physical effects, and, therefore, our approach is referred to as a phenomenon-based approach. The state of the art of this approach is discussed and some techniques of calculations are demonstrated.

The main advantages of the multiscale approach follow from its
architecture, its fundamentality, and its versatility. The
approach evaluates the relative contributions and significance of a
variety of phenomena; it elucidates a complex multiscale scenario in
sufficient detail and has a solid predictive power. It is
structurally simple and inclusive, and allows for modifications and
extensions by including new effects on different scales and
improvements on the way.

There are several areas in which major developments are expected. First, as it has been shown, the empirical models for calculation of survival curves can be improved using the multiscale approach. This can be implemented in the optimization codes for clinical purposes. The curves and radiation strategy will depend on the kind of cells and type of radiation. Moreover, the modality can also be optimized for particular cases. In order to achieve this, a more thorough comparison with experiments should be established and parameters such as $\Gamma_{SSB}$ should be tuned. The criteria for cell death should also be understood in contact with biologists. Since the radicals play a significant role in radiation damage, the aerobic/hypoxyc conditions of the target will determine the degree of inclusion of radicals in the calculations. 

The second area of development is related to the modification of the medium. The use of nanoparticles such as gold nanoparticles (GNP) as sensitizers has been discussed both theoretically and experimentally~\cite{Prise11,SancheGNP}, in order to boost the production of secondary electrons near the target and thus increase the RBE. The use of nanoparticles is considered for different modalities. Such a modification of the medium should be feasible within the multiscale approach. The relevant cross sections of secondary electron production and their energy spectrum will define their effect on nearby biomolecules.

The third area is the modification of modality. Ions heavier than carbon require a better understanding of thermomechanical effects discussed above, since the shock waves initiated in the Bragg peak area would be more pronounced. The use of these ions may not necessarily be therapeutic. Rather, the understanding of the mechanisms of radiation damage at very high values of LET will help the assessment of the hazards of exposure to such ions during space missions or elsewhere. Also, the targets may not necessarily be biological, e.g., the assessment of radiation damage of electronics or other equipment can be done in a similar fashion.
Another aspect, which can also be regarded as a modality modification, is the series of effects related to irradiation with ion beams produced
by high-power lasers~\cite{Bulanov}. In these conditions, the beam is much more dense and the tracks substantially interfere. The application of the multiscale approach for the calculation of survival curves may be especially beneficial in this case.

{The fourth area is related to the change of the field of science related to radiation damage with ions. The development of the multiscale approach is shifting a paradigm as in the case with thermomechanical damage. However, in some cases, where analytical methods turned out to be successful, a new understanding was obtained. Some common terms, such as dose, are shown to have limitations when describing radiation damage with ions. This means that the science of radiation damage is evolving. 

The future development of the multiscale approach will make a worthwhile tool for the assessment of radiation damage
on the molecular level. While there is more work to be done to make it practical, its fundamental basis and depth related to atomic/molecular physics is becoming more and more evident.

\section*{Acknowledgements}
We are grateful to J.~S. Payson and B. Roth who critically read the manuscript, R. Garcia-Molina, M. Niklas, I.~M. Solovyeva, I.~A. Solov'yov, P. de Vera for the assistance with figures, important advice, and insight,  Center for Scientific Computing of Goethe University, and the support of COST Action MP1002 ``Nano-scale insights in ion beam cancer therapy.''

\begin{appendix}
\section{Calculations of fluence for different geometries}

Here it is shown how to calculate the fluence of secondary electrons applied to different geometries.

\subsubsection{Fluence for the radial dose calculation}
\label{sec.app.raddose}

In order to adjust Eq.~(\ref{mult4}) for the calculation of the
radial dose, $d{\bf A}$ is chosen to be an element of the surface of
a cylinder of radius $\rho$, coaxial to the ion's path, $d{\bf A}=dA
{\bf n_\rho}$, where ${\bf n_\rho}$ is a unit vector in the radial
direction toward the element. Then, the square of the distance from
any point on the ion path to the area element is given by
$r^2=\rho^2+\zeta^2$, where $\rho$ is the radius of the cylinder from
the ion path and $\zeta$ is the coordinate along the ion's path and
${\bf n_\rho}\cdot{\bf n_r}=\frac{\rho}{\sqrt{\rho^2+\zeta^2}}$.
Finally, if the element is a belt of radius $\rho$ and width
$\delta$, $dA=2 \pi \rho \delta$, and the number of secondary
electrons incident on this belt is given by the integral of
Eq.~(\ref{mult4}) over the whole $\zeta$-axis:
\begin{eqnarray}
N_\delta(\rho)=2 \pi \rho \delta \frac{dN}{d\zeta} \int_{-\infty}^\infty
d\zeta \frac{\rho}{\sqrt{\rho^2+\zeta^2}} \int_{r/l}^\infty d k~~~~~~~~~~~~ \nonumber\\
\times \frac{r}{2 k}
     \left(\frac{3}{2 \pi k l^2}\right)^{3/2}
\exp\left[-\frac{3 r^2}{2 k l^2}-\gamma k\right]~~~~~~~~~~~~~
\nonumber
\\
=2 \pi \rho \delta \frac{dN}{d\zeta}  \int_{\rho/l}^\infty d k
\frac{\rho}{2 k}
     \left(\frac{3}{2 \pi k l^2}\right)^{3/2}~~~~~~~~~~~~~~~~~~~~\nonumber\\
\times \exp\left(-\frac{3 \rho^2}{2 k l^2}-\gamma k\right)
 \int_{-\sqrt{k^2l^2-\rho^2}}^{\sqrt{k^2l^2-\rho^2}} d\zeta
\exp\left[-\frac{3 \zeta^2}{2 k l^2}\right]\nonumber
\\
=2 \pi \rho \delta \frac{dN}{d\zeta}  \int_{\rho/l}^\infty d k
\frac{3\rho}{4 \pi k^2 l^2}
    ~~~~~~~~~~~~~~~~~~~~~~~~~~~~~~\nonumber\\
\times \exp\left[-\frac{3 \rho^2}{2 k l^2}-\gamma k\right]
 {\rm erf}\left[\frac{3}{2}\left(k-\frac{\rho^2}{k
 l^2}\right)\right]
 \nonumber
\\
=2 \pi \rho \delta \frac{dN}{d\zeta}{\cal
Q}(\rho/l,\gamma)~,~~~\label{mult4y}
\end{eqnarray}
where
\begin{eqnarray}
{\cal Q}(\rho/l,\gamma)=\frac{3\rho}{4 \pi l^2} ~~~~~~~~~~~~~~~~~~~~~~~~~~~~~~~~~~~~~~~~~~~\nonumber\\
\times \int_{\rho/l}^\infty \frac{d k }{k^2}\exp\left[-\frac{3
\rho^2}{2 k l^2}-\gamma k\right]
 {\rm erf}\left[\frac{3}{2}\left(k-\frac{\rho^2}{k
 l^2}\right)\right]
 \label{mult5y}
\end{eqnarray}
has been introduced. Equation~(\ref{mult4y}) with (\ref{mult5y}) provide the general expression for the number of secondary electrons incident on a
cylindrical belt coaxial with the ion's path.

\subsubsection{Cylinders enwraping a DNA twist}
\label{sec.geo.cyl}

The geometry for the calculations of the fluence through a cylinder enwraping a DNA twist is shown in the inset of Fig.~\ref{fig.fluence}.
The vector ${\bf r}$ from the point of origin of a secondary electron on the ion's path to a point on the surface of the cylinder enwraping a DNA twist is given by,
\begin{eqnarray}
{\bf r}=(a \cos \varphi - \rho){\bf \hat i} + (a \sin \varphi - \zeta \sin \alpha){\bf \hat j}\nonumber \\  + (z-z_0 -\zeta \cos \alpha){\bf \hat k}~,
\label{r.vec}
\end{eqnarray}
where $z$ is the coordinate along the cylinder.
Equation (\ref{mult4}) has to be integrated over the length of the path $\zeta$ and the area of the cylinder. It is reasonable to compare two limiting cases of the cylinder: that of the cylinder being parallel and perpendicular to the path for $z_0=0$. The perpendicular case corresponds to $\alpha=\pi/2$ and $r^2$ and the dot product of Eq.~(\ref{mult4}) with (\ref{r.vec}) are as follows:
\begin{eqnarray}
{r^2}=(a \cos \varphi - \rho)^2 + (a \sin \varphi - \zeta)^2  + z^2,~~~~~{\rm and}~~~~~~\nonumber\\
d{\bf A}\cdot {\bf n_r}=a d \varphi d z {\bf n_A}\cdot {\bf n_r}\nonumber~~~~~~~~~~~~~~~~~~~~~~~~~~~~~~~~~~~ \\=a d \varphi d z \frac{(a \cos \varphi - \rho)\cos \varphi + (a \sin \varphi - \zeta)\sin \varphi}{r}.~~~
\label{r2.perp}
\end{eqnarray}
The parallel case corresponds to $\alpha=0$ and the similar expressions are given by
\begin{eqnarray}
{r^2}=(a \cos \varphi - \rho)^2 + (a \sin \varphi)^2 + (z -\zeta)^2~,~~{\rm and}~~~\nonumber\\
d{\bf A}\cdot {\bf n_r}=a d \varphi d z {\bf n_A}\cdot {\bf n_r}\nonumber ~~~~~~~~~~~~~~~~~~~~~~~~~~~~~~~~~~ \\=a d \varphi d z \frac{(a \cos \varphi - \rho)\cos \varphi + (a \sin \varphi - \zeta)\sin \varphi}{r}.~~~
\label{r2.paral}
\end{eqnarray}
In the parallel case, one has to also include the bases of the cylinder, which correspond to:
\begin{eqnarray}
{r^2}=(r' \cos \varphi - \rho)^2 + (r' \sin \varphi)^2 + (z -\zeta)^2~,~~{\rm and}~~~\nonumber\\
d{\bf A}\cdot {\bf n_r}=r' d r' d \varphi {\bf n_A}\cdot {\bf n_r}~~~~~~~~~~~~~~~~~~~~~~~~~~~~~~~~~~ \nonumber \\=r' d r' d \varphi \frac{(r' \cos \varphi - \rho)\cos \varphi + (r' \sin \varphi - \zeta)\sin \varphi}{r'}.~~~
\label{r2.base}
\end{eqnarray}

\subsubsection{DNA on the surface of a nucleosome for complex damage calculation}
\label{sec.geo.complex}

In the case of the calculation of complex damage the electrons are incident on a cylindrical surface of a nucleosome. The corresponding $r^2$ and ${\bf n_r\cdot n_A}$ are given by
\begin{eqnarray}
{r^2}=(a_n \cos \varphi - \rho)^2 + (a_n \sin \varphi - \zeta)^2  + z^2,~~~~{\rm and}~~~~~~\nonumber\\
d{\bf A_n}\cdot {\bf n_r}=a_n d \varphi d z {\bf n_A}\cdot {\bf n_r}\nonumber~~~~~~~~~~~~~~~~~~~~~~~~~~~~~~~~~~~ \\=a_n d \varphi d z \frac{(a_n \cos \varphi - \rho)\cos \varphi + (a \sin \varphi - \zeta)\sin \varphi}{r}~~~
\label{nucl.perp}
\end{eqnarray}
and the integration is done over $z$ in the limits from $-1.15$ nm to $+1.15$ and over $\varphi$ from \newline $\max\left[-\psi, \arctan \frac{\zeta}{\rho}-\arccos \frac{a_n}{\zeta^2+\rho^2}\right] $ to \newline $\min\left[\psi, \arctan \frac{\zeta}{\rho}+\arccos \frac{a_n}{\zeta^2+\rho^2}\right] $, where $\psi$ is the ratio of the length of a twist (3.4~nm) to the radius of a nucleosome (5.75~nm)~\cite{nucleosome,precomplex}.


\end{appendix}

\begin{thebibliography}{134}
\expandafter\ifx\csname natexlab\endcsname\relax\def\natexlab#1{#1}\fi
\expandafter\ifx\csname bibnamefont\endcsname\relax
  \def\bibnamefont#1{#1}\fi
\expandafter\ifx\csname bibfnamefont\endcsname\relax
  \def\bibfnamefont#1{#1}\fi
\expandafter\ifx\csname citenamefont\endcsname\relax
  \def\citenamefont#1{#1}\fi
\expandafter\ifx\csname url\endcsname\relax
  \def\url#1{\texttt{#1}}\fi
\expandafter\ifx\csname urlprefix\endcsname\relax\def\urlprefix{URL }\fi
\providecommand{\bibinfo}[2]{#2}
\providecommand{\eprint}[2][]{\url{#2}}

\bibitem[{\citenamefont{Surdutovich and
  Solov'yov}(2012{\natexlab{a}})}]{MSAreview}
\bibinfo{author}{\bibfnamefont{E.}~\bibnamefont{Surdutovich}} \bibnamefont{and}
  \bibinfo{author}{\bibfnamefont{A.}~\bibnamefont{Solov'yov}},
  \bibinfo{journal}{J. Phys.: Conf. Ser.} \textbf{\bibinfo{volume}{373}},
  \bibinfo{pages}{012001} (\bibinfo{year}{2012}{\natexlab{a}}).

\bibitem[{\citenamefont{Baccarelli et~al.}(2010)\citenamefont{Baccarelli,
  Gianturco, Scifoni, Solov'yov, and Surdutovich}}]{Radam09editorial}
\bibinfo{author}{\bibfnamefont{I.}~\bibnamefont{Baccarelli}},
  \bibinfo{author}{\bibfnamefont{F.}~\bibnamefont{Gianturco}},
  \bibinfo{author}{\bibfnamefont{E.}~\bibnamefont{Scifoni}},
  \bibinfo{author}{\bibfnamefont{A.}~\bibnamefont{Solov'yov}},
  \bibnamefont{and}
  \bibinfo{author}{\bibfnamefont{E.}~\bibnamefont{Surdutovich}},
  \bibinfo{journal}{Eur. Phys. J. D} \textbf{\bibinfo{volume}{60}},
  \bibinfo{pages}{1} (\bibinfo{year}{2010}).

\bibitem[{\citenamefont{Amaldi and Kraft}(2007)}]{Kraft07}
\bibinfo{author}{\bibfnamefont{U.}~\bibnamefont{Amaldi}} \bibnamefont{and}
  \bibinfo{author}{\bibfnamefont{G.}~\bibnamefont{Kraft}}, \bibinfo{journal}{J.
  Radiat. Res.} \textbf{\bibinfo{volume}{48}}, \bibinfo{pages}{A27}
  (\bibinfo{year}{2007}).

\bibitem[{\citenamefont{Schardt et~al.}(2010)\citenamefont{Schardt,
  Els{\"a}sser, and Schulz-Ertner}}]{SchardtRMP10}
\bibinfo{author}{\bibfnamefont{D.}~\bibnamefont{Schardt}},
  \bibinfo{author}{\bibfnamefont{T.}~\bibnamefont{Els{\"a}sser}},
  \bibnamefont{and}
  \bibinfo{author}{\bibfnamefont{D.}~\bibnamefont{Schulz-Ertner}},
  \bibinfo{journal}{Rev. Mod. Phys.} \textbf{\bibinfo{volume}{82}},
  \bibinfo{pages}{383} (\bibinfo{year}{2010}).

\bibitem[{\citenamefont{Durante and Loeffler}(2010)}]{Durante10}
\bibinfo{author}{\bibfnamefont{M.}~\bibnamefont{Durante}} \bibnamefont{and}
  \bibinfo{author}{\bibfnamefont{J.}~\bibnamefont{Loeffler}},
  \bibinfo{journal}{Nat. Rev. Clin. Oncol.} \textbf{\bibinfo{volume}{7}},
  \bibinfo{pages}{37} (\bibinfo{year}{2010}).

\bibitem[{wik({\natexlab{a}})}]{wikiCentres}
\emph{\bibinfo{title}{Proton therapy -- wikipedia}}, \bibinfo{note}{accessed on
  {09/2013}}, \urlprefix\url{http://en.wikipedia.org/wiki/Proton{\_}therapy}.

\bibitem[{wik({\natexlab{b}})}]{wikiCarbon}
\emph{\bibinfo{title}{Particle therapy -- wikipedia}}, \bibinfo{note}{accessed
  on {09/2013}},
  \urlprefix\url{http://en.wikipedia.org/wiki/Particle{\_}therapy}.

\bibitem[{\citenamefont{Haettner et~al.}(2006)\citenamefont{Haettner, Iwase,
  and Schardt}}]{Schardt}
\bibinfo{author}{\bibfnamefont{E.}~\bibnamefont{Haettner}},
  \bibinfo{author}{\bibfnamefont{H.}~\bibnamefont{Iwase}}, \bibnamefont{and}
  \bibinfo{author}{\bibfnamefont{D.}~\bibnamefont{Schardt}},
  \bibinfo{journal}{Rad. Protec. Dosim.} \textbf{\bibinfo{volume}{122}},
  \bibinfo{pages}{485} (\bibinfo{year}{2006}).

\bibitem[{\citenamefont{Sihver et~al.}(1998)\citenamefont{Sihver, Schardt, and
  Kanai}}]{Schardt4}
\bibinfo{author}{\bibfnamefont{L.}~\bibnamefont{Sihver}},
  \bibinfo{author}{\bibfnamefont{D.}~\bibnamefont{Schardt}}, \bibnamefont{and}
  \bibinfo{author}{\bibfnamefont{T.}~\bibnamefont{Kanai}},
  \bibinfo{journal}{Jpn. J. Med. Phys.} \textbf{\bibinfo{volume}{18}},
  \bibinfo{pages}{1} (\bibinfo{year}{1998}).

\bibitem[{\citenamefont{Pshenichnov et~al.}(2008)\citenamefont{Pshenichnov,
  Mishustin, and Greiner}}]{Pshen}
\bibinfo{author}{\bibfnamefont{I.}~\bibnamefont{Pshenichnov}},
  \bibinfo{author}{\bibfnamefont{I.}~\bibnamefont{Mishustin}},
  \bibnamefont{and} \bibinfo{author}{\bibfnamefont{W.}~\bibnamefont{Greiner}},
  \bibinfo{journal}{Nucl. Inst. Meth. B} \textbf{\bibinfo{volume}{266}},
  \bibinfo{pages}{1094} (\bibinfo{year}{2008}).

\bibitem[{\citenamefont{Surdutovich et~al.}(2009)\citenamefont{Surdutovich,
  Obolensky, Scifoni, Pshenichnov, Mishustin, Solov'yov, and Greiner}}]{epjd}
\bibinfo{author}{\bibfnamefont{E.}~\bibnamefont{Surdutovich}},
  \bibinfo{author}{\bibfnamefont{O.}~\bibnamefont{Obolensky}},
  \bibinfo{author}{\bibfnamefont{E.}~\bibnamefont{Scifoni}},
  \bibinfo{author}{\bibfnamefont{I.}~\bibnamefont{Pshenichnov}},
  \bibinfo{author}{\bibfnamefont{I.}~\bibnamefont{Mishustin}},
  \bibinfo{author}{\bibfnamefont{A.}~\bibnamefont{Solov'yov}},
  \bibnamefont{and} \bibinfo{author}{\bibfnamefont{W.}~\bibnamefont{Greiner}},
  \bibinfo{journal}{Eur. Phys. J. D} \textbf{\bibinfo{volume}{51}},
  \bibinfo{pages}{63} (\bibinfo{year}{2009}).

\bibitem[{\citenamefont{Scifoni et~al.}(2010)\citenamefont{Scifoni,
  Surdutovich, and Solovyov}}]{Scif}
\bibinfo{author}{\bibfnamefont{E.}~\bibnamefont{Scifoni}},
  \bibinfo{author}{\bibfnamefont{E.}~\bibnamefont{Surdutovich}},
  \bibnamefont{and} \bibinfo{author}{\bibfnamefont{A.}~\bibnamefont{Solovyov}},
  \bibinfo{journal}{Phys Rev. E} \textbf{\bibinfo{volume}{81}},
  \bibinfo{pages}{021903} (\bibinfo{year}{2010}).

\bibitem[{\citenamefont{Chatterjee and Holley}(1993)}]{Chatterjee93}
\bibinfo{author}{\bibfnamefont{A.}~\bibnamefont{Chatterjee}} \bibnamefont{and}
  \bibinfo{author}{\bibfnamefont{W.~R.} \bibnamefont{Holley}},
  \bibinfo{journal}{Adv. Radiat. Biol.} \textbf{\bibinfo{volume}{17}},
  \bibinfo{pages}{181} (\bibinfo{year}{1993}).

\bibitem[{\citenamefont{von Sonntag}(1987)}]{hyd2}
\bibinfo{author}{\bibfnamefont{C.}~\bibnamefont{von Sonntag}},
  \emph{\bibinfo{title}{The chemical basis of radiation biology}}
  (\bibinfo{publisher}{Taylor \& Francis}, \bibinfo{address}{London},
  \bibinfo{year}{1987}).

\bibitem[{\citenamefont{Surdutovich et~al.}(2011)\citenamefont{Surdutovich,
  Gallagher, and Solov'yov}}]{precomplex}
\bibinfo{author}{\bibfnamefont{E.}~\bibnamefont{Surdutovich}},
  \bibinfo{author}{\bibfnamefont{D.~C.} \bibnamefont{Gallagher}},
  \bibnamefont{and} \bibinfo{author}{\bibfnamefont{A.~V.}
  \bibnamefont{Solov'yov}}, \bibinfo{journal}{Phys. Rev. E}
  \textbf{\bibinfo{volume}{84}}, \bibinfo{pages}{051918}
  (\bibinfo{year}{2011}).

\bibitem[{\citenamefont{Sanche}(2005)}]{Sanche05}
\bibinfo{author}{\bibfnamefont{L.}~\bibnamefont{Sanche}},
  \bibinfo{journal}{Eur. Phys. J. D} \textbf{\bibinfo{volume}{35}},
  \bibinfo{pages}{367} (\bibinfo{year}{2005}).

\bibitem[{\citenamefont{Bouda{\"i}ffa et~al.}(2000)\citenamefont{Bouda{\"i}ffa,
  Cloutier, Hunting, Huels, and Sanche}}]{DNA2}
\bibinfo{author}{\bibfnamefont{B.}~\bibnamefont{Bouda{\"i}ffa}},
  \bibinfo{author}{\bibfnamefont{P.}~\bibnamefont{Cloutier}},
  \bibinfo{author}{\bibfnamefont{D.}~\bibnamefont{Hunting}},
  \bibinfo{author}{\bibfnamefont{M.~A.} \bibnamefont{Huels}}, \bibnamefont{and}
  \bibinfo{author}{\bibfnamefont{L.}~\bibnamefont{Sanche}},
  \bibinfo{journal}{Science} \textbf{\bibinfo{volume}{287}},
  \bibinfo{pages}{1658} (\bibinfo{year}{2000}).

\bibitem[{\citenamefont{Huels et~al.}(2003)\citenamefont{Huels, Bouda{\"i}ffa,
  Cloutier, Hunting, and Sanche}}]{DNA3}
\bibinfo{author}{\bibfnamefont{M.~A.} \bibnamefont{Huels}},
  \bibinfo{author}{\bibfnamefont{B.}~\bibnamefont{Bouda{\"i}ffa}},
  \bibinfo{author}{\bibfnamefont{P.}~\bibnamefont{Cloutier}},
  \bibinfo{author}{\bibfnamefont{D.}~\bibnamefont{Hunting}}, \bibnamefont{and}
  \bibinfo{author}{\bibfnamefont{L.}~\bibnamefont{Sanche}},
  \bibinfo{journal}{JACS} \textbf{\bibinfo{volume}{125}}, \bibinfo{pages}{4467}
  (\bibinfo{year}{2003}).

\bibitem[{\citenamefont{Sanche}(2010)}]{SancheCh9.2012}
\bibinfo{author}{\bibfnamefont{L.}~\bibnamefont{Sanche}}, in
  \emph{\bibinfo{booktitle}{Radical and radical ion reactivity in nucleic acid
  chemistry}}, edited by
  \bibinfo{editor}{\bibfnamefont{M.}~\bibnamefont{Greenberg}}
  (\bibinfo{publisher}{J. Wiley \& Sons, Inc.}, \bibinfo{address}{New York},
  \bibinfo{year}{2010}), p. \bibinfo{pages}{239}.

\bibitem[{\citenamefont{Surdutovich and
  Solov'yov}(2012{\natexlab{b}})}]{prauger}
\bibinfo{author}{\bibfnamefont{E.}~\bibnamefont{Surdutovich}} \bibnamefont{and}
  \bibinfo{author}{\bibfnamefont{A.~V.} \bibnamefont{Solov'yov}},
  \bibinfo{journal}{Eur. Phys. J. D} \textbf{\bibinfo{volume}{66}},
  \bibinfo{pages}{206} (\bibinfo{year}{2012}{\natexlab{b}}).

\bibitem[{\citenamefont{McMahon et~al.}(2011)\citenamefont{McMahon, Hyland,
  Muir, Coulter, Jain, Butterworth, Schettino, Dickson, Hounsell, O'Sullivan
  et~al.}}]{Prise11}
\bibinfo{author}{\bibfnamefont{S.}~\bibnamefont{McMahon}},
  \bibinfo{author}{\bibfnamefont{W.}~\bibnamefont{Hyland}},
  \bibinfo{author}{\bibfnamefont{M.}~\bibnamefont{Muir}},
  \bibinfo{author}{\bibfnamefont{J.}~\bibnamefont{Coulter}},
  \bibinfo{author}{\bibfnamefont{S.}~\bibnamefont{Jain}},
  \bibinfo{author}{\bibfnamefont{K.}~\bibnamefont{Butterworth}},
  \bibinfo{author}{\bibfnamefont{G.}~\bibnamefont{Schettino}},
  \bibinfo{author}{\bibfnamefont{G.}~\bibnamefont{Dickson}},
  \bibinfo{author}{\bibfnamefont{A.}~\bibnamefont{Hounsell}},
  \bibinfo{author}{\bibfnamefont{J.}~\bibnamefont{O'Sullivan}},
  \bibnamefont{et~al.}, \bibinfo{journal}{Sci. Rep.}
  \textbf{\bibinfo{volume}{1}}, \bibinfo{pages}{18} (\bibinfo{year}{2011}).

\bibitem[{\citenamefont{Solov'yov et~al.}(2009)\citenamefont{Solov'yov,
  Surdutovich, Scifoni, Mishustin, and Greiner}}]{pre}
\bibinfo{author}{\bibfnamefont{A.}~\bibnamefont{Solov'yov}},
  \bibinfo{author}{\bibfnamefont{E.}~\bibnamefont{Surdutovich}},
  \bibinfo{author}{\bibfnamefont{E.}~\bibnamefont{Scifoni}},
  \bibinfo{author}{\bibfnamefont{I.}~\bibnamefont{Mishustin}},
  \bibnamefont{and} \bibinfo{author}{\bibfnamefont{W.}~\bibnamefont{Greiner}},
  \bibinfo{journal}{Phys. Rev.} \textbf{\bibinfo{volume}{E79}},
  \bibinfo{pages}{011909} (\bibinfo{year}{2009}).

\bibitem[{nan()}]{nanoIBCT}
\emph{\bibinfo{title}{Cost nano-ibct -- nanoscale insights into ion beam cancer
  therapy}}, \bibinfo{note}{accessed on {01/2013}},
  \urlprefix\url{http://www.cost.eu/domains{\textunderscore}actions/mpns/Actions/nano-ibct/}.

\bibitem[{\citenamefont{Surdutovich and Solov'yov}(2009)}]{epn}
\bibinfo{author}{\bibfnamefont{E.}~\bibnamefont{Surdutovich}} \bibnamefont{and}
  \bibinfo{author}{\bibfnamefont{A.}~\bibnamefont{Solov'yov}},
  \bibinfo{journal}{Europhys. News} \textbf{\bibinfo{volume}{40/2}},
  \bibinfo{pages}{21} (\bibinfo{year}{2009}).

\bibitem[{\citenamefont{Surdutovich
  et~al.}(2010{\natexlab{a}})\citenamefont{Surdutovich, Scifoni, , and
  Solov'yov}}]{mutat}
\bibinfo{author}{\bibfnamefont{E.}~\bibnamefont{Surdutovich}},
  \bibinfo{author}{\bibfnamefont{E.}~\bibnamefont{Scifoni}}, ,
  \bibnamefont{and}
  \bibinfo{author}{\bibfnamefont{A.}~\bibnamefont{Solov'yov}},
  \bibinfo{journal}{Mutat. Res.} \textbf{\bibinfo{volume}{704}},
  \bibinfo{pages}{206} (\bibinfo{year}{2010}{\natexlab{a}}).

\bibitem[{\citenamefont{Toulemonde et~al.}(2009)\citenamefont{Toulemonde,
  Surdutovich, and Solov'yov}}]{preheat}
\bibinfo{author}{\bibfnamefont{M.}~\bibnamefont{Toulemonde}},
  \bibinfo{author}{\bibfnamefont{E.}~\bibnamefont{Surdutovich}},
  \bibnamefont{and}
  \bibinfo{author}{\bibfnamefont{A.}~\bibnamefont{Solov'yov}},
  \bibinfo{journal}{Phys. Rev. E} \textbf{\bibinfo{volume}{80}},
  \bibinfo{pages}{031913} (\bibinfo{year}{2009}).

\bibitem[{\citenamefont{Surdutovich
  et~al.}(2010{\natexlab{b}})\citenamefont{Surdutovich, Yakubovich, and
  Solov'yov}}]{SYS}
\bibinfo{author}{\bibfnamefont{E.}~\bibnamefont{Surdutovich}},
  \bibinfo{author}{\bibfnamefont{A.}~\bibnamefont{Yakubovich}},
  \bibnamefont{and}
  \bibinfo{author}{\bibfnamefont{A.}~\bibnamefont{Solov'yov}},
  \bibinfo{journal}{Eur. Phys. J. D} \textbf{\bibinfo{volume}{60}},
  \bibinfo{pages}{101} (\bibinfo{year}{2010}{\natexlab{b}}).

\bibitem[{\citenamefont{Surdutovich and Solov'yov}(2010)}]{prehydro}
\bibinfo{author}{\bibfnamefont{E.}~\bibnamefont{Surdutovich}} \bibnamefont{and}
  \bibinfo{author}{\bibfnamefont{A.}~\bibnamefont{Solov'yov}},
  \bibinfo{journal}{Phys. Rev. E} \textbf{\bibinfo{volume}{82}},
  \bibinfo{pages}{051915} (\bibinfo{year}{2010}).

\bibitem[{\citenamefont{Surdutovich et~al.}(2013)\citenamefont{Surdutovich,
  Yakubovich, and Solov'yov}}]{natnuke}
\bibinfo{author}{\bibfnamefont{E.}~\bibnamefont{Surdutovich}},
  \bibinfo{author}{\bibfnamefont{A.~V.} \bibnamefont{Yakubovich}},
  \bibnamefont{and} \bibinfo{author}{\bibfnamefont{A.~V.}
  \bibnamefont{Solov'yov}}, \bibinfo{journal}{Sci. Rep.}
  \textbf{\bibinfo{volume}{3}}, \bibinfo{pages}{1289} (\bibinfo{year}{2013}).

\bibitem[{\citenamefont{de~Vera et~al.}(2013)\citenamefont{de~Vera,
  Garcia-Molina, Abril, and Solov'yov}}]{Pablo2012}
\bibinfo{author}{\bibfnamefont{P.}~\bibnamefont{de~Vera}},
  \bibinfo{author}{\bibfnamefont{R.}~\bibnamefont{Garcia-Molina}},
  \bibinfo{author}{\bibfnamefont{I.}~\bibnamefont{Abril}}, \bibnamefont{and}
  \bibinfo{author}{\bibfnamefont{A.~V.} \bibnamefont{Solov'yov}},
  \bibinfo{journal}{Phys. Rev. Lett.} \textbf{\bibinfo{volume}{110}},
  \bibinfo{pages}{148104} (\bibinfo{year}{2013}).

\bibitem[{\citenamefont{Alpen}(1998)}]{Alpen}
\bibinfo{author}{\bibfnamefont{E.~L.} \bibnamefont{Alpen}},
  \emph{\bibinfo{title}{Radiation Biophysics}} (\bibinfo{publisher}{Academic
  Press}, \bibinfo{address}{San Diego, London, Boston, New York, Sydney, Tokyo,
  Toronto}, \bibinfo{year}{1998}).

\bibitem[{\citenamefont{Hall and Giaccia}(2012)}]{Hall}
\bibinfo{author}{\bibfnamefont{E.~J.} \bibnamefont{Hall}} \bibnamefont{and}
  \bibinfo{author}{\bibfnamefont{A.~J.} \bibnamefont{Giaccia}},
  \emph{\bibinfo{title}{Radiobiology for Radiologist}}
  (\bibinfo{publisher}{Lippincott Williams \& Wilkins},
  \bibinfo{address}{Philadelphia, Baltimore, New York, London},
  \bibinfo{year}{2012}).

\bibitem[{\citenamefont{Pimblott and Siebbeles}(2002)}]{Pimblott2}
\bibinfo{author}{\bibfnamefont{S.}~\bibnamefont{Pimblott}} \bibnamefont{and}
  \bibinfo{author}{\bibfnamefont{L.}~\bibnamefont{Siebbeles}},
  \bibinfo{journal}{Nucl. Inst. Meth. B} \textbf{\bibinfo{volume}{194}},
  \bibinfo{pages}{237} (\bibinfo{year}{2002}).

\bibitem[{\citenamefont{Pimblott et~al.}(1996)\citenamefont{Pimblott, LaVerne,
  and Mozumder}}]{Pimblott3}
\bibinfo{author}{\bibfnamefont{S.}~\bibnamefont{Pimblott}},
  \bibinfo{author}{\bibfnamefont{J.}~\bibnamefont{LaVerne}}, \bibnamefont{and}
  \bibinfo{author}{\bibfnamefont{A.}~\bibnamefont{Mozumder}},
  \bibinfo{journal}{J. Phys. Chem} \textbf{\bibinfo{volume}{100}},
  \bibinfo{pages}{8595} (\bibinfo{year}{1996}).

\bibitem[{\citenamefont{Pimblott and LaVerne}(2007)}]{Pimblott}
\bibinfo{author}{\bibfnamefont{S.}~\bibnamefont{Pimblott}} \bibnamefont{and}
  \bibinfo{author}{\bibfnamefont{J.}~\bibnamefont{LaVerne}},
  \bibinfo{journal}{Rad. Phys. Chem.} \textbf{\bibinfo{volume}{76}},
  \bibinfo{pages}{1244} (\bibinfo{year}{2007}).

\bibitem[{\citenamefont{Meesungnoen et~al.}(2002)\citenamefont{Meesungnoen,
  Jay-Gerin, Filali-Mouhim, and Mankhetkorn}}]{Meesungnoen02}
\bibinfo{author}{\bibfnamefont{J.}~\bibnamefont{Meesungnoen}},
  \bibinfo{author}{\bibfnamefont{J.-P.} \bibnamefont{Jay-Gerin}},
  \bibinfo{author}{\bibfnamefont{A.}~\bibnamefont{Filali-Mouhim}},
  \bibnamefont{and}
  \bibinfo{author}{\bibfnamefont{S.}~\bibnamefont{Mankhetkorn}},
  \bibinfo{journal}{Radiat. Res.} \textbf{\bibinfo{volume}{158}},
  \bibinfo{pages}{657} (\bibinfo{year}{2002}).

\bibitem[{\citenamefont{Nikjoo et~al.}(2006)\citenamefont{Nikjoo, Uehara,
  Emfietzoglou, and Cucinotta}}]{Nikjoo06}
\bibinfo{author}{\bibfnamefont{H.}~\bibnamefont{Nikjoo}},
  \bibinfo{author}{\bibfnamefont{S.}~\bibnamefont{Uehara}},
  \bibinfo{author}{\bibfnamefont{D.}~\bibnamefont{Emfietzoglou}},
  \bibnamefont{and} \bibinfo{author}{\bibfnamefont{F.~A.}
  \bibnamefont{Cucinotta}}, \bibinfo{journal}{Radiat. Meas.}
  \textbf{\bibinfo{volume}{41}}, \bibinfo{pages}{1052} (\bibinfo{year}{2006}).

\bibitem[{\citenamefont{Surdutovich and
  Solov'yov}(2012{\natexlab{c}})}]{epjdisacc2011}
\bibinfo{author}{\bibfnamefont{E.}~\bibnamefont{Surdutovich}} \bibnamefont{and}
  \bibinfo{author}{\bibfnamefont{A.~V.} \bibnamefont{Solov'yov}},
  \bibinfo{journal}{Eur. Phys. J. D} \textbf{\bibinfo{volume}{66}},
  \bibinfo{pages}{245} (\bibinfo{year}{2012}{\natexlab{c}}).

\bibitem[{\citenamefont{Bug et~al.}(2012)\citenamefont{Bug, Surdutovich, Rabus,
  Rosenfeld, and Solov'yov}}]{epjdmarion}
\bibinfo{author}{\bibfnamefont{M.}~\bibnamefont{Bug}},
  \bibinfo{author}{\bibfnamefont{E.}~\bibnamefont{Surdutovich}},
  \bibinfo{author}{\bibfnamefont{H.}~\bibnamefont{Rabus}},
  \bibinfo{author}{\bibfnamefont{A.~B.} \bibnamefont{Rosenfeld}},
  \bibnamefont{and} \bibinfo{author}{\bibfnamefont{A.~V.}
  \bibnamefont{Solov'yov}}, \bibinfo{journal}{Eur. Phys. J. D}
  \textbf{\bibinfo{volume}{66}}, \bibinfo{pages}{291} (\bibinfo{year}{2012}).

\bibitem[{\citenamefont{Park et~al.}(2011)\citenamefont{Park, Li, Cloutier,
  Sanche, and Wagner}}]{Sanche11}
\bibinfo{author}{\bibfnamefont{Y.}~\bibnamefont{Park}},
  \bibinfo{author}{\bibfnamefont{Z.}~\bibnamefont{Li}},
  \bibinfo{author}{\bibfnamefont{P.}~\bibnamefont{Cloutier}},
  \bibinfo{author}{\bibfnamefont{L.}~\bibnamefont{Sanche}}, \bibnamefont{and}
  \bibinfo{author}{\bibfnamefont{J.}~\bibnamefont{Wagner}},
  \bibinfo{journal}{Radiat. Res.} \textbf{\bibinfo{volume}{175}},
  \bibinfo{pages}{240} (\bibinfo{year}{2011}).

\bibitem[{\citenamefont{Smyth and Kohanoff}(2012)}]{Kohanoff2012}
\bibinfo{author}{\bibfnamefont{M.}~\bibnamefont{Smyth}} \bibnamefont{and}
  \bibinfo{author}{\bibfnamefont{J.}~\bibnamefont{Kohanoff}},
  \bibinfo{journal}{J. Am. Chem. Soc.} \textbf{\bibinfo{volume}{134}},
  \bibinfo{pages}{9122} (\bibinfo{year}{2012}).

\bibitem[{\citenamefont{Becker et~al.}(2010)\citenamefont{Becker, Adhikary, and
  Sevilla}}]{SevReview}
\bibinfo{author}{\bibfnamefont{D.}~\bibnamefont{Becker}},
  \bibinfo{author}{\bibfnamefont{A.}~\bibnamefont{Adhikary}}, \bibnamefont{and}
  \bibinfo{author}{\bibfnamefont{M.}~\bibnamefont{Sevilla}}, in
  \emph{\bibinfo{booktitle}{Charged Particle and Photon Interactions with
  Matter Recent Advances, Applications, and Interfaces}}
  (\bibinfo{publisher}{CRC Press, Taylor \& Francis}, \bibinfo{address}{Boca
  Raton}, \bibinfo{year}{2010}).

\bibitem[{\citenamefont{Bethe}(1930)}]{Bethe}
\bibinfo{author}{\bibfnamefont{H.}~\bibnamefont{Bethe}}, \bibinfo{journal}{Ann.
  Phys.} \textbf{\bibinfo{volume}{397}}, \bibinfo{pages}{325}
  (\bibinfo{year}{1930}).

\bibitem[{\citenamefont{Bloch}(1933{\natexlab{a}})}]{Bloch1}
\bibinfo{author}{\bibfnamefont{F.}~\bibnamefont{Bloch}}, \bibinfo{journal}{Z.
  Phys. A: Hadrons Nucl.} \textbf{\bibinfo{volume}{81}}, \bibinfo{pages}{363}
  (\bibinfo{year}{1933}{\natexlab{a}}).

\bibitem[{\citenamefont{Bloch}(1933{\natexlab{b}})}]{Bloch2}
\bibinfo{author}{\bibfnamefont{F.}~\bibnamefont{Bloch}}, \bibinfo{journal}{Ann.
  Phys.} \textbf{\bibinfo{volume}{408}}, \bibinfo{pages}{285–}
  (\bibinfo{year}{1933}{\natexlab{b}}).

\bibitem[{\citenamefont{Abril et~al.}(2011)\citenamefont{Abril, Garcia-Molina,
  Denton, Kyriakou, and Emfietzoglou}}]{Molina}
\bibinfo{author}{\bibfnamefont{I.}~\bibnamefont{Abril}},
  \bibinfo{author}{\bibfnamefont{R.}~\bibnamefont{Garcia-Molina}},
  \bibinfo{author}{\bibfnamefont{C.}~\bibnamefont{Denton}},
  \bibinfo{author}{\bibfnamefont{I.}~\bibnamefont{Kyriakou}}, \bibnamefont{and}
  \bibinfo{author}{\bibfnamefont{D.}~\bibnamefont{Emfietzoglou}},
  \bibinfo{journal}{Radiat. Res.} \textbf{\bibinfo{volume}{175}},
  \bibinfo{pages}{247} (\bibinfo{year}{2011}).

\bibitem[{\citenamefont{Obolensky et~al.}(2008)\citenamefont{Obolensky,
  Surdutovich, Pshenichnov, Mishustin, Solov'yov, and Greiner}}]{nimb}
\bibinfo{author}{\bibfnamefont{O.}~\bibnamefont{Obolensky}},
  \bibinfo{author}{\bibfnamefont{E.}~\bibnamefont{Surdutovich}},
  \bibinfo{author}{\bibfnamefont{I.}~\bibnamefont{Pshenichnov}},
  \bibinfo{author}{\bibfnamefont{I.}~\bibnamefont{Mishustin}},
  \bibinfo{author}{\bibfnamefont{A.}~\bibnamefont{Solov'yov}},
  \bibnamefont{and} \bibinfo{author}{\bibfnamefont{W.}~\bibnamefont{Greiner}},
  \bibinfo{journal}{Nucl. Inst. Meth. B} \textbf{\bibinfo{volume}{266}},
  \bibinfo{pages}{1623} (\bibinfo{year}{2008}).

\bibitem[{\citenamefont{Rudd et~al.}(1992)\citenamefont{Rudd, Kim, Madison, and
  Gay}}]{Rudd92}
\bibinfo{author}{\bibfnamefont{M.~E.} \bibnamefont{Rudd}},
  \bibinfo{author}{\bibfnamefont{Y.-K.} \bibnamefont{Kim}},
  \bibinfo{author}{\bibfnamefont{D.~H.} \bibnamefont{Madison}},
  \bibnamefont{and} \bibinfo{author}{\bibfnamefont{T.}~\bibnamefont{Gay}},
  \bibinfo{journal}{Rev. Mod. Phys.} \textbf{\bibinfo{volume}{64}},
  \bibinfo{pages}{441} (\bibinfo{year}{1992}).

\bibitem[{\citenamefont{Landau et~al.}(1984)\citenamefont{Landau, Lifshitz, and
  Pitaevskii}}]{landau8}
\bibinfo{author}{\bibfnamefont{L.}~\bibnamefont{Landau}},
  \bibinfo{author}{\bibfnamefont{E.}~\bibnamefont{Lifshitz}}, \bibnamefont{and}
  \bibinfo{author}{\bibfnamefont{L.}~\bibnamefont{Pitaevskii}},
  \emph{\bibinfo{title}{Electrodynamics of Continuous Media, Second Edition:
  Volume 8}} (\bibinfo{publisher}{Butterworth-Heinemann},
  \bibinfo{address}{Burlington}, \bibinfo{year}{1984}).

\bibitem[{\citenamefont{Lindhard and Dan}(1954)}]{Lindhard54}
\bibinfo{author}{\bibfnamefont{J.}~\bibnamefont{Lindhard}} \bibnamefont{and}
  \bibinfo{author}{\bibfnamefont{K.}~\bibnamefont{Dan}},
  \bibinfo{journal}{Vidensk. Selsk. Mat. Fys. Medd.}
  \textbf{\bibinfo{volume}{28}}, \bibinfo{pages}{8} (\bibinfo{year}{1954}).

\bibitem[{\citenamefont{Tan et~al.}(2004)\citenamefont{Tan, Xia, Zhao, Liu, Li,
  Huang, and Ji}}]{Tan04}
\bibinfo{author}{\bibfnamefont{Z.}~\bibnamefont{Tan}},
  \bibinfo{author}{\bibfnamefont{Y.}~\bibnamefont{Xia}},
  \bibinfo{author}{\bibfnamefont{M.}~\bibnamefont{Zhao}},
  \bibinfo{author}{\bibfnamefont{X.}~\bibnamefont{Liu}},
  \bibinfo{author}{\bibfnamefont{F.}~\bibnamefont{Li}},
  \bibinfo{author}{\bibfnamefont{B.}~\bibnamefont{Huang}}, \bibnamefont{and}
  \bibinfo{author}{\bibfnamefont{Y.}~\bibnamefont{Ji}}, \bibinfo{journal}{Nucl.
  Instrum. Methods Phys. Res. B} \textbf{\bibinfo{volume}{222}},
  \bibinfo{pages}{27} (\bibinfo{year}{2004}).

\bibitem[{\citenamefont{Altarelli and Smith}(1974)}]{Altarelli74}
\bibinfo{author}{\bibfnamefont{M.}~\bibnamefont{Altarelli}} \bibnamefont{and}
  \bibinfo{author}{\bibfnamefont{D.}~\bibnamefont{Smith}},
  \bibinfo{journal}{Phys. Rev. B} \textbf{\bibinfo{volume}{9}},
  \bibinfo{pages}{1290} (\bibinfo{year}{1974}).

\bibitem[{\citenamefont{Garcia-Molina
  et~al.}(2012{\natexlab{a}})\citenamefont{Garcia-Molina, Abril, Kyriakou, and
  Emfietzoglou}}]{RGM12}
\bibinfo{author}{\bibfnamefont{R.}~\bibnamefont{Garcia-Molina}},
  \bibinfo{author}{\bibfnamefont{I.}~\bibnamefont{Abril}},
  \bibinfo{author}{\bibfnamefont{I.}~\bibnamefont{Kyriakou}}, \bibnamefont{and}
  \bibinfo{author}{\bibfnamefont{D.}~\bibnamefont{Emfietzoglou}}, in
  \emph{\bibinfo{booktitle}{Radiation Damage in Biomolecular Systems}}, edited
  by \bibinfo{editor}{\bibfnamefont{G.~G.} \bibnamefont{G{\'o}mez-Tejedor}}
  \bibnamefont{and} \bibinfo{editor}{\bibfnamefont{M.~C.} \bibnamefont{Fuss}}
  (\bibinfo{publisher}{Springer}, \bibinfo{address}{Dordrecht},
  \bibinfo{year}{2012}{\natexlab{a}}), p. \bibinfo{pages}{Chap. 15}.

\bibitem[{\citenamefont{Ritchie and Howie}(1977)}]{Ritchie}
\bibinfo{author}{\bibfnamefont{R.~H.} \bibnamefont{Ritchie}} \bibnamefont{and}
  \bibinfo{author}{\bibfnamefont{A.}~\bibnamefont{Howie}},
  \bibinfo{journal}{Philos. Mag.} \textbf{\bibinfo{volume}{36}},
  \bibinfo{pages}{436} (\bibinfo{year}{1977}).

\bibitem[{\citenamefont{Dingfelder et~al.}(1999)\citenamefont{Dingfelder,
  Hantke, Inokuti, and Paretzke}}]{Dingfelder}
\bibinfo{author}{\bibfnamefont{M.}~\bibnamefont{Dingfelder}},
  \bibinfo{author}{\bibfnamefont{D.}~\bibnamefont{Hantke}},
  \bibinfo{author}{\bibfnamefont{M.}~\bibnamefont{Inokuti}}, \bibnamefont{and}
  \bibinfo{author}{\bibfnamefont{H.}~\bibnamefont{Paretzke}},
  \bibinfo{journal}{Radiat. Phys. Chem.} \textbf{\bibinfo{volume}{53}},
  \bibinfo{pages}{1} (\bibinfo{year}{1999}).

\bibitem[{\citenamefont{Emfietzoglou}(2003)}]{Emfietzoglou03}
\bibinfo{author}{\bibfnamefont{D.}~\bibnamefont{Emfietzoglou}},
  \bibinfo{journal}{Radiat. Phys. Chem.} \textbf{\bibinfo{volume}{66}},
  \bibinfo{pages}{373} (\bibinfo{year}{2003}).

\bibitem[{\citenamefont{Bernhardt and Paretzke}(2003)}]{Paretzke03}
\bibinfo{author}{\bibfnamefont{P.}~\bibnamefont{Bernhardt}} \bibnamefont{and}
  \bibinfo{author}{\bibfnamefont{H.~G.} \bibnamefont{Paretzke}},
  \bibinfo{journal}{Int. J. Mass Spectrom.}
  \textbf{\bibinfo{volume}{223-–224}}, \bibinfo{pages}{599}
  (\bibinfo{year}{2003}).

\bibitem[{\citenamefont{Peudon et~al.}(2006)\citenamefont{Peudon, Edel, and
  Terrisol}}]{Terrisol06}
\bibinfo{author}{\bibfnamefont{A.}~\bibnamefont{Peudon}},
  \bibinfo{author}{\bibfnamefont{S.}~\bibnamefont{Edel}}, \bibnamefont{and}
  \bibinfo{author}{\bibfnamefont{M.}~\bibnamefont{Terrisol}},
  \bibinfo{journal}{Radiat. Prot. Dosim.} \textbf{\bibinfo{volume}{122}},
  \bibinfo{pages}{128} (\bibinfo{year}{2006}).

\bibitem[{\citenamefont{Kim et~al.}(2004)}]{YKKim04}
\bibinfo{author}{\bibfnamefont{Y.-K.} \bibnamefont{Kim}} \bibnamefont{et~al.},
  \emph{\bibinfo{title}{Electron-impact ionization cross section for ionization
  and excitation database (version 3.0)}} (\bibinfo{year}{2004}),
  \urlprefix\url{http://www.nist.gov/pml/data/ ionization/index.cfm}.

\bibitem[{\citenamefont{White et~al.}(1992)\citenamefont{White, Griffith, and
  Wilson}}]{ICRU46}
\bibinfo{author}{\bibfnamefont{D.~R.} \bibnamefont{White}},
  \bibinfo{author}{\bibfnamefont{R.~V.} \bibnamefont{Griffith}},
  \bibnamefont{and} \bibinfo{author}{\bibfnamefont{I.~J.}
  \bibnamefont{Wilson}}, \emph{\bibinfo{title}{Photon, Electron, Proton and
  Neutron Interaction Data for Body Tissues}}
  (\bibinfo{publisher}{International Commission on Radiation Units and
  Measurements (ICRU 46)}, \bibinfo{address}{Bethesda, MD},
  \bibinfo{year}{1992}).

\bibitem[{\citenamefont{Wilson et~al.}(1984)\citenamefont{Wilson, Miller,
  Toburen, and Manson}}]{Wilson84}
\bibinfo{author}{\bibfnamefont{W.~E.} \bibnamefont{Wilson}},
  \bibinfo{author}{\bibfnamefont{J.~H.} \bibnamefont{Miller}},
  \bibinfo{author}{\bibfnamefont{L.~H.} \bibnamefont{Toburen}},
  \bibnamefont{and} \bibinfo{author}{\bibfnamefont{S.~T.}
  \bibnamefont{Manson}}, \bibinfo{journal}{J. Chem. Phys.}
  \textbf{\bibinfo{volume}{80}}, \bibinfo{pages}{5631} (\bibinfo{year}{1984}).

\bibitem[{\citenamefont{Rudd et~al.}(1985)\citenamefont{Rudd, Goffe, DuBois,
  and Toburen}}]{Rudd85}
\bibinfo{author}{\bibfnamefont{M.}~\bibnamefont{Rudd}},
  \bibinfo{author}{\bibfnamefont{T.}~\bibnamefont{Goffe}},
  \bibinfo{author}{\bibfnamefont{R.}~\bibnamefont{DuBois}}, \bibnamefont{and}
  \bibinfo{author}{\bibfnamefont{L.}~\bibnamefont{Toburen}},
  \bibinfo{journal}{Phys. Rev.} \textbf{\bibinfo{volume}{A31}},
  \bibinfo{pages}{492} (\bibinfo{year}{1985}).

\bibitem[{\citenamefont{Bolorizadeh and Rudd}(1986)}]{Rudd86}
\bibinfo{author}{\bibfnamefont{M.~A.} \bibnamefont{Bolorizadeh}}
  \bibnamefont{and} \bibinfo{author}{\bibfnamefont{M.~E.} \bibnamefont{Rudd}},
  \bibinfo{journal}{Phys. Rev. A} \textbf{\bibinfo{volume}{33}},
  \bibinfo{pages}{888} (\bibinfo{year}{1986}).

\bibitem[{\citenamefont{Iriki et~al.}(2011)\citenamefont{Iriki, Kikuchi, Imai,
  and Itoh}}]{Itoh11}
\bibinfo{author}{\bibfnamefont{Y.}~\bibnamefont{Iriki}},
  \bibinfo{author}{\bibfnamefont{Y.}~\bibnamefont{Kikuchi}},
  \bibinfo{author}{\bibfnamefont{M.}~\bibnamefont{Imai}}, \bibnamefont{and}
  \bibinfo{author}{\bibfnamefont{A.}~\bibnamefont{Itoh}},
  \bibinfo{journal}{Phys. Rev. A} \textbf{\bibinfo{volume}{84}},
  \bibinfo{pages}{052719} (\bibinfo{year}{2011}).

\bibitem[{\citenamefont{Simons}(2007)}]{Simons07}
\bibinfo{author}{\bibfnamefont{J.}~\bibnamefont{Simons}},
  \bibinfo{journal}{Adv. Quantum Chem.} \textbf{\bibinfo{volume}{52}},
  \bibinfo{pages}{171} (\bibinfo{year}{2007}).

\bibitem[{\citenamefont{Dingfelder et~al.}(2000)\citenamefont{Dingfelder,
  Inokuti, and Paretzke}}]{Dingfelder2000}
\bibinfo{author}{\bibfnamefont{M.}~\bibnamefont{Dingfelder}},
  \bibinfo{author}{\bibfnamefont{M.}~\bibnamefont{Inokuti}}, \bibnamefont{and}
  \bibinfo{author}{\bibfnamefont{H.}~\bibnamefont{Paretzke}},
  \bibinfo{journal}{Rad. Phys. Chem.} \textbf{\bibinfo{volume}{59}},
  \bibinfo{pages}{255} (\bibinfo{year}{2000}).

\bibitem[{\citenamefont{Garcia-Molina
  et~al.}(2012{\natexlab{b}})\citenamefont{Garcia-Molina, Abril, de~Vera,
  Kyriakou, and Emfietzoglou}}]{RafaIBCT2011}
\bibinfo{author}{\bibfnamefont{R.}~\bibnamefont{Garcia-Molina}},
  \bibinfo{author}{\bibfnamefont{I.}~\bibnamefont{Abril}},
  \bibinfo{author}{\bibfnamefont{P.}~\bibnamefont{de~Vera}},
  \bibinfo{author}{\bibfnamefont{I.}~\bibnamefont{Kyriakou}}, \bibnamefont{and}
  \bibinfo{author}{\bibfnamefont{D.}~\bibnamefont{Emfietzoglou}},
  \bibinfo{journal}{J. Phys.: Conf. Ser.} \textbf{\bibinfo{volume}{373}},
  \bibinfo{pages}{012015} (\bibinfo{year}{2012}{\natexlab{b}}).

\bibitem[{\citenamefont{Barkas}(1963)}]{Barkas63}
\bibinfo{author}{\bibfnamefont{W.~H.} \bibnamefont{Barkas}},
  \emph{\bibinfo{title}{Nuclear Research Emulsions I. Techniques and Theory}},
  vol.~\bibinfo{volume}{1} (\bibinfo{publisher}{Academic Press},
  \bibinfo{address}{New York, London}, \bibinfo{year}{1963}).

\bibitem[{\citenamefont{Schiwietz and Grande}(2001)}]{ChargeFluct}
\bibinfo{author}{\bibfnamefont{G.}~\bibnamefont{Schiwietz}} \bibnamefont{and}
  \bibinfo{author}{\bibfnamefont{P.~L.} \bibnamefont{Grande}},
  \bibinfo{journal}{Nucl. Instr. Meth. B} \textbf{\bibinfo{volume}{175-177}},
  \bibinfo{pages}{125} (\bibinfo{year}{2001}).

\bibitem[{\citenamefont{Kundrat}(2007)}]{Kundrat07}
\bibinfo{author}{\bibfnamefont{P.}~\bibnamefont{Kundrat}},
  \bibinfo{journal}{Phys. Med. Biol.} \textbf{\bibinfo{volume}{52}},
  \bibinfo{pages}{6813} (\bibinfo{year}{2007}).

\bibitem[{\citenamefont{Hollmark et~al.}(2004)\citenamefont{Hollmark, Uhrdin,
  Belkic, Gudowska, and Brahme}}]{Hollmark04}
\bibinfo{author}{\bibfnamefont{M.}~\bibnamefont{Hollmark}},
  \bibinfo{author}{\bibfnamefont{J.}~\bibnamefont{Uhrdin}},
  \bibinfo{author}{\bibfnamefont{D.}~\bibnamefont{Belkic}},
  \bibinfo{author}{\bibfnamefont{I.}~\bibnamefont{Gudowska}}, \bibnamefont{and}
  \bibinfo{author}{\bibfnamefont{A.}~\bibnamefont{Brahme}},
  \bibinfo{journal}{Phys. Med. Biol.} \textbf{\bibinfo{volume}{49}},
  \bibinfo{pages}{3247} (\bibinfo{year}{2004}).

\bibitem[{\citenamefont{Inokuti}(1971)}]{Inokuti}
\bibinfo{author}{\bibfnamefont{M.}~\bibnamefont{Inokuti}},
  \bibinfo{journal}{Rev. Mod. Phys.} \textbf{\bibinfo{volume}{43}},
  \bibinfo{pages}{297} (\bibinfo{year}{1971}).

\bibitem[{\citenamefont{Schmidt-B{\"o}cking
  et~al.}(2004)\citenamefont{Schmidt-B{\"o}cking, Schmidt, Weber, Mergel,
  Jagutzki, Czasch, Hagmann, Doerner, Demkov, Jahnke et~al.}}]{Doerner2004}
\bibinfo{author}{\bibfnamefont{H.}~\bibnamefont{Schmidt-B{\"o}cking}},
  \bibinfo{author}{\bibfnamefont{L.}~\bibnamefont{Schmidt}},
  \bibinfo{author}{\bibfnamefont{T.}~\bibnamefont{Weber}},
  \bibinfo{author}{\bibfnamefont{V.}~\bibnamefont{Mergel}},
  \bibinfo{author}{\bibfnamefont{O.}~\bibnamefont{Jagutzki}},
  \bibinfo{author}{\bibfnamefont{A.}~\bibnamefont{Czasch}},
  \bibinfo{author}{\bibfnamefont{S.}~\bibnamefont{Hagmann}},
  \bibinfo{author}{\bibfnamefont{R.}~\bibnamefont{Doerner}},
  \bibinfo{author}{\bibfnamefont{Y.}~\bibnamefont{Demkov}},
  \bibinfo{author}{\bibfnamefont{T.}~\bibnamefont{Jahnke}},
  \bibnamefont{et~al.}, \bibinfo{journal}{Radiat. Phys. Chem.}
  \textbf{\bibinfo{volume}{71}}, \bibinfo{pages}{627} (\bibinfo{year}{2004}).

\bibitem[{\citenamefont{Tung et~al.}(2007)\citenamefont{Tung, Chao, Hsieh, and
  Chan}}]{Tung}
\bibinfo{author}{\bibfnamefont{C.}~\bibnamefont{Tung}},
  \bibinfo{author}{\bibfnamefont{T.}~\bibnamefont{Chao}},
  \bibinfo{author}{\bibfnamefont{H.}~\bibnamefont{Hsieh}}, \bibnamefont{and}
  \bibinfo{author}{\bibfnamefont{W.}~\bibnamefont{Chan}},
  \bibinfo{journal}{Nucl. Inst. Meth B} \textbf{\bibinfo{volume}{262}},
  \bibinfo{pages}{231} (\bibinfo{year}{2007}).

\bibitem[{\citenamefont{Chandrasekhar}(1943)}]{chandra}
\bibinfo{author}{\bibfnamefont{S.}~\bibnamefont{Chandrasekhar}},
  \bibinfo{journal}{Rev. Mod. Phys.} \textbf{\bibinfo{volume}{15}},
  \bibinfo{pages}{1} (\bibinfo{year}{1943}).

\bibitem[{\citenamefont{Butts and Katz}(1967)}]{Katz67}
\bibinfo{author}{\bibfnamefont{J.~J.} \bibnamefont{Butts}} \bibnamefont{and}
  \bibinfo{author}{\bibfnamefont{R.}~\bibnamefont{Katz}},
  \bibinfo{journal}{Radiat. Res.} \textbf{\bibinfo{volume}{30}},
  \bibinfo{pages}{855} (\bibinfo{year}{1967}).

\bibitem[{\citenamefont{Katz et~al.}(1971)\citenamefont{Katz, Ackerson,
  Homayoonfar, and Sharma}}]{Katz71}
\bibinfo{author}{\bibfnamefont{R.}~\bibnamefont{Katz}},
  \bibinfo{author}{\bibfnamefont{B.}~\bibnamefont{Ackerson}},
  \bibinfo{author}{\bibfnamefont{M.}~\bibnamefont{Homayoonfar}},
  \bibnamefont{and} \bibinfo{author}{\bibfnamefont{S.~C.}
  \bibnamefont{Sharma}}, \bibinfo{journal}{Radiat. Res.}
  \textbf{\bibinfo{volume}{47}}, \bibinfo{pages}{402} (\bibinfo{year}{1971}).

\bibitem[{\citenamefont{Korcyl and Walig{\'o}rski}(2009)}]{KatzRBE}
\bibinfo{author}{\bibfnamefont{M.}~\bibnamefont{Korcyl}} \bibnamefont{and}
  \bibinfo{author}{\bibfnamefont{M.}~\bibnamefont{Walig{\'o}rski}},
  \bibinfo{journal}{Int. J. Radiat. Biol.} \textbf{\bibinfo{volume}{85}},
  \bibinfo{pages}{1101} (\bibinfo{year}{2009}).

\bibitem[{\citenamefont{Waligorski et~al.}(1986)\citenamefont{Waligorski, Hamm,
  and Katz}}]{31}
\bibinfo{author}{\bibfnamefont{M.}~\bibnamefont{Waligorski}},
  \bibinfo{author}{\bibfnamefont{R.}~\bibnamefont{Hamm}}, \bibnamefont{and}
  \bibinfo{author}{\bibfnamefont{R.}~\bibnamefont{Katz}},
  \bibinfo{journal}{Nucl. Tracks Radiat. Meas.} \textbf{\bibinfo{volume}{11}},
  \bibinfo{pages}{309} (\bibinfo{year}{1986}).

\bibitem[{\citenamefont{Cucinotta et~al.}(1998)\citenamefont{Cucinotta, Katz,
  and Wilson}}]{Cucinotta98}
\bibinfo{author}{\bibfnamefont{F.~A.} \bibnamefont{Cucinotta}},
  \bibinfo{author}{\bibfnamefont{R.}~\bibnamefont{Katz}}, \bibnamefont{and}
  \bibinfo{author}{\bibfnamefont{J.~W.} \bibnamefont{Wilson}},
  \bibinfo{journal}{Radiat. Environ. Biophys.} \textbf{\bibinfo{volume}{37}},
  \bibinfo{pages}{259} (\bibinfo{year}{1998}).

\bibitem[{\citenamefont{Plante and Cucinotta}(2010)}]{Plante}
\bibinfo{author}{\bibfnamefont{I.}~\bibnamefont{Plante}} \bibnamefont{and}
  \bibinfo{author}{\bibfnamefont{F.}~\bibnamefont{Cucinotta}},
  \bibinfo{journal}{Radiat. Environ. Biophys.} \textbf{\bibinfo{volume}{49}},
  \bibinfo{pages}{5} (\bibinfo{year}{2010}).

\bibitem[{\citenamefont{Ward}(1995)}]{Ward2}
\bibinfo{author}{\bibfnamefont{J.}~\bibnamefont{Ward}},
  \bibinfo{journal}{Radiat. Res.} \textbf{\bibinfo{volume}{142}},
  \bibinfo{pages}{362} (\bibinfo{year}{1995}).

\bibitem[{\citenamefont{Fabrikant et~al.}(2012)\citenamefont{Fabrikant,
  Caprasecca, Gallup, and Gorfinkiel}}]{Fabrikant12}
\bibinfo{author}{\bibfnamefont{I.~I.} \bibnamefont{Fabrikant}},
  \bibinfo{author}{\bibfnamefont{S.}~\bibnamefont{Caprasecca}},
  \bibinfo{author}{\bibfnamefont{G.~A.} \bibnamefont{Gallup}},
  \bibnamefont{and} \bibinfo{author}{\bibfnamefont{J.~D.}
  \bibnamefont{Gorfinkiel}}, \bibinfo{journal}{J. Chem. Phys.}
  \textbf{\bibinfo{volume}{136}}, \bibinfo{pages}{184301}
  (\bibinfo{year}{2012}).

\bibitem[{\citenamefont{Becker and Sevilla}(1993)}]{SevDQD}
\bibinfo{author}{\bibfnamefont{D.}~\bibnamefont{Becker}} \bibnamefont{and}
  \bibinfo{author}{\bibfnamefont{M.}~\bibnamefont{Sevilla}}, in
  \emph{\bibinfo{booktitle}{Advances in Radiation Biology}}, edited by
  \bibinfo{editor}{\bibfnamefont{J.}~\bibnamefont{Lett}}
  (\bibinfo{publisher}{Acad. Press}, \bibinfo{year}{1993}),
  vol.~\bibinfo{volume}{17}, pp. \bibinfo{pages}{121--180}.

\bibitem[{\citenamefont{Gianturco et~al.}(2008)\citenamefont{Gianturco,
  Sebastianelli, Lucchese, Baccarelli, and Sanna}}]{Gianturco1}
\bibinfo{author}{\bibfnamefont{F.~A.} \bibnamefont{Gianturco}},
  \bibinfo{author}{\bibfnamefont{F.}~\bibnamefont{Sebastianelli}},
  \bibinfo{author}{\bibfnamefont{R.~R.} \bibnamefont{Lucchese}},
  \bibinfo{author}{\bibfnamefont{I.}~\bibnamefont{Baccarelli}},
  \bibnamefont{and} \bibinfo{author}{\bibfnamefont{N.}~\bibnamefont{Sanna}},
  \bibinfo{journal}{J. Chem. Phys.} \textbf{\bibinfo{volume}{128}},
  \bibinfo{pages}{174302} (\bibinfo{year}{2008}).

\bibitem[{\citenamefont{Sanche}(2012)}]{SancheCH1}
\bibinfo{author}{\bibfnamefont{L.}~\bibnamefont{Sanche}}, in
  \emph{\bibinfo{booktitle}{Radiation Damage in Biomolecular Systems}}, edited
  by \bibinfo{editor}{\bibfnamefont{G.}~\bibnamefont{Garcia}} \bibnamefont{and}
  \bibinfo{editor}{\bibfnamefont{M.~C.} \bibnamefont{Fuss}}
  (\bibinfo{publisher}{Springer}, \bibinfo{year}{2012}).

\bibitem[{\citenamefont{Panajotovic et~al.}(2006)\citenamefont{Panajotovic,
  Martin, Cloutier, Hunting, and Sanche}}]{Sanche2006}
\bibinfo{author}{\bibfnamefont{R.}~\bibnamefont{Panajotovic}},
  \bibinfo{author}{\bibfnamefont{F.}~\bibnamefont{Martin}},
  \bibinfo{author}{\bibfnamefont{P.}~\bibnamefont{Cloutier}},
  \bibinfo{author}{\bibfnamefont{D.}~\bibnamefont{Hunting}}, \bibnamefont{and}
  \bibinfo{author}{\bibfnamefont{L.}~\bibnamefont{Sanche}},
  \bibinfo{journal}{Radiat. Res.} \textbf{\bibinfo{volume}{165}},
  \bibinfo{pages}{452} (\bibinfo{year}{2006}).

\bibitem[{\citenamefont{Mucke et~al.}(2010)\citenamefont{Mucke, Braune, Barth,
  F{\"o}rstel, Lischke, Ulrich, Arion, Becker, Bradshaw, and
  Hergenhahn}}]{Becker}
\bibinfo{author}{\bibfnamefont{M.}~\bibnamefont{Mucke}},
  \bibinfo{author}{\bibfnamefont{M.}~\bibnamefont{Braune}},
  \bibinfo{author}{\bibfnamefont{S.}~\bibnamefont{Barth}},
  \bibinfo{author}{\bibfnamefont{M.}~\bibnamefont{F{\"o}rstel}},
  \bibinfo{author}{\bibfnamefont{T.}~\bibnamefont{Lischke}},
  \bibinfo{author}{\bibfnamefont{V.}~\bibnamefont{Ulrich}},
  \bibinfo{author}{\bibfnamefont{T.}~\bibnamefont{Arion}},
  \bibinfo{author}{\bibfnamefont{U.}~\bibnamefont{Becker}},
  \bibinfo{author}{\bibfnamefont{A.}~\bibnamefont{Bradshaw}}, \bibnamefont{and}
  \bibinfo{author}{\bibfnamefont{U.}~\bibnamefont{Hergenhahn}},
  \bibinfo{journal}{Nat. Phys.} \textbf{\bibinfo{volume}{6}},
  \bibinfo{pages}{143} (\bibinfo{year}{2010}).

\bibitem[{\citenamefont{Bulanov et~al.}(2008)\citenamefont{Bulanov, Brantov,
  Bychenkov, Chvykov, Kalinchenko, Matsuoka, Rousseau, Reed, Yanovsky,
  Krushelnick et~al.}}]{Bulanov}
\bibinfo{author}{\bibfnamefont{S.~S.} \bibnamefont{Bulanov}},
  \bibinfo{author}{\bibfnamefont{A.}~\bibnamefont{Brantov}},
  \bibinfo{author}{\bibfnamefont{V.~Y.} \bibnamefont{Bychenkov}},
  \bibinfo{author}{\bibfnamefont{V.}~\bibnamefont{Chvykov}},
  \bibinfo{author}{\bibfnamefont{G.}~\bibnamefont{Kalinchenko}},
  \bibinfo{author}{\bibfnamefont{T.}~\bibnamefont{Matsuoka}},
  \bibinfo{author}{\bibfnamefont{P.}~\bibnamefont{Rousseau}},
  \bibinfo{author}{\bibfnamefont{S.}~\bibnamefont{Reed}},
  \bibinfo{author}{\bibfnamefont{V.}~\bibnamefont{Yanovsky}},
  \bibinfo{author}{\bibfnamefont{K.}~\bibnamefont{Krushelnick}},
  \bibnamefont{et~al.}, \bibinfo{journal}{Med. Phys.}
  \textbf{\bibinfo{volume}{35}}, \bibinfo{pages}{1770} (\bibinfo{year}{2008}).

\bibitem[{\citenamefont{Adamcik et~al.}(2012)\citenamefont{Adamcik, Jeon,
  Karczewski, Metzler, and Dietler}}]{plasmid}
\bibinfo{author}{\bibfnamefont{J.}~\bibnamefont{Adamcik}},
  \bibinfo{author}{\bibfnamefont{J.-H.} \bibnamefont{Jeon}},
  \bibinfo{author}{\bibfnamefont{K.~J.} \bibnamefont{Karczewski}},
  \bibinfo{author}{\bibfnamefont{R.}~\bibnamefont{Metzler}}, \bibnamefont{and}
  \bibinfo{author}{\bibfnamefont{G.}~\bibnamefont{Dietler}},
  \bibinfo{journal}{Soft Matt.} \textbf{\bibinfo{volume}{8}},
  \bibinfo{pages}{8651} (\bibinfo{year}{2012}).

\bibitem[{\citenamefont{Dang et~al.}(2011)\citenamefont{Dang, Goethem, Graaf,
  Brandenburg, Hoekstra, and Schlath{\"o}lter}}]{Thomas1}
\bibinfo{author}{\bibfnamefont{H.~M.} \bibnamefont{Dang}},
  \bibinfo{author}{\bibfnamefont{M.~J.~V.} \bibnamefont{Goethem}},
  \bibinfo{author}{\bibfnamefont{E.~R. V.~D.} \bibnamefont{Graaf}},
  \bibinfo{author}{\bibfnamefont{S.}~\bibnamefont{Brandenburg}},
  \bibinfo{author}{\bibfnamefont{R.}~\bibnamefont{Hoekstra}}, \bibnamefont{and}
  \bibinfo{author}{\bibfnamefont{T.}~\bibnamefont{Schlath{\"o}lter}},
  \bibinfo{journal}{Eur. Phys. J. D} \textbf{\bibinfo{volume}{63}},
  \bibinfo{pages}{359} (\bibinfo{year}{2011}).

\bibitem[{\citenamefont{Sevilla et~al.}(1997)\citenamefont{Sevilla, Becker, and
  Razskazovskii}}]{sev97}
\bibinfo{author}{\bibfnamefont{M.}~\bibnamefont{Sevilla}},
  \bibinfo{author}{\bibfnamefont{D.}~\bibnamefont{Becker}}, \bibnamefont{and}
  \bibinfo{author}{\bibfnamefont{Y.}~\bibnamefont{Razskazovskii}},
  \bibinfo{journal}{Nucleonika} \textbf{\bibinfo{volume}{42}},
  \bibinfo{pages}{283} (\bibinfo{year}{1997}).

\bibitem[{\citenamefont{Niklas et~al.}(2013)\citenamefont{Niklas, Abdollahi,
  Akselrod, Debus, J{\"a}kel, and Greilich}}]{Niklas}
\bibinfo{author}{\bibfnamefont{M.}~\bibnamefont{Niklas}},
  \bibinfo{author}{\bibfnamefont{A.}~\bibnamefont{Abdollahi}},
  \bibinfo{author}{\bibfnamefont{M.}~\bibnamefont{Akselrod}},
  \bibinfo{author}{\bibfnamefont{J.}~\bibnamefont{Debus}},
  \bibinfo{author}{\bibfnamefont{O.}~\bibnamefont{J{\"a}kel}},
  \bibnamefont{and} \bibinfo{author}{\bibfnamefont{S.}~\bibnamefont{Greilich}},
  \bibinfo{journal}{Int. J. Radiat. Oncol.} p. \bibinfo{pages}{in press}
  (\bibinfo{year}{2013}).

\bibitem[{\citenamefont{de~Jiang et~al.}(2010)\citenamefont{de~Jiang, Shen, and
  Piao}}]{A549}
\bibinfo{author}{\bibfnamefont{R.}~\bibnamefont{de~Jiang}},
  \bibinfo{author}{\bibfnamefont{H.}~\bibnamefont{Shen}}, \bibnamefont{and}
  \bibinfo{author}{\bibfnamefont{Y.-J.} \bibnamefont{Piao}},
  \bibinfo{journal}{Roman. J. Morphol. Embryol.}
  \textbf{\bibinfo{volume}{51(4)}}, \bibinfo{pages}{663–}
  (\bibinfo{year}{2010}).

\bibitem[{\citenamefont{Toulemonde et~al.}(2000)\citenamefont{Toulemonde,
  Dufour, Meftah, and Paumier}}]{Toul01}
\bibinfo{author}{\bibfnamefont{M.}~\bibnamefont{Toulemonde}},
  \bibinfo{author}{\bibfnamefont{C.}~\bibnamefont{Dufour}},
  \bibinfo{author}{\bibfnamefont{A.}~\bibnamefont{Meftah}}, \bibnamefont{and}
  \bibinfo{author}{\bibfnamefont{E.}~\bibnamefont{Paumier}},
  \bibinfo{journal}{Nucl. Inst. Meth. B} \textbf{\bibinfo{volume}{166-167}},
  \bibinfo{pages}{903} (\bibinfo{year}{2000}).

\bibitem[{\citenamefont{Toulemonde et~al.}(2004)\citenamefont{Toulemonde,
  Trautmann, Balanzat, Hjort, and Weidinger}}]{Toul02}
\bibinfo{author}{\bibfnamefont{M.}~\bibnamefont{Toulemonde}},
  \bibinfo{author}{\bibfnamefont{C.}~\bibnamefont{Trautmann}},
  \bibinfo{author}{\bibfnamefont{E.}~\bibnamefont{Balanzat}},
  \bibinfo{author}{\bibfnamefont{K.}~\bibnamefont{Hjort}}, \bibnamefont{and}
  \bibinfo{author}{\bibfnamefont{A.}~\bibnamefont{Weidinger}},
  \bibinfo{journal}{Nucl. Inst. Meth. B} \textbf{\bibinfo{volume}{216}},
  \bibinfo{pages}{1} (\bibinfo{year}{2004}).

\bibitem[{\citenamefont{Skupinski et~al.}(2005)\citenamefont{Skupinski,
  Toulemonde, Lindeberg, and Hjort}}]{Toul03}
\bibinfo{author}{\bibfnamefont{M.}~\bibnamefont{Skupinski}},
  \bibinfo{author}{\bibfnamefont{M.}~\bibnamefont{Toulemonde}},
  \bibinfo{author}{\bibfnamefont{M.}~\bibnamefont{Lindeberg}},
  \bibnamefont{and} \bibinfo{author}{\bibfnamefont{K.}~\bibnamefont{Hjort}},
  \bibinfo{journal}{Nucl. Inst. Meth. B} \textbf{\bibinfo{volume}{240}},
  \bibinfo{pages}{681} (\bibinfo{year}{2005}).

\bibitem[{\citenamefont{Pawlak et~al.}(1999)\citenamefont{Pawlak, Dufour,
  Laurent, Paumier, Perri{\`e}re, Stoquert, and Toulemonde}}]{Toul04}
\bibinfo{author}{\bibfnamefont{F.}~\bibnamefont{Pawlak}},
  \bibinfo{author}{\bibfnamefont{C.}~\bibnamefont{Dufour}},
  \bibinfo{author}{\bibfnamefont{A.}~\bibnamefont{Laurent}},
  \bibinfo{author}{\bibfnamefont{E.}~\bibnamefont{Paumier}},
  \bibinfo{author}{\bibfnamefont{J.}~\bibnamefont{Perri{\`e}re}},
  \bibinfo{author}{\bibfnamefont{J.~P.} \bibnamefont{Stoquert}},
  \bibnamefont{and}
  \bibinfo{author}{\bibfnamefont{M.}~\bibnamefont{Toulemonde}},
  \bibinfo{journal}{Nucl. Inst. Meth. B} \textbf{\bibinfo{volume}{151}},
  \bibinfo{pages}{140} (\bibinfo{year}{1999}).

\bibitem[{\citenamefont{Toulemonde et~al.}(2006)\citenamefont{Toulemonde,
  Assmann, Dufour, Meftah, Studer, and Trautmann}}]{Toul05}
\bibinfo{author}{\bibfnamefont{M.}~\bibnamefont{Toulemonde}},
  \bibinfo{author}{\bibfnamefont{W.}~\bibnamefont{Assmann}},
  \bibinfo{author}{\bibfnamefont{C.}~\bibnamefont{Dufour}},
  \bibinfo{author}{\bibfnamefont{A.}~\bibnamefont{Meftah}},
  \bibinfo{author}{\bibfnamefont{F.}~\bibnamefont{Studer}}, \bibnamefont{and}
  \bibinfo{author}{\bibfnamefont{C.}~\bibnamefont{Trautmann}},
  \bibinfo{journal}{Mat. Fys. Medd.} \textbf{\bibinfo{volume}{52}},
  \bibinfo{pages}{263} (\bibinfo{year}{2006}).

\bibitem[{\citenamefont{Dammak et~al.}(1993)\citenamefont{Dammak, Lesueur,
  Dunlop, Legrand, and Morillo}}]{20}
\bibinfo{author}{\bibfnamefont{H.}~\bibnamefont{Dammak}},
  \bibinfo{author}{\bibfnamefont{D.}~\bibnamefont{Lesueur}},
  \bibinfo{author}{\bibfnamefont{A.}~\bibnamefont{Dunlop}},
  \bibinfo{author}{\bibfnamefont{P.}~\bibnamefont{Legrand}}, \bibnamefont{and}
  \bibinfo{author}{\bibfnamefont{J.}~\bibnamefont{Morillo}},
  \bibinfo{journal}{Radiat. Eff. Defect. Sol.} \textbf{\bibinfo{volume}{126}},
  \bibinfo{pages}{111} (\bibinfo{year}{1993}).

\bibitem[{\citenamefont{Mieskes et~al.}(2003)\citenamefont{Mieskes, Assmann,
  Gr{\"u}ner, Kucal, Wang, and Toulemonde}}]{22}
\bibinfo{author}{\bibfnamefont{H.~D.} \bibnamefont{Mieskes}},
  \bibinfo{author}{\bibfnamefont{W.}~\bibnamefont{Assmann}},
  \bibinfo{author}{\bibfnamefont{F.}~\bibnamefont{Gr{\"u}ner}},
  \bibinfo{author}{\bibfnamefont{H.}~\bibnamefont{Kucal}},
  \bibinfo{author}{\bibfnamefont{Z.~G.} \bibnamefont{Wang}}, \bibnamefont{and}
  \bibinfo{author}{\bibfnamefont{M.}~\bibnamefont{Toulemonde}},
  \bibinfo{journal}{Phys. Rev. B} \textbf{\bibinfo{volume}{67}},
  \bibinfo{pages}{155414} (\bibinfo{year}{2003}).

\bibitem[{\citenamefont{Meftah et~al.}(1998)\citenamefont{Meftah, Djebara,
  Khalfaoui, and Toulemonde}}]{24}
\bibinfo{author}{\bibfnamefont{A.}~\bibnamefont{Meftah}},
  \bibinfo{author}{\bibfnamefont{M.}~\bibnamefont{Djebara}},
  \bibinfo{author}{\bibfnamefont{N.}~\bibnamefont{Khalfaoui}},
  \bibnamefont{and}
  \bibinfo{author}{\bibfnamefont{M.}~\bibnamefont{Toulemonde}},
  \bibinfo{journal}{Nucl. Instr. Meth. B} \textbf{\bibinfo{volume}{146}},
  \bibinfo{pages}{431} (\bibinfo{year}{1998}).

\bibitem[{\citenamefont{Katin et~al.}(1987)\citenamefont{Katin, Martinenko, and
  Yavlinskii}}]{26}
\bibinfo{author}{\bibfnamefont{V.}~\bibnamefont{Katin}},
  \bibinfo{author}{\bibfnamefont{Y.}~\bibnamefont{Martinenko}},
  \bibnamefont{and}
  \bibinfo{author}{\bibfnamefont{Y.}~\bibnamefont{Yavlinskii}},
  \bibinfo{journal}{Sov. Techn. Phys. Lett.} \textbf{\bibinfo{volume}{13}},
  \bibinfo{pages}{276} (\bibinfo{year}{1987}).

\bibitem[{\citenamefont{Toulemonde et~al.}(2002)\citenamefont{Toulemonde,
  Assmann, Trautmann, and Gr{\"u}ner}}]{27}
\bibinfo{author}{\bibfnamefont{M.}~\bibnamefont{Toulemonde}},
  \bibinfo{author}{\bibfnamefont{W.}~\bibnamefont{Assmann}},
  \bibinfo{author}{\bibfnamefont{C.}~\bibnamefont{Trautmann}},
  \bibnamefont{and}
  \bibinfo{author}{\bibfnamefont{F.}~\bibnamefont{Gr{\"u}ner}},
  \bibinfo{journal}{Phys. Rev. Lett.} \textbf{\bibinfo{volume}{88}},
  \bibinfo{pages}{057602} (\bibinfo{year}{2002}).

\bibitem[{\citenamefont{Meftah et~al.}(1993)\citenamefont{Meftah, Brisard,
  Costantini, Hage-Ali, Stoquert, Studer, and Toulemonde}}]{28}
\bibinfo{author}{\bibfnamefont{A.}~\bibnamefont{Meftah}},
  \bibinfo{author}{\bibfnamefont{F.}~\bibnamefont{Brisard}},
  \bibinfo{author}{\bibfnamefont{J.}~\bibnamefont{Costantini}},
  \bibinfo{author}{\bibfnamefont{M.}~\bibnamefont{Hage-Ali}},
  \bibinfo{author}{\bibfnamefont{J.}~\bibnamefont{Stoquert}},
  \bibinfo{author}{\bibfnamefont{F.}~\bibnamefont{Studer}}, \bibnamefont{and}
  \bibinfo{author}{\bibfnamefont{M.}~\bibnamefont{Toulemonde}},
  \bibinfo{journal}{Phys. Rev.} \textbf{\bibinfo{volume}{B48}},
  \bibinfo{pages}{920} (\bibinfo{year}{1993}).

\bibitem[{\citenamefont{Yakubovich et~al.}(2011)\citenamefont{Yakubovich,
  Surdutovich, and Solov'yov}}]{Vilnius}
\bibinfo{author}{\bibfnamefont{A.~V.} \bibnamefont{Yakubovich}},
  \bibinfo{author}{\bibfnamefont{E.}~\bibnamefont{Surdutovich}},
  \bibnamefont{and} \bibinfo{author}{\bibfnamefont{A.~V.}
  \bibnamefont{Solov'yov}}, \bibinfo{journal}{AIP Conf. Proc.}
  \textbf{\bibinfo{volume}{1344}}, \bibinfo{pages}{230} (\bibinfo{year}{2011}).

\bibitem[{\citenamefont{Yakubovich et~al.}(2012)\citenamefont{Yakubovich,
  Surdutovich, and Solov'yov}}]{nimbnuke}
\bibinfo{author}{\bibfnamefont{A.~V.} \bibnamefont{Yakubovich}},
  \bibinfo{author}{\bibfnamefont{E.}~\bibnamefont{Surdutovich}},
  \bibnamefont{and} \bibinfo{author}{\bibfnamefont{A.~V.}
  \bibnamefont{Solov'yov}}, \bibinfo{journal}{Nucl. Instr. Meth. B}
  \textbf{\bibinfo{volume}{279}}, \bibinfo{pages}{135} (\bibinfo{year}{2012}).

\bibitem[{\citenamefont{Landau and Lifshitz}(1987)}]{LL6}
\bibinfo{author}{\bibfnamefont{L.}~\bibnamefont{Landau}} \bibnamefont{and}
  \bibinfo{author}{\bibfnamefont{E.}~\bibnamefont{Lifshitz}},
  \emph{\bibinfo{title}{Fluid dynamics, Second Edition: Volume 6}}
  (\bibinfo{publisher}{Reed-Elsevier}, \bibinfo{address}{Oxford, Boston,
  Johannesburg}, \bibinfo{year}{1987}).

\bibitem[{\citenamefont{Zeldovich and Raiser}(1966)}]{Zeldovich}
\bibinfo{author}{\bibfnamefont{Y.}~\bibnamefont{Zeldovich}} \bibnamefont{and}
  \bibinfo{author}{\bibfnamefont{Y.}~\bibnamefont{Raiser}},
  \emph{\bibinfo{title}{Physics of Shock Waves and High-Temperature
  Hydrodynamic Phenomena (Volume 1)}} (\bibinfo{publisher}{Oxford},
  \bibinfo{address}{New York}, \bibinfo{year}{1966}).

\bibitem[{\citenamefont{Chernyj}(1994)}]{Chernyj}
\bibinfo{author}{\bibfnamefont{G.}~\bibnamefont{Chernyj}},
  \emph{\bibinfo{title}{Gas dynamics}} (\bibinfo{publisher}{Nauka},
  \bibinfo{address}{Moscow}, \bibinfo{year}{1994}).

\bibitem[{\citenamefont{Solov'yov et~al.}(2013)\citenamefont{Solov'yov,
  Yakubovich, Surdutovich, and Solov'yov}}]{membraneilia}
\bibinfo{author}{\bibfnamefont{I.~A.} \bibnamefont{Solov'yov}},
  \bibinfo{author}{\bibfnamefont{A.~V.} \bibnamefont{Yakubovich}},
  \bibinfo{author}{\bibfnamefont{E.}~\bibnamefont{Surdutovich}},
  \bibnamefont{and} \bibinfo{author}{\bibfnamefont{A.~V.}
  \bibnamefont{Solov'yov}}, \bibinfo{journal}{Unpublished}
  (\bibinfo{year}{2013}).

\bibitem[{\citenamefont{Range et~al.}(2004)\citenamefont{Range, McGrath, Lopez,
  and York}}]{Range}
\bibinfo{author}{\bibfnamefont{K.}~\bibnamefont{Range}},
  \bibinfo{author}{\bibfnamefont{M.~J.} \bibnamefont{McGrath}},
  \bibinfo{author}{\bibfnamefont{X.}~\bibnamefont{Lopez}}, \bibnamefont{and}
  \bibinfo{author}{\bibfnamefont{D.~M.} \bibnamefont{York}},
  \bibinfo{journal}{J. Am. Chem. Soc.} \textbf{\bibinfo{volume}{126}},
  \bibinfo{pages}{1654} (\bibinfo{year}{2004}).

\bibitem[{\citenamefont{LaVerne}(1989)}]{laverne89}
\bibinfo{author}{\bibfnamefont{J.}~\bibnamefont{LaVerne}},
  \bibinfo{journal}{Radiat. Phys. Chem.} \textbf{\bibinfo{volume}{34}},
  \bibinfo{pages}{135} (\bibinfo{year}{1989}).

\bibitem[{\citenamefont{Jakob et~al.}(2003)\citenamefont{Jakob, Scholz, and
  Taucher-Scholz}}]{Jakob}
\bibinfo{author}{\bibfnamefont{B.}~\bibnamefont{Jakob}},
  \bibinfo{author}{\bibfnamefont{M.}~\bibnamefont{Scholz}}, \bibnamefont{and}
  \bibinfo{author}{\bibfnamefont{G.}~\bibnamefont{Taucher-Scholz}},
  \bibinfo{journal}{Radiat. Res.} \textbf{\bibinfo{volume}{159}},
  \bibinfo{pages}{676} (\bibinfo{year}{2003}).

\bibitem[{\citenamefont{Tobias et~al.}(2010)\citenamefont{Tobias, Durante,
  Taucher-Scholz, and Jakob}}]{Jakob1}
\bibinfo{author}{\bibfnamefont{F.}~\bibnamefont{Tobias}},
  \bibinfo{author}{\bibfnamefont{M.}~\bibnamefont{Durante}},
  \bibinfo{author}{\bibfnamefont{G.}~\bibnamefont{Taucher-Scholz}},
  \bibnamefont{and} \bibinfo{author}{\bibfnamefont{B.}~\bibnamefont{Jakob}},
  \bibinfo{journal}{Mutat. Res.} \textbf{\bibinfo{volume}{704}},
  \bibinfo{pages}{54} (\bibinfo{year}{2010}).

\bibitem[{\citenamefont{Roos and Kaina}(2006)}]{apoptosis06}
\bibinfo{author}{\bibfnamefont{W.~P.} \bibnamefont{Roos}} \bibnamefont{and}
  \bibinfo{author}{\bibfnamefont{B.}~\bibnamefont{Kaina}},
  \bibinfo{journal}{Trends Mol. Med.} \textbf{\bibinfo{volume}{12}},
  \bibinfo{pages}{440} (\bibinfo{year}{2006}).

\bibitem[{\citenamefont{Ward}(1988)}]{Ward1}
\bibinfo{author}{\bibfnamefont{J.}~\bibnamefont{Ward}}, \bibinfo{journal}{Prog.
  Nucleic Acid. Res. Mol. biol.} \textbf{\bibinfo{volume}{35}},
  \bibinfo{pages}{95} (\bibinfo{year}{1988}).

\bibitem[{\citenamefont{Goodhead}(1994)}]{Goodhead94}
\bibinfo{author}{\bibfnamefont{D.~T.} \bibnamefont{Goodhead}},
  \bibinfo{journal}{Int. J. Radiat. Biol.} \textbf{\bibinfo{volume}{65}},
  \bibinfo{pages}{7} (\bibinfo{year}{1994}).

\bibitem[{\citenamefont{Malyarchuk et~al.}(2009)\citenamefont{Malyarchuk,
  Castore, and Harrison}}]{Lynn1}
\bibinfo{author}{\bibfnamefont{S.}~\bibnamefont{Malyarchuk}},
  \bibinfo{author}{\bibfnamefont{R.}~\bibnamefont{Castore}}, \bibnamefont{and}
  \bibinfo{author}{\bibfnamefont{L.}~\bibnamefont{Harrison}},
  \bibinfo{journal}{DNA Repair} \textbf{\bibinfo{volume}{8}},
  \bibinfo{pages}{1343} (\bibinfo{year}{2009}).

\bibitem[{\citenamefont{Malyarchuk et~al.}(2008)\citenamefont{Malyarchuk,
  Castore, and Harrison}}]{Lynn2}
\bibinfo{author}{\bibfnamefont{S.}~\bibnamefont{Malyarchuk}},
  \bibinfo{author}{\bibfnamefont{R.}~\bibnamefont{Castore}}, \bibnamefont{and}
  \bibinfo{author}{\bibfnamefont{L.}~\bibnamefont{Harrison}},
  \bibinfo{journal}{Nucleic Acids Res.} \textbf{\bibinfo{volume}{36}},
  \bibinfo{pages}{4872} (\bibinfo{year}{2008}).

\bibitem[{\citenamefont{Sage and Harrison}(2011)}]{Lynn11}
\bibinfo{author}{\bibfnamefont{E.}~\bibnamefont{Sage}} \bibnamefont{and}
  \bibinfo{author}{\bibfnamefont{L.}~\bibnamefont{Harrison}},
  \bibinfo{journal}{Mutat. Res.} \textbf{\bibinfo{volume}{711}},
  \bibinfo{pages}{123} (\bibinfo{year}{2011}).

\bibitem[{\citenamefont{Heuskin et~al.}(2013)\citenamefont{Heuskin, Michiels,
  and Lucas}}]{Heuskin}
\bibinfo{author}{\bibfnamefont{A.~C.} \bibnamefont{Heuskin}},
  \bibinfo{author}{\bibfnamefont{C.}~\bibnamefont{Michiels}}, \bibnamefont{and}
  \bibinfo{author}{\bibfnamefont{S.}~\bibnamefont{Lucas}},
  \bibinfo{journal}{Phys. Med. Biol.} \textbf{\bibinfo{volume}{58}},
  \bibinfo{pages}{6495–} (\bibinfo{year}{2013}).

\bibitem[{\citenamefont{Scholz et~al.}(1997)\citenamefont{Scholz, Kellerer,
  Kraft-Weyrather, and Kraft}}]{Elsaesser}
\bibinfo{author}{\bibfnamefont{M.}~\bibnamefont{Scholz}},
  \bibinfo{author}{\bibfnamefont{A.}~\bibnamefont{Kellerer}},
  \bibinfo{author}{\bibfnamefont{W.}~\bibnamefont{Kraft-Weyrather}},
  \bibnamefont{and} \bibinfo{author}{\bibfnamefont{G.}~\bibnamefont{Kraft}},
  \bibinfo{journal}{Radiat. Environ. Biophys.} \textbf{\bibinfo{volume}{36}},
  \bibinfo{pages}{59} (\bibinfo{year}{1997}).

\bibitem[{\citenamefont{Weyrather et~al.}(1999)\citenamefont{Weyrather, Ritter,
  Scholz, and Kraft}}]{Weyrather99}
\bibinfo{author}{\bibfnamefont{W.~K.} \bibnamefont{Weyrather}},
  \bibinfo{author}{\bibfnamefont{S.}~\bibnamefont{Ritter}},
  \bibinfo{author}{\bibfnamefont{M.}~\bibnamefont{Scholz}}, \bibnamefont{and}
  \bibinfo{author}{\bibfnamefont{G.}~\bibnamefont{Kraft}},
  \bibinfo{journal}{Int. J. Rad. Biol.} \textbf{\bibinfo{volume}{75}},
  \bibinfo{pages}{1357} (\bibinfo{year}{1999}).

\bibitem[{\citenamefont{Hawkins}(1996)}]{MKModel}
\bibinfo{author}{\bibfnamefont{R.}~\bibnamefont{Hawkins}},
  \bibinfo{journal}{Int. J. Radiat. Biol.} \textbf{\bibinfo{volume}{69}},
  \bibinfo{pages}{739} (\bibinfo{year}{1996}).

\bibitem[{\citenamefont{Hawkins}(2009)}]{MKModel09}
\bibinfo{author}{\bibfnamefont{R.}~\bibnamefont{Hawkins}},
  \bibinfo{journal}{Radiat. Res.} \textbf{\bibinfo{volume}{172}},
  \bibinfo{pages}{761} (\bibinfo{year}{2009}).

\bibitem[{\citenamefont{Frese et~al.}(2012)\citenamefont{Frese, Yu, Stewart,
  and Carlson}}]{RobertMech}
\bibinfo{author}{\bibfnamefont{M.~C.} \bibnamefont{Frese}},
  \bibinfo{author}{\bibfnamefont{V.~K.} \bibnamefont{Yu}},
  \bibinfo{author}{\bibfnamefont{R.~D.} \bibnamefont{Stewart}},
  \bibnamefont{and} \bibinfo{author}{\bibfnamefont{D.~J.}
  \bibnamefont{Carlson}}, \bibinfo{journal}{Int. J. Radiat. Oncol.}
  \textbf{\bibinfo{volume}{83}}, \bibinfo{pages}{442–} (\bibinfo{year}{2012}).

\bibitem[{\citenamefont{Goodhead et~al.}(1993)\citenamefont{Goodhead, Thacker,
  and Cox}}]{Goodhead93}
\bibinfo{author}{\bibfnamefont{D.}~\bibnamefont{Goodhead}},
  \bibinfo{author}{\bibfnamefont{J.}~\bibnamefont{Thacker}}, \bibnamefont{and}
  \bibinfo{author}{\bibfnamefont{R.}~\bibnamefont{Cox}}, \bibinfo{journal}{Int.
  J. Radiat. Biol.} \textbf{\bibinfo{volume}{63}}, \bibinfo{pages}{543}
  (\bibinfo{year}{1993}).

\bibitem[{\citenamefont{Nikjoo et~al.}(2002)\citenamefont{Nikjoo, Bolton,
  Watanabe, Terrisol, P.O'Neill, and Goodhead}}]{DNA5}
\bibinfo{author}{\bibfnamefont{H.}~\bibnamefont{Nikjoo}},
  \bibinfo{author}{\bibfnamefont{C.}~\bibnamefont{Bolton}},
  \bibinfo{author}{\bibfnamefont{R.}~\bibnamefont{Watanabe}},
  \bibinfo{author}{\bibfnamefont{M.}~\bibnamefont{Terrisol}},
  \bibinfo{author}{\bibnamefont{P.O'Neill}}, \bibnamefont{and}
  \bibinfo{author}{\bibfnamefont{D.}~\bibnamefont{Goodhead}},
  \bibinfo{journal}{Radiat. Prot. Dosim.} \textbf{\bibinfo{volume}{99}},
  \bibinfo{pages}{77} (\bibinfo{year}{2002}).

\bibitem[{\citenamefont{Goodhead}(2006)}]{Goodhead06}
\bibinfo{author}{\bibfnamefont{D.}~\bibnamefont{Goodhead}},
  \bibinfo{journal}{Radiat. Prot. Dosim.} \textbf{\bibinfo{volume}{122}},
  \bibinfo{pages}{3} (\bibinfo{year}{2006}).

\bibitem[{\citenamefont{Cucinotta et~al.}(1999)\citenamefont{Cucinotta, Nikjoo,
  and Goodhead}}]{Cucinotta99}
\bibinfo{author}{\bibfnamefont{F.}~\bibnamefont{Cucinotta}},
  \bibinfo{author}{\bibfnamefont{H.}~\bibnamefont{Nikjoo}}, \bibnamefont{and}
  \bibinfo{author}{\bibfnamefont{D.}~\bibnamefont{Goodhead}},
  \bibinfo{journal}{Radiat. Environ. Biophys.} \textbf{\bibinfo{volume}{38}},
  \bibinfo{pages}{81} (\bibinfo{year}{1999}).

\bibitem[{\citenamefont{Scholz and Kraft}(1996)}]{LEM96}
\bibinfo{author}{\bibfnamefont{M.}~\bibnamefont{Scholz}} \bibnamefont{and}
  \bibinfo{author}{\bibfnamefont{G.}~\bibnamefont{Kraft}},
  \bibinfo{journal}{Adv. Space Res.} \textbf{\bibinfo{volume}{18}},
  \bibinfo{pages}{5–} (\bibinfo{year}{1996}).

\bibitem[{\citenamefont{Zheng et~al.}(2008)\citenamefont{Zheng, Hunting,
  Ayotte, and Sanche}}]{SancheGNP}
\bibinfo{author}{\bibfnamefont{Y.}~\bibnamefont{Zheng}},
  \bibinfo{author}{\bibfnamefont{D.~J.} \bibnamefont{Hunting}},
  \bibinfo{author}{\bibfnamefont{P.}~\bibnamefont{Ayotte}}, \bibnamefont{and}
  \bibinfo{author}{\bibfnamefont{L.}~\bibnamefont{Sanche}},
  \bibinfo{journal}{Radiat. Res.} \textbf{\bibinfo{volume}{169}},
  \bibinfo{pages}{19} (\bibinfo{year}{2008}).

\bibitem[{\citenamefont{Depken and Schiessel}(2009)}]{nucleosome}
\bibinfo{author}{\bibfnamefont{M.}~\bibnamefont{Depken}} \bibnamefont{and}
  \bibinfo{author}{\bibfnamefont{H.}~\bibnamefont{Schiessel}},
  \bibinfo{journal}{Biophys. J.} \textbf{\bibinfo{volume}{96}},
  \bibinfo{pages}{777} (\bibinfo{year}{2009}).

\end{thebibliography}

\end{document}